\documentclass{article}

\PassOptionsToPackage{numbers, compress}{natbib}

\usepackage[eandd,final]{neurips_2026}
\usepackage{graphicx}
\usepackage{longtable}
\usepackage{booktabs}
\usepackage{tabularx}
\usepackage{multirow}
\usepackage{caption}
\usepackage{array}
\usepackage{amsmath}
\usepackage{pdflscape}
\usepackage{placeins}
\usepackage{float}
\usepackage{collcell}
\usepackage{threeparttable}
\usepackage{booktabs}
\usepackage{tcolorbox}
\usepackage{inconsolata}
\usepackage{dirtree}
\usepackage{enumitem}

\usepackage[utf8]{inputenc} % allow utf-8 input
\usepackage[T1]{fontenc}    % use 8-bit T1 fonts
\usepackage{hyperref}       % hyperlinks
\usepackage{url}            % simple URL typesetting
\usepackage{booktabs}       % professional-quality tables
\usepackage{amsfonts}       % blackboard math symbols
\usepackage{nicefrac}       % compact symbols for 1/2, etc.
\usepackage{microtype}      % microtypography
\usepackage{xcolor}         % colors

\newcommand{\modeldesc}[2]{\smallskip\noindent\textbf{#1:} #2\par}

\title{EmoPhone: A Multi-Wave Dataset for In-the-Wild Mobile and Wearable Affect Sensing}

\author{%
  \textbf{Panyu Zhang$^{1}$\thanks{These authors contributed equally.}, Minseo Park$^{1}$\footnotemark[1], Soowon Kang$^{2}$, Ismatzoda Tomiris$^{1}$} \\
  \textbf{Azizbek Mustafakulov$^{3,7}$, Otabek Najimov$^{3,7}$, Woohyeok Choi$^{4}$} \\
  \textbf{Jumabek Alikhanov$^{5,7}$, Surjya Ghosh$^{6}$, Uichin Lee$^{1}$\thanks{Corresponding author.}} \\[0.3em]
  {\normalfont $^{1}$KAIST \quad $^{2}$Samsung \quad $^{3}$Inha University} \\
  {\normalfont $^{4}$Kangwon National University \quad $^{5}$Gachon University} \\
  {\normalfont $^{6}$BITS Pilani, K. K. Birla Goa Campus \quad $^{7}$HumbleBeeAI} \\
  {\normalfont\texttt{\{panyu,uclee\}@kaist.ac.kr}}
}

\begin{document}

\maketitle

\begin{abstract}

% We introduce a three-wave, in-the-wild multimodal dataset for affect sensing that integrates smartphone sensing, wearable sensing, and dense experience-sampling-method (ESM) labels collected annually from 2020 to 2022. The dataset supports moment-level affect modeling through a shared dimensional label core across all waves, with additional affective descriptors available in the third wave (D-3). We begin with a systematic characterization of the resource, analyzing its temporal density, sensing coverage, and cross-wave and cross-user heterogeneity. Drawing on these insights, we define a three-setting benchmark spanning temporal prediction from within-user history, within-wave cross-user generalization, and cross-wave generalization in which each wave is treated as a separate dataset. Our benchmark results show that the strongest method family depends on the evaluation setting: baseline models perform best in the temporal setting, domain adaptation is strongest overall in the within-wave cross-user setting, and cross-wave evaluation is more mixed, with no single family dominating uniformly across source-target directions. These findings indicate that robust mobile affective computing is constrained not only by label availability but also by substantial participant-level variability and realistic cross-wave differences inherent in longitudinal in-situ deployments.
We introduce a three-wave, in-the-wild repeated-cohort dataset for affect sensing that integrates smartphone sensing, wearable sensing, and dense experience-sampling-method (ESM) labels collected annually from 2020 to 2022 (297 participant-wave records from 282 distinct individuals, including 14 verified recurring participants). The dataset supports moment-level affect modeling through a shared dimensional label core across all waves, with eight custom affect-intensity items available only in the third wave (D-3). We describe the resource in terms of study design, temporal density of in-situ labels, and sensing and label coverage across waves. To support evaluation within this resource, we define an initial three-setting benchmark spanning chronological prediction using earlier within-user sensing observations, within-wave cross-user generalization, and transfer to a held-out annual wave. In participant-level paired bootstrap analyses, domain-generalization (DG) and unsupervised domain-adaptation (UDA) objectives improve their MLP backbone performances, but do not consistently outperform the strongest supervised models. Across waves, DG gives small gains over the MLP, UDA is less reliable, and neither family exceeds the tree-based supervised models. These results illustrate a challenging benchmark with modest predictive signal; held-out-wave performance reflects multiple entangled cohort, device, protocol, and labeling differences, not isolated temporal drift.

\begin{tcolorbox}[
    colback=white,
    colframe=black,
    arc=2pt,
    outer arc=2pt,
    boxrule=0.5pt,
    left=2pt,
    right=2pt,
    top=2pt,
    bottom=2pt
]
\centering \small
\textbf{EmoPhone Resources:}
\href{https://emophone.iclab.dev/}{Website}
\quad | \quad
\href{https://doi.org/10.7910/DVN/QGRVU8}{Dataset}
\quad | \quad
\href{https://github.com/Kaist-ICLab/EmoPhone}{Codebase}
\end{tcolorbox}

\end{abstract}

\section{Introduction}

Passive sensing from smartphones and wearables has created new opportunities for modeling human states and behaviors in everyday life~\cite{Trifan2019, Ruensuk2020, Xu2021, Hu2021, Sheikh2021, Swain2022}. Recent longitudinal mobile sensing benchmarks have shown the value of multi-year collection for studying robustness and generalization, but their benchmark targets are typically broad behavioral or mental-health outcomes collected at relatively sparse intervals.

Existing mobile-sensing resources typically emphasize either long-term behavioral and mental-health outcomes, large-scale in-situ labels, or generalization benchmarking, but rarely combine dense moment-level affect labels, smartphone-plus-wearable sensing, multi-wave collection, and a structured benchmark in one resource. This gap matters because affective states are inherently dynamic, short-horizon, and person-dependent~\cite{Mohr2017, Jean2025}.

Our resource is a three-wave, in-the-wild repeated-cohort dataset (D-1: 2020; D-2: 2021; D-3: 2022) integrating smartphone sensing, wearable sensing, and dense in-situ affective labels collected via an in-situ experience sampling method (ESM)~\cite{Hektner2007}. Across waves, valence, arousal, stress, and disturbance form a shared benchmarkable label core, while D-3 adds eight custom D-3-specific affect-intensity items (not a PANAS/PANAS-X instrument) for finer-grained emotion analysis. The waves contain 297 participant-wave records from 282 distinct individuals, including 14 verified recurring participants. The student cohort limits some background variation but also restricts generalizability~\cite{Xu2021}.

We pair the dataset release with a characterization-first benchmark that can be applied to other datasets. After documenting temporal density, sensing and label coverage, participation, and participant/wave heterogeneity, we evaluate three complementary settings: within-user temporal prediction, within-wave generalization to unseen users, and cross-wave generalization to held-out waves. The held-out-wave setting captures combined cohort, device, protocol, sensing, and label differences; it does not isolate time-related drift. Together, these evaluations support studying affect prediction and robustness in in-the-wild mobile sensing.

\section{Related Work \& Positioning}

\emph{Longitudinal mobile sensing and affect benchmarks:} Recent longitudinal mobile sensing benchmarks emphasize evaluation under realistic distribution shift rather than only within-dataset prediction~\cite{Xu2023, Nanchen2023, Meegahapola2024}. GLOBEM is a relevant multi-year benchmark precedent: it releases multi-year mobile and wearable sensing datasets and evaluates cross-dataset generalization for depression-related behavior modeling~\cite{Xu2022}. Its targets, however, are sparse weekly or end-of-term mental-health outcomes, whereas EmoPhone focuses on dense moment-level affect labels. K-EmoPhone is the closest single-wave predecessor, providing real-world smartphone and wearable sensing with in-situ emotion, stress, and attention labels~\cite{Kang2023}. 
Rather than extending K-EmoPhone itself, we introduce three newly collected dataset waves and benchmark settings for user- and wave-level generalization, with denser labels and richer D-3 affect descriptors.
%EmoPhone is different from this dataset in that we extend this line of research by collecting new datasets and further provide a three-wave benchmark with denser labels and richer D-3 affect descriptors. 
DiversityOne, the College Experience Study, and BRIGHTEN provide important daily-life or remote mobile-sensing resources~\cite{Busso2025, Nepal2024, Pratap2022}, with different combinations of sensing modalities, label frequencies, and study populations. FEEL further motivates our heterogeneity-aware benchmarking framework by showing that experimental setting, device type, and labeling strategy shape generalization in physiological emotion recognition~\cite{Singh2025}. Our comparison focuses on documented naturalistic mobile/wearable sensing resources with repeated affect or mental-health self-reports and reported study-design attributes, rather than assigning subjective quality scores. Table~\ref{tab:dataset_comparison} summarizes these differences.

% \emph{Wearable representation learning:}
% Recent advances in wearable foundation models have explored
% self-supervised representation learning from large-scale physiological
% signals. Abbaspourazad et al.~\cite{Abbaspourazad2024}
% develop pretrained representations for PPG and ECG signals,
% while Narayanswamy et al.~\cite{Narayanswamy2025}
% investigate scaling multimodal wearable foundation models.
% These studies emphasize learning transferable representations
% from physiological signals, whereas our initial benchmark
% evaluates generalization using extracted tabular features
% from smartphone and wearable sensing.

% \emph{Generalization and realistic evaluation:}
% Prior work has also investigated domain adaptation and robust
% evaluation for mobile and wearable sensing.
% Meegahapola et al.~\cite{Meegahapola2024} propose
% multimodal unsupervised domain adaptation, while
% Merrill et al.~\cite{Merrill2023} introduce Homekit2020,
% a wearable time-series benchmark emphasizing realistic
% deployment-oriented evaluation and statistical uncertainty.
% Complementing these efforts, EmoPhone provides three
% evaluation settings: within-user chronological prediction,
% within-wave generalization to unseen participants,
% and transfer to a held-out annual wave.
\emph{Recent methodological advances.}
Prior work investigates transferable biosignal
representations~\cite{Abbaspourazad2024,Narayanswamy2025},
multimodal domain adaptation~\cite{Meegahapola2024},
and realistic wearable time-series
evaluation~\cite{Merrill2023}.
Complementing these efforts, EmoPhone provides an initial
extracted-feature benchmark spanning within-user
chronological prediction, generalization to unseen
participants, and held-out-wave transfer, rather than
evaluating raw-signal pretraining or sequence architectures.

% Camera-ready: numerical attributes instead of subjective High/Low and Yes/No scores.
\begin{table*}[t]
\centering
\footnotesize
\setlength{\tabcolsep}{2.2pt}
\renewcommand{\arraystretch}{1.06}
\caption{Selected naturalistic sensing resources with repeated self-reports. Counts and schedules follow the cited sources and are not directly comparable across studies.}
\label{tab:dataset_comparison}
\begin{tabular}{@{}>{\raggedright\arraybackslash}p{0.145\textwidth}>{\raggedright\arraybackslash}p{0.182\textwidth}>{\raggedright\arraybackslash}p{0.187\textwidth}>{\raggedright\arraybackslash}p{0.233\textwidth}>{\raggedright\arraybackslash}p{0.201\textwidth}@{}}
\toprule
\textbf{Dataset} & \textbf{Waves; people} & \textbf{Duration; self-reports} & \textbf{Affect / well-being measures} & \textbf{Sensing; setting} \\
\midrule
GLOBEM~\cite{Xu2022} & 4 annual; 497 distinct (705 person-years) & 10 weeks/wave; weekly EMA & PHQ-4, PSS-4, PANAS (by wave) & Phone + Fitbit; 1 US university \\
DiversityOne~\cite{Busso2025} & 1 multi-site; 782 & 28 days; $>$350k self-reports & Mood/valence, activity/social context & Android phone; 8 countries \\
College Experience~\cite{Nepal2024} & 2 cohorts; 215 final (123 full follow-up) & 4 years/cohort; weekly EMA & Mental health and well-being & Phone; 1 US university \\
BRIGHTEN~\cite{Pratap2022} & 2 remote trials; 2,193 enrolled & 12 weeks; daily PHQ-2, others weekly/biweekly & Depression, anxiety, functioning & Phone; US-wide \\
K-EmoPhone~\cite{Kang2023} & 1 wave; 77 & 7 days; 5,582 ESM reports & Valence, arousal, stress, attention, etc. & Android + wrist wearable; Korea \\
\textbf{EmoPhone (ours)} & \textbf{3 annual; 282 distinct (297 participant-waves)} & \textbf{22.5/31.7/30.2 days mean; 53,139 ESM} & \textbf{4 shared + 8 D-3-only affect items} & \textbf{Android + Fitbit; 1 Korean university} \\
\bottomrule
\end{tabular}
\end{table*}

\section{EmoPhone Dataset: Design and Characterization}
\label{sec:dataset}
This section describes the three-wave EmoPhone resource, its study protocol, released data, temporal label density, and sensing/label coverage. We use descriptive analyses here to clarify what the dataset contains and what kinds of benchmark claims it can support. We reserve the term ``validation'' for the benchmark evaluation in later sections, where models are tested under temporal, cross-user, and cross-wave settings.

\subsection{Three-wave resource overview}

As shown in Table~\ref{tab:dataset_overview}, our resource consists of three annually collected, in-the-wild multimodal waves, denoted D-1 (2020), D-2 (2021), and D-3 (2022), with longitudinal observations within each wave. Each wave was conducted with university students over approximately four weeks and combined repeated experience-sampling labels with passive sensing from participants' own Android phones and wearable devices. The three waves share a common collection backbone while differing in several wave-specific design choices, enabling both unified benchmarking and analysis of cross-wave variation.

The retained waves represent 297 participant-wave records from 282 distinct people; 14 participants are confirmed to recur across waves (13 in exactly two waves, one in all three). The resource is substantial in both participant count and label volume. D-1, D-2, and D-3 include 102, 112, and 114 recruited participants, with 92, 99, and 106 retained after quality control based on ESM response rate and smartphone/wearable data coverage thresholds (Appendix~\ref{appendix:study_demographics}). The corresponding ESM totals are 10,259, 21,042, and 21,838, with mean per-participant counts of 111.51, 212.55, and 206.02. Because participant observation windows vary across waves, we report ESM frequency over active days rather than over a fixed calendar window: D-1, D-2, and D-3 yield 6.70, 6.99, and 7.03 responses per active day, respectively. These values show that the resource is not only multi-wave but also densely labeled at the moment level.

The waves remain comparable enough for shared-label benchmarking while preserving meaningful cross-wave heterogeneity. Across D-1, D-2, and D-3, the resource differs in questionnaire structure, sensing availability, and label coverage. For example, keystroke sensing is available from D-2 onward, and D-3 includes additional affective descriptors beyond the shared core labels (valence, arousal, stress, disturbance), including Happy, Relaxed, Cheerful, Content, Sad, Angry, Anxious, and Depressed. This label design makes the resource relevant to a range of tasks in affective computing, including emotion recognition, stress detection, and attention-related modeling~\cite{Park2020, Kang2023}.

\begin{table*}[t]
\centering
\caption{Overview of dataset characteristics across the three collection waves. Active day denotes a day with at least one valid ESM response.}
\label{tab:dataset_overview}
\renewcommand{\arraystretch}{1.2}
\resizebox{\textwidth}{!}{%
\begin{tabular}{lcccc}
\toprule
\textbf{Characteristic} & \textbf{D-1 (2020)} & \textbf{D-2 (2021)} & \textbf{D-3 (2022)} & \textbf{Total / Notes} \\
\midrule
\multicolumn{5}{l}{\textit{Demographics}} \\
\quad Recruited ($N$) & 102 & 112 & 114 & 328 \\
\quad Final sample ($N$) & 92 & 99 & 106 & 297 \\
\quad Gender (M/F) & 64/28 & 75/24 & 67/39 & 206/91 \\
\quad Mean age, SD & 22.9 (3.4) & 24.0 (3.6) & 23.6 (3.5) & — \\
\quad Psych.\ baselines & PSS-10, GHQ-12, PHQ-9 & PSS-10, GHQ-12, PHQ-9 & PSS-10, GHQ-12 & PHQ-9 absent in D\#3 \\
\midrule
\multicolumn{5}{l}{\textit{Data Collection}} \\
\quad Nominal study window & $\sim$4 weeks & $\sim$4--5 weeks & $\sim$4--5 weeks & participant span varies \\
\quad Mean span (days)/person & 22.51 & 31.67 & 30.18 & realized observation span \\
\quad Mean active days/person & 17.97 & 30.36 & 29.66 & active day = $\geq$1 valid ESM \\
\quad ESM sent freq.\ (per day) & 15 & 12 & 10 & $\downarrow$ intentional reduction \\
\quad Avg.\ ESM responses per active day & 6.70 & 6.99 & 7.03 & computed over active days only \\
\quad Total ESM responses & 10,259 & 21,042 & 21,838 & 53,139 \\
\quad ESM avg/person, SD & 111.51 (51.09) & 212.55 (44.05) & 206.02 (24.72) & — \\
\midrule
\multicolumn{5}{l}{\textit{Sensors \& Instruments}} \\
\quad Smartphone sensing & $\checkmark$ & $\checkmark$ & $\checkmark$ & Continuous \\
\quad Wearable (wrist) & Fitbit HR & Fitbit HR & Fitbit HR & Continuous \\
\quad Keystroke sensing & — & $\checkmark$ & $\checkmark$ & Added from D\#2 onward \\

\bottomrule
\end{tabular}}
\end{table*}

\begin{figure}[tb]
\centering
\includegraphics[width=0.95\columnwidth]{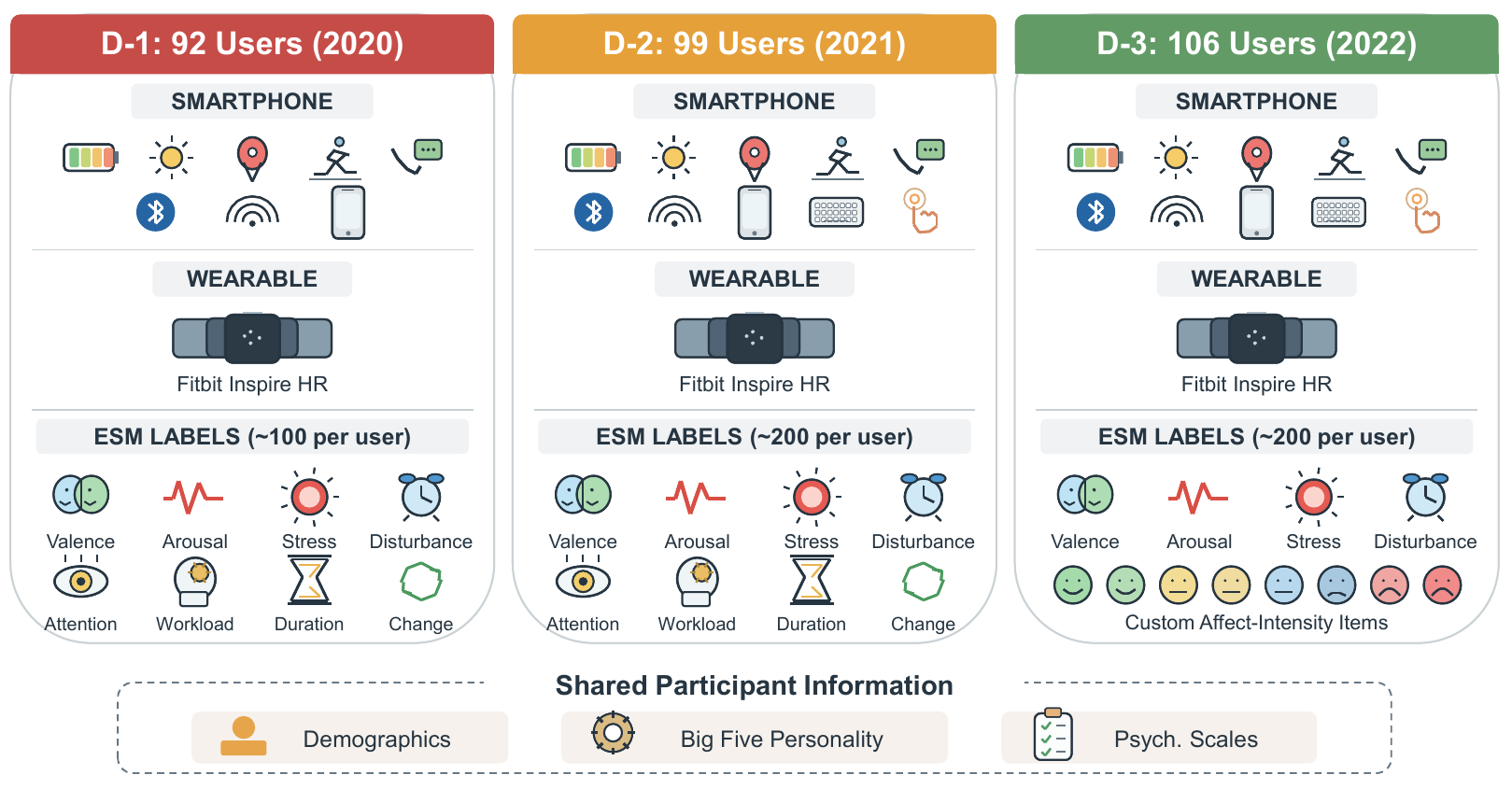}
\caption{Overview of the three annual collection waves, smartphone/Fitbit sensing, and ESM instruments. Participant counts are retained participant-wave records rather than a multi-year individual panel.}
\label{fig:study_design_protocol}
\end{figure}

\subsection{Study protocol and released data}

The study protocol was designed to capture affective states in naturalistic daily life rather than under laboratory control. Participants were recruited from the same university-student population under an Institutional Review Board (IRB)-approved protocol, provided informed consent, installed ABC Logger~\cite{Kang2023} on their own Android phones, and wore the designated Fitbit device during the study period. This student-centered cohort limits some broad demographic variation but cannot isolate participant effects from device, protocol, and contextual differences across waves.

Across waves, the study collected three main streams during daily life: repeated ESM prompts for momentary affective and cognitive states, passive smartphone sensing through ABC Logger~\cite{Kang2023}, and wearable physiological and mobility streams from Fitbit HR. Participant-level metadata, including demographics, personality traits, and psychological scales, were also collected to support cohort characterization and secondary analyses. The released resource includes these ESM labels, smartphone-derived behavioral and contextual signals, wearable streams, and participant-level metadata; detailed file schemas, sensor fields, preprocessing notes, and release structure are provided in Appendix~\ref{appendix:study_documents}.

\subsection{Temporal density of in-situ affect labels}

% We begin by showing that the resource is dense enough to support meaningful analysis and benchmarking. 

% Across the three waves, the dataset contains substantial numbers of participants and ESM responses. D-1, D-2, and D-3 include 102, 112, and 114 recruited participants, with retained analysis samples of 92, 99, and 106 participants, respectively. As summarized in Table~\ref{tab:dataset_overview}, the three waves provide dense moment-level supervision, with mean per-participant ESM counts of 111.51, 212.55, and 206.02, corresponding to 6.70, 6.99, and 7.03 responses per active day. Overall, these numbers show that the resource provides dense moment-level affect supervision across all three waves, making it suitable for both dataset characterization and later benchmark evaluation.

\begin{figure}[tb]
\centerline{\includegraphics[width=0.6\columnwidth]{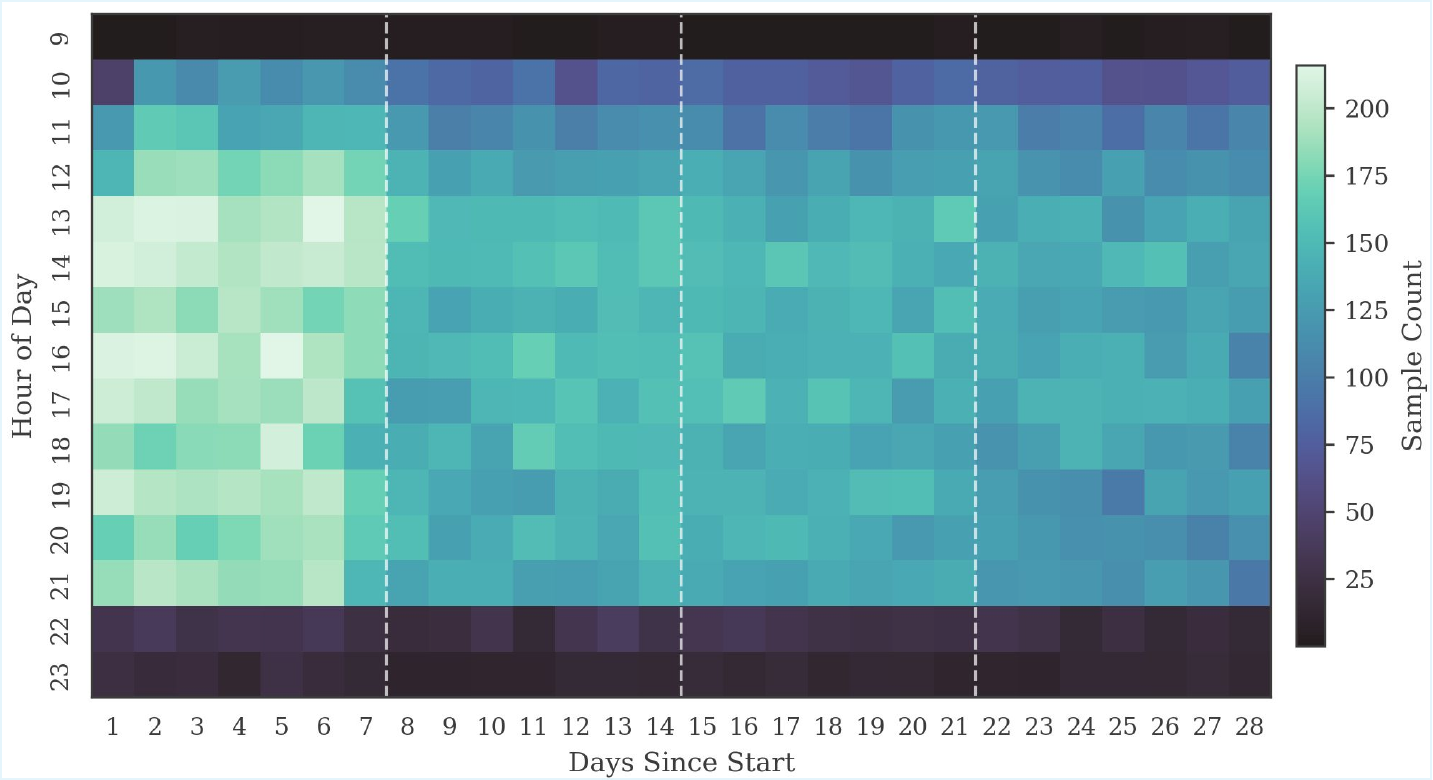}}
% \caption{Temporal distribution of ESM labels across the full three-wave resource.
% Heatmap of aggregated ESM label counts for all participants across D-1, D-2, and D-3, indexed by day since study start and hour of day. The figure shows that labels are concentrated during waking hours and remain distributed across the full collection period, indicating that the resource provides sustained, dense moment-level supervision rather than labels concentrated in only a few days or narrow time windows.}
\caption{Temporal distribution of ESM labels across D-1, D-2, and D-3 by day since study start and hour of day. Labels concentrate during waking hours while remaining distributed across the collection period, supporting sustained moment-level supervision.}
\label{fig:temporal_distribution_ESM_labels}
\end{figure}

Figure~\ref{fig:temporal_distribution_ESM_labels} complements the summary statistics in Table~\ref{tab:dataset_overview} by showing when ESM labels are observed over the study horizon. Responses are distributed across many study days with clear daytime concentration, indicating sustained engagement rather than short bursts of labeling, and supporting the temporal and cross-user evaluations used below.

% The daily ESM heatmap in Figure~\ref{fig:temporal_distribution_ESM_labels} complements these summary statistics by showing when ESM label data are actually observed over the study horizon. Rather than being concentrated in a short burst, participant activity is distributed across many days with clear daytime concentration, indicating sustained engagement over the course of the study. This temporal density is important for the benchmark design developed later in the paper: the resource is not merely large in total count, but also sufficiently distributed over time to support within-user temporal prediction and cross-user evaluation. In this sense, density should be understood not only as the number of labels, but also as the temporal continuity and practical usability of the collected observations.

\subsection{Sensing and label coverage across waves}

% Dense labels alone are not sufficient for a credible benchmark if passive sensing is highly incomplete or inconsistent across waves, and rich sensing alone is not enough if labels are not aligned across datasets.

% We characterize the resource jointly in terms of sensing coverage and label coverage as shown in Figure~\ref{fig:sensor_label_matrix}, since together they determine what kinds of analyses and benchmarks the dataset can meaningfully support.

\begin{figure}[tb]
\centerline{\includegraphics[width=0.99\columnwidth]{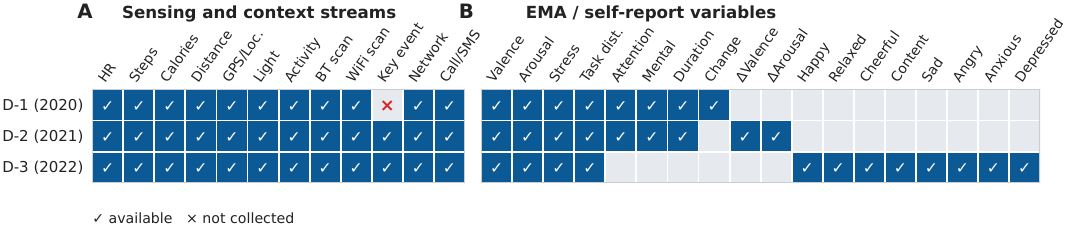}}
% \caption{Sensor Modality and  Label Coverage Matrix. A) Panel A shows the sensor modality matrix across the three study waves, indicating a substantial shared sensing backbone together with wave-specific hardware and protocol differences. In particular, keystroke interaction sensing is available from the middle wave onward. B) Panel B shows the label coverage matrix across waves. Valence, arousal, stress, and task disturbance form the shared core label space used for unified benchmarking, while D-3 extends this common backbone with a richer affective descriptor layer, including Happy, Relaxed, Cheerful, Content, Sad, Angry, Anxious, and Depressed.}
\caption{Sensor and label coverage across waves. A) Sensor and context streams, including shared smartphone/Fitbit streams and wave-specific additions such as key stroke events from D-2 onward. B) ESM/self-report variables, with valence, arousal, stress, and task disturbance forming the shared benchmark label core, and D-3 adding richer affect descriptors (e.g., happy, angry, depressed).}
\label{fig:sensor_label_matrix}
\end{figure}

Figure~\ref{fig:sensor_label_matrix} summarizes the protocol-level structure of the resource. D-3 added eight custom, non-standardized affect-intensity descriptors (happy, relaxed, cheerful, content, sad, anxious, depressed, angry), not the original PANAS/PANAS-X and informed by affect-differentiation research~\cite{Dejonckheere2019}; the earlier waves did not collect these items, so they cannot be retrospectively recovered as measured responses. At the sensing level, the waves share a common backbone to support harmonized evaluation, while still preserving realistic cross-wave heterogeneity. At the label level, the dataset combines a shared benchmarkable core with wave-specific enrichment. This dual structure is central to the later benchmark design: the shared sensing and label backbone supports common evaluation across waves, whereas the wave-specific sensing and richer D-3 descriptors make cross-wave generalization and fine-grained affective analysis meaningfully challenging.

%Figure~\ref{fig:sensor_label_matrix} summarizes this structure with two complementary views.  Figure~\ref{fig:sensor_label_matrix}(a) presents the sensor modality matrix, showing that D-1, D-2, and D-3 share a substantial common sensing backbone, including core Fitbit streams and multiple smartphone-derived modalities, while also exhibiting realistic hardware and protocol differences across waves. For example, keystroke interaction sensing was introduced from D-2 onward. This pattern is important because it shows that the three waves are similar enough to support shared benchmarking, yet different enough to make cross-wave generalization nontrivial.

%Figure~\ref{fig:sensor_label_matrix}(b) presents the corresponding label coverage matrix. It shows that valence, arousal, stress, and disturbance are shared across D-1, D-2, and D-3, forming the common label space used later for unified benchmarking. It also shows how earlier waves include auxiliary constructs such as attention, mental load, emotion duration, and emotion-change measures, while D-3 introduces a richer affective descriptor layer, including Happy, Relaxed, Cheerful, Content, Sad, Angry, Anxious, and Depressed. Importantly, this expansion is additive rather than inconsistent: the richer D-3 descriptors extend the shared core instead of replacing it.

% Taken together, these two views define the structure of the resource. 

\section{Benchmark framework: tasks and experimental design}

\subsection{Motivation: cross-user and cross-wave heterogeneity}

% The dataset characterization above shows that EmoPhone has both a shared sensing/label backbone and realistic heterogeneity. Figure~\ref{fig:TSNE_cross_user} provides an exploratory view of participant-level variability within each wave, where each point represents a participant-level aggregate of preprocessed sensor features. The spread of participants within each wave indicates that random sample-level splits are insufficient for evaluating generalization. We therefore organize the benchmark around three settings that progressively separate temporal learnability, cross-user generalization, and cross-wave robustness.
The dataset characterization above shows that EmoPhone has both a shared backbone and realistic heterogeneity. The shared sensing and label structure enables common evaluation across waves, while participant-level and wave-level differences make random sample-level evaluation insufficient. We therefore evaluate three complementary settings: chronological prediction, within-wave transfer to unseen users, and held-out-wave robustness.

\subsection{Benchmark overview}

We organize the benchmark as an initial three-setting evaluation suite (temporal, within-wave, and cross-wave) that evaluates chronological prediction, within-wave transfer to unseen users, and held-out-wave transfer~\cite{Singh2025, Xu2022}. Cross-wave transfer jointly reflects differences in cohort, devices, sensing, protocol, and labeling; it does not isolate the effect of calendar time.

% This layered design is motivated by the two main sources of heterogeneity characterized in Section~\ref{sec:dataset_characterization}, namely cross-user and cross-wave differences. In this sense, the benchmark is aligned with the structure of variation present in the resource itself.

Accordingly, the \emph{temporal} benchmark asks whether affective states are predictable from a user’s own earlier observations, the \emph{within-wave} benchmark asks whether models trained on some users generalize to unseen users within the same wave, and the \emph{cross-wave} benchmark asks whether models trained on one or more waves generalize to a held-out wave under a shared feature and label space. Framing the benchmark this way makes clear that the three settings are complementary components of one evaluation suite rather than three disconnected experiments. Table~\ref{tab:benchmark_model_selection} summarizes the evaluation settings and model families.

\newcolumntype{L}[1]{>{\raggedright\arraybackslash}p{#1}}

\begin{table*}[t]
\centering
\scriptsize
\setlength{\tabcolsep}{4pt}
\renewcommand{\arraystretch}{1.15}
\captionsetup{skip=6pt}
\caption{Overview of the three benchmark settings. Full model inventories and hyperparameter spaces are reported in Table~\ref{tab:appendix_model_inventory} in Appendix~C.2.}
\label{tab:benchmark_model_selection}
\begin{tabular}{
  L{0.12\textwidth}
  L{0.29\textwidth}
  L{0.18\textwidth}
  L{0.22\textwidth}
  L{0.12\textwidth}
}
\toprule
\textbf{Benchmark} &
\textbf{Evaluation setting} &
\textbf{Targets} &
\textbf{Domain / adaptation protocol} &
\textbf{Model families} \\
\midrule

Temporal &
Per-user chronological 60/20/20 train/validation/test split within the first 30 days; users are excluded if any split is empty or lacks both classes. &
Shared labels in D-1/D-2/D-3; D-3-specific affect labels in D-3. &
Within-user chronological prediction (not sequence-based modeling). &
Baselines only \\
\midrule

Within-wave &
User-grouped stratified 5-fold evaluation within each wave; no user overlap between train and test folds; feature normalization fitted on training data only. &
Shared labels in D-1/D-2/D-3; D-3-specific affect labels in D-3. &
User identity defines DG domains. UDA uses unlabeled held-out-user features but no target labels. &
Baselines, DG, UDA \\
\midrule

Cross-wave &
Held-out-wave transfer in six one-source and three two-source directions; the shared feature intersection is used. Verified participant recurrence means original splits are not fully identity-disjoint. &
Shared labels only, so source and target label spaces remain aligned. &
Train/test split is wave-level. For DG, each source-wave participant is treated as a separate domain. UDA uses unlabeled target-wave features but no target labels. &
Baselines, DG, UDA \\
\bottomrule
\end{tabular}
\end{table*}

Sensor features are aligned to each ESM prompt $t$ using only observations at or before $t$ (including 15/30-minute rolling windows, day/previous-day epoch summaries, and point-in-time descriptors). Appendix~A.7 describes the exact windows and missing-data rules; earlier ESM labels are not benchmark inputs.

\subsection{Prediction targets}

The benchmark is built around a shared core label space consisting of valence, arousal, stress, and disturbance, because these four targets can be aligned consistently across D-1, D-2, and D-3 and therefore support unified comparison across waves. Consequently, they serve as the common evaluation targets whenever cross-wave comparability is required.

In addition to this common label layer, D-3 provides a rich semantic descriptor layer consisting of happy, relaxed, cheerful, content, sad, anxious, depressed, and angry, which enables a more semantically fine-grained view of affect than the shared dimensional labels alone. These richer targets are valuable because they allow the benchmark to examine whether some affective concepts are easier to predict or generalize than others once the label space moves beyond the common core.

The benchmark targets follow directly from this design. The temporal and within-wave benchmarks evaluate the shared labels across all applicable waves, and they also include the richer D-3-specific labels when those targets are available. The cross-wave benchmark, on the contrary, is restricted to the shared labels so that comparison across datasets remains strictly aligned. This design preserves a common label space for cross-wave evaluation while still allowing richer semantic analysis where the label set supports it.

For binary classification, HIGH/LOW targets use the label-wise median of the labeled training data only (per-user training slices for the temporal setting), applied unchanged to validation/test responses; Appendix~A.7.6 gives details.

\subsection{Temporal benchmark}
\label{sec:tier_a}

The temporal benchmark measures within-user predictability by testing whether future affective states can be predicted from earlier self-observations. It provides a within-user reference without evaluating all possible sequence models or using earlier ESM labels as benchmark inputs.

\textbf{Task construction:} Tasks are defined at the (label, dataset) level. We evaluate the four shared labels in each wave (D-1, D-2, and D-3) as separate binary classification problems; in D-3, we additionally evaluate the eight affect labels as D-3-only tasks.

\textbf{Splitting:} For each user, we consider the first 30 days of samples, order them chronologically, and split them into 60\%/20\%/20\% train/validation/test partitions; the middle 20\% serves as the temporal validation slice. We then concatenate splits across users while preserving temporal order within each user.

\textbf{Models:} We evaluate the supervised baseline family, which includes both tree-based methods (XGBoost~\cite{Chen2016}, LightGBM~\cite{Ke2017}) and basic and tabular neural networks (MLP, ResNet~\cite{He2016}, TabNet~\cite{Arik2021}, DCN~\cite{Wang2021}, SAINT~\cite{Somepalli2021}, TabTransformer~\cite{Huang2020}, FTTransformer~\cite{Gorishniy2021}). We do not include domain generalization (DG) or domain adaptation (DA) methods in this benchmark because source and target distributions coincide by construction.

\subsection{Within-wave benchmark}
\label{sec:tier_b}

The within-wave benchmark evaluates user-independent generalization within each wave. Models must generalize from seen users to unseen users with the wave-specific collection protocol held fixed; the evaluation does not causally isolate participant effects.

\textbf{Task construction:} Within-wave tasks are again defined at the level (label, dataset). The four shared labels are evaluated in each of D-1, D-2, and D-3, and the eight D-3-only affect labels are evaluated in D-3 only. Feature $z$-score statistics are fitted only on the training split and applied unchanged to held-out users (Appendix~C.1.3).

\textbf{Splitting and preprocessing:} We use stratified group 5-fold cross-validation with user identity as the grouping variable so that each user's samples appear in exactly one fold. Appendix~C.1.2 provides the exact split definition.

\textbf{Models:} We compare three method families under a shared training loop, the temporal-benchmark baselines, DG methods (user as domain), and UDA methods. In the within-wave setting, the domain variable is user identity; Appendix~C.2.1 details how domain labels are instantiated for DG/UDA methods.

\subsection{Cross-wave benchmark}
\label{sec:tier_c}

The cross-wave benchmark evaluates cross-wave generalization by treating each wave as a distinct dataset. By combining unseen-user heterogeneity with shifts in study conditions, feature availability, and label distributions, this setting tests transfer under several entangled sources of real-world heterogeneity. Because a small verified subset recurs across waves, the primary evaluation is not fully identity-disjoint (Appendix~C.3.3).

\textbf{Task construction:} We follow a leave-one-dataset-out protocol, with each wave taking a turn as the target. This yields six one-to-one generalization directions (e.g. \mbox{D-1}$\rightarrow$\mbox{D-2}) and three two-to-one directions (e.g. \mbox{D-1+D-2}$\rightarrow$\mbox{D-3}). The cross-wave benchmark is restricted to the four shared labels so that source and target label spaces are aligned.

\textbf{Splitting and preprocessing:} For each generalization direction, we train on the source wave(s) and evaluate on the held-out target wave. We align feature spaces using a fixed three-way intersection across D-1/D-2/D-3; Appendix~C.1.1 details the alignment procedure.

% \textbf{Models:} As in the within-wave benchmark, we compare baseline models, DG methods, and unsupervised domain adaptation (UDA) methods. In the cross-wave setting, the domain variable is the wave rather than the user; Appendix~C.2.1 details how DG/UDA methods instantiate wave domains. UDA methods use unlabeled target-wave features during adaptation, but target labels are never used for training, preprocessing, validation, or model selection.

\textbf{Models:} As in the within-wave benchmark, we compare baseline models, DG methods, and unsupervised domain adaptation (UDA) methods. The cross-wave split is defined at the wave level: models are trained on one or more source waves and evaluated on a held-out target wave. For DG training, however, domains are instantiated at the participant level, using each source-wave participant as a separate domain. Thus, in both single-source and multi-source cross-wave settings, DG methods learn from multiple user domains within the source training pool. UDA methods use unlabeled target-wave features during adaptation, but target labels are never used for training, validation, or model selection.

\subsection{Cross-method comparability}
\label{sec:comparability}

The within-wave and cross-wave benchmarks compare three method families (baselines, DG, and UDA), so the three sources of unfairness in such comparisons are explicitly controlled. The corresponding values (optimizer, loss, batch size, schedule, seeds, and per-method search spaces) are documented in Appendix~C.2.3 and Tables~\ref{tab:appendix_hyperparameter_grids_part1}--\ref{tab:appendix_hyperparameter_grids_part3}.

\textbf{Data treatment.} ID-linked and temporal identifiers (participant IDs, timestamps) are excluded from the feature set so that user identity cannot leak into the inputs; standardization statistics are computed only on the training split; and multi-source training pools are aligned via feature-space intersection rather than padding (Appendix~C.1). These rules apply uniformly to every method, so no model is given a different feature set or split.

\textbf{Capacity matching.} All DG and UDA methods share the same MLP encoder as the MLP baseline. Performance gaps between families therefore reflect the training objective and adaptation strategy rather than backbone capacity.

\textbf{Tuning budget.} Every method is tuned with $30$ Optuna~\cite{optuna_2019} trials per (benchmark, label, dataset, model) configuration, selected on validation AUROC, and trained under a single shared training loop and schedule, so no method is favored or penalized by under-tuning or schedule mismatch. Concretely, all neural methods use AdamW with cross-entropy, batch size $512$, up to $50$ epochs with patience $20$ on validation AUROC, and seed $42$; the full per-method search spaces are listed in Appendix~C.2.3.

\section{Benchmark Results}

We summarize the main results in this section, with full details in Appendix~\ref{appendix:benchmark}. We report results for the three benchmark settings (temporal, within-wave, and cross-wave) and then interpret them as a label-dependent difficulty ladder.

\subsection{Within-User Temporal Prediction}

The temporal benchmark establishes a within-user predictability reference for each target (Table~\ref{tab:tier_a_summary}; the full model-by-(label, dataset) matrix is provided in Appendix~\ref{appendix:benchmark}). Predictability is strongly label-dependent: on the shared labels, stress is the most predictable (mean AUROC $0.627\pm0.016$), followed by arousal ($0.609\pm0.035$) and valence ($0.601\pm0.031$), whereas disturbance lies closest to chance ($0.555\pm0.005$). The D-3-only affect labels span a comparably wide range, indicating that within-user signal varies substantially across affect dimensions even when training and test distributions coincide.

% \begin{table}[t]
% \centering
% \small
% \caption{Within-user temporal benchmark summary. For shared labels, we report the best model and mean test AUROC $\pm$ std across D-1, D-2, and D-3. For D-3-specific labels, we report the best-performing model and its test AUROC.}
% \label{tab:tier_a_summary}
% \setlength{\tabcolsep}{5pt}
% \begin{tabular}{lll}
% \toprule
% Label & Best Model & Test AUROC \\
% \midrule
% \multicolumn{3}{l}{\textbf{Shared labels (mean across datasets)}} \\
% Stress      & SAINT & 0.580 $\pm$ 0.021 \\
% Arousal     & XGB   & 0.609 $\pm$ 0.035 \\
% Valence     & XGB   & 0.601 $\pm$ 0.031 \\
% Disturbance & SAINT & 0.555 $\pm$ 0.005 \\
% \midrule
% Mean $\pm$ Std across shared labels & -- & 0.586 $\pm$ 0.024 \\
% \midrule
% \multicolumn{3}{l}{\textbf{D-3 rich labels}} \\
% Angry     & XGB            & 0.673 \\
% Sad       & LGB            & 0.607 \\
% Content   & XGB            & 0.601 \\
% Cheerful  & LGB            & 0.599 \\
% Happy     & LGB            & 0.593 \\
% Relaxed   & XGB            & 0.585 \\
% Anxious   & SAINT          & 0.575 \\
% Depressed & FTTransformer  & 0.548 \\
% \bottomrule
% \end{tabular}
% \end{table}
\begin{table}[t]
\centering
\footnotesize
\caption{Within-user temporal benchmark summary. Shared labels report the best model and mean test AUROC $\pm$ std across D-1/D-2/D-3; D-3-only labels report the best model and AUROC on D-3. Full results are in Appendix~C (Table~\ref{tab:appendix_tier_a_full_results}).}
\label{tab:tier_a_summary}
\setlength{\tabcolsep}{3pt}
\renewcommand{\arraystretch}{0.8}
\begin{tabular}{lll}
\toprule
Label & Best Model & Test AUROC \\
\midrule
\multicolumn{3}{l}{\textbf{Shared labels (mean across datasets)}} \\
Stress      & XGB           & 0.627 $\pm$ 0.016 \\
Arousal     & XGB           & 0.609 $\pm$ 0.035 \\
Valence     & XGB           & 0.601 $\pm$ 0.031 \\
Disturbance & SAINT         & 0.555 $\pm$ 0.005 \\
\midrule
Mean $\pm$ Std across shared labels & -- & 0.598 $\pm$ 0.031 \\
\midrule
\multicolumn{3}{l}{\textbf{D-3 rich labels}} \\
Angry       & XGB           & 0.673 \\
Sad         & LGB           & 0.607 \\
Content     & XGB           & 0.601 \\
Cheerful    & LGB           & 0.599 \\
Happy       & LGB           & 0.593 \\
Relaxed     & XGB           & 0.585 \\
Anxious     & SAINT         & 0.575 \\
Depressed   & FTTransformer & 0.548 \\
\bottomrule
\end{tabular}
\end{table}

\subsection{Within-Wave Cross-User Generalization}

The within-wave benchmark tests cross-user generalization via a stratified 5-fold group split by participant. 
Table~\ref{tab:shared_label_family_summary} summarizes family-level results on the shared labels, and Appendix~\ref{appendix:benchmark} reports full per-task matrices and DG/UDA breakdowns.

\noindent\textbf{DG and UDA objectives improve their MLP backbone performances.}
UDA has the highest family-best fold-mean AUROC on the shared labels ($0.571$ vs.\ $0.561$ for DG and $0.557$ for the baseline family) and on seven of the eight D-3-only labels, but these are point differences. In participant-level paired bootstrap contrasts (Appendix~C.3.2), family-best DG and UDA exceed the MLP backbone by $+0.018$ and $+0.012$ AUROC (significant after Holm adjustment), while neither family consistently exceeds the strongest supervised model. The best UDA model varies by label, and no individual DG or UDA method differs from MLP after adjustment.

\subsection{Cross-Wave Generalization}

The cross-wave benchmark evaluates one-to-one and two-to-one generalization across waves with feature spaces aligned by intersection (Table~\ref{tab:shared_label_family_summary}; per-direction results in Appendix~\ref{appendix:benchmark}). Family-level differences narrow sharply relative to the within-wave setting, with baseline, DG, and UDA means falling within a $0.013$ AUROC band ($0.538$, $0.542$, and $0.529$). The best DG method differs by label (MASF, IRM, Fish, and CSD in Table~\ref{tab:key_shared_results}). In paired contrasts (Appendix~C.3.3), DG gives small gains over the MLP backbone (at most $+0.007$), UDA is less reliable, and neither family exceeds the tree-based supervised models. UDA, in contrast, loses $0.042$ AUROC on average relative to the within-wave benchmark, suggesting that these adaptation objectives do not reliably improve the composite held-out-wave task; performance also varies considerably by source-target direction, consistent with heterogeneous mixtures of covariate and label shift across wave pairs.

% \begin{table}[tb]
% \centering
% \footnotesize
% \caption{Shared-label AUROC across Tier B and Tier C by method family. Each cell reports the best AUROC within a method family for each shared label. The last row reports the mean and standard deviation across shared labels.}
% \label{tab:shared_label_family_summary}
% \resizebox{\columnwidth}{!}{%
% \begin{tabular}{lcccccc}
% \toprule
% & \multicolumn{3}{c}{Within-wave} & \multicolumn{3}{c}{Cross-wave} \\
% \cmidrule(lr){2-4} \cmidrule(lr){5-7}
% Label & Baseline & DG & DA & Baseline & DG & DA \\
% \midrule
% Stress      & 0.566 & 0.532 & \textbf{0.612} & 0.531 & 0.537 & \textbf{0.539} \\
% Arousal     & 0.581 & 0.588 & \textbf{0.593} & 0.582 & \textbf{0.595} & 0.558 \\
% Valence     & 0.562 & 0.563 & \textbf{0.579} & \textbf{0.562} & 0.554 & 0.561 \\
% Disturbance & 0.554 & 0.552 & \textbf{0.564} & \textbf{0.565} & 0.556 & 0.551 \\
% \midrule
% Mean $\pm$ Std & 0.566 $\pm$ 0.012 & 0.559 $\pm$ 0.023 & \textbf{0.587 $\pm$ 0.020} & 0.560 $\pm$ 0.021 & \textbf{0.561 $\pm$ 0.025} & 0.552 $\pm$ 0.010 \\
% \bottomrule
% \end{tabular}%
% }
% \end{table}

\begin{table}[tb]
\centering
\scriptsize
\caption{Family-best shared-label AUROC for within-wave and cross-wave benchmarks. Each cell reports the best model within a family, averaged across datasets or transfer directions. Bold denotes the highest value in each row (paired tests in Appendix~C.3). For within-wave evaluation, we select the best-performing model
within each family separately for each wave and average the
resulting AUROCs across waves. For cross-wave evaluation, we
select the single model within each family with the highest
average AUROC across transfer directions.}
\label{tab:shared_label_family_summary}
\setlength{\tabcolsep}{3pt}
\begin{tabular}{lcccccc}
\toprule
& \multicolumn{3}{c}{Within-wave} & \multicolumn{3}{c}{Cross-wave} \\
\cmidrule(lr){2-4} \cmidrule(lr){5-7}
Label & Baseline & DG & UDA & Baseline & DG & UDA \\
\midrule
Stress      & 0.549 & \textbf{0.558} & 0.558 & \textbf{0.525} & 0.525 & 0.522 \\
Arousal     & 0.576 & 0.574 & \textbf{0.587} & 0.552 & \textbf{0.564} & 0.538 \\
Valence     & 0.560 & 0.561 & \textbf{0.577} & 0.540 & \textbf{0.541} & 0.525 \\
Disturbance & 0.543 & 0.550 & \textbf{0.563} & 0.535 & \textbf{0.537} & 0.532 \\
\midrule
Mean $\pm$ Std & 0.557 $\pm$ 0.014 & 0.561 $\pm$ 0.010 & \textbf{0.571 $\pm$ 0.013} & 0.538 $\pm$ 0.011 & \textbf{0.542 $\pm$ 0.016} & 0.529 $\pm$ 0.007 \\
\bottomrule
\end{tabular}
\end{table}

\begin{table}[tb]
\centering
\scriptsize
% \caption{Key shared-label results across benchmarks. For each (benchmark, label) pair we report the best single model and its mean test AUROC $\pm$ std -- across the three datasets in the temporal and within-wave benchmarks, and across all nine cross-wave source-target directions (single-source and multi-source combined) in the cross-wave benchmark. The last row reports the mean and standard deviation across the four shared labels.}
\caption{Best-achieved shared-label AUROC across benchmarks. Values are mean $\pm$ std across datasets for temporal/within-wave settings and across nine transfer directions for cross-wave.}
\label{tab:key_shared_results}
\setlength{\tabcolsep}{4pt}
\begin{tabular}{lcccccc}
\toprule
& \multicolumn{2}{c}{Temporal} & \multicolumn{2}{c}{Within-wave} & \multicolumn{2}{c}{Cross-wave} \\
\cmidrule(lr){2-3} \cmidrule(lr){4-5} \cmidrule(lr){6-7}
Label & Model & AUROC & Model & AUROC & Model & AUROC \\
\midrule
Stress      & XGB   & 0.627 $\pm$ 0.016 & SHOT & 0.554 $\pm$ 0.018 & MASF & 0.525 $\pm$ 0.013 \\
Arousal     & XGB   & 0.609 $\pm$ 0.035 & MCD  & 0.584 $\pm$ 0.022 & IRM  & 0.564 $\pm$ 0.020 \\
Valence     & XGB   & 0.601 $\pm$ 0.031 & MCC  & 0.576 $\pm$ 0.010 & Fish & 0.541 $\pm$ 0.008 \\
Disturbance & SAINT & 0.555 $\pm$ 0.005 & SHOT & 0.558 $\pm$ 0.009 & CSD  & 0.537 $\pm$ 0.012 \\
\midrule
Mean $\pm$ Std & -- & 0.598 $\pm$ 0.031 & -- & 0.568 $\pm$ 0.014 & -- & 0.542 $\pm$ 0.016 \\
\bottomrule
\end{tabular}
\end{table}

\subsection{Benchmark Ladder}

Taken together, the three benchmarks describe a label-dependent set of evaluation outcomes rather than a universal ranking (Table~\ref{tab:key_shared_results}). Averaged across the four shared labels, the temporal setting performs best ($0.598$), the within-wave setting drops to $0.568$, and the cross-wave setting drops further to $0.542$ and has the lowest aggregate AUROC in our experiments. The drop is uneven across labels: stress shows the biggest fall from temporal to cross-wave ($0.103$ AUROC), followed by valence ($0.060$) and arousal ($0.045$), while disturbance remains within a narrow band ($0.537$--$0.558$ across all three settings). No single setting therefore characterizes difficulty across all shared targets.

\section{Discussion}

% \subsection{Dataset contribution}
% This resource brings together three strengths that are still uncommon in affective sensing: dense moment-level affect supervision, multi-year naturalistic sensing, and smartphone-plus-wearable multimodality~\cite{Amin2025, Meegahapola2024, Wu2026}. Across three annual in-the-wild waves, participants used their own Android phones while contributing wearable streams and repeated in-situ ESM labels, yielding 10,259, 21,042, and 21,838 post-QC ESM responses in D-1, D-2, and D-3, respectively. The resource also expands semantically across waves. While all three waves support shared tasks on valence, arousal, stress, and disturbance, D-3 additionally introduces richer affect descriptors such as Happy, Relaxed, Cheerful, Content, Sad, Anxious, Depressed, and Angry. As a result, the dataset supports both unified comparison on a shared label space and finer-grained affect modeling where richer semantic supervision is available.

\subsection{Implications of the benchmark findings}
The three-setting benchmark shows that benchmark conclusions depend strongly on the evaluation setting. Within-user evaluation provides modest predictive signal for some labels using earlier passive-sensing features, while within-wave evaluation shows that DG and UDA objectives improve their MLP backbone without consistently outperforming the strongest supervised models. Held-out-wave evaluation yields lower aggregate AUROC: DG gives small gains over the MLP backbone, UDA is less reliable, and neither family exceeds the tree-based supervised models (Appendix~C.3). These results indicate that user-level adaptation does not automatically translate into wave-level robustness, and that no single score adequately summarizes performance across within-user, cross-user, and cross-wave settings. 

\subsection{Recommended use cases}
This resource is well suited to moment-level affect prediction from passive sensing, user-aware or personalized affect modeling, across-user generalization within a cohort, and cross-dataset generalization studies involving baseline, DG, and UDA methods~\cite{Kang2023, Xu2023, Singh2025, Meegahapola2024}. It is also a useful foundation for richer affect representation learning, especially when models need to connect broad affective dimensions such as valence and arousal with more specific descriptors available in D-3~\cite{Huo2024, Toisoul2021}. For users primarily interested in robust method comparison, the within-wave and cross-wave benchmarks are the most informative settings. For users interested in target construction or label semantics, D-3 provides the richest starting point. 

\subsection{Limitations}
The dataset should also be read with clear but proportionate boundaries. First, the cohort is student-centered and relatively homogeneous in age and life context, which is useful for reducing some confounds but limits broader population coverage. Second, the Android-only, bring-your-own-phone design
excludes iOS users; device and OS heterogeneity may also
affect sensing availability and reliability
(see Appendix~A.3.3). Third, the three waves are not perfectly identical in sensing and questionnaire design, which improves the resource in some respects but also introduces real cross-wave differences that affect direct comparability. Fourth, modality overlap is strong but not complete: some streams are partial or wave-specific. Fifth, training-median label binarization creates relative HIGH/LOW benchmark targets rather than absolute affect severity, so they should not be interpreted as clinical or population-level affect measures. Finally, the eight custom (non-PANAS) affect-intensity items appear only in D-3, so fine-grained emotion analyses cannot yet be carried out symmetrically across all three waves. These are not flaws to hide; they are the practical boundaries that define what kinds of claims the dataset can support most cleanly. 

Accordingly, the benchmark should be interpreted as a scoped evaluation suite for the EmoPhone three-wave resource, not as a definitive benchmark for all populations, device ecosystems, or mobile affect-sensing deployments.

\section{Conclusion}

We presented a three-wave, in-the-wild multimodal resource for affective sensing that combines smartphone sensing, wearable sensing, and dense ESM labels collected from 2020 to 2022. By pairing a shared dimensional label core with eight custom D-3-specific affect descriptors, the resource provides an initial benchmark for complementary chronological, cross-user, and held-out-wave evaluation settings. Our initial benchmark results show that method rankings vary across evaluation settings. Taken together, these findings suggest that performance remains modest and the observed cross-wave differences cannot be attributed to a single cause. More broadly, this work offers a characterization-first template for future dataset design in mobile affective computing.

% AUTHOR SIGN-OFF (required): verify the competing-interests disclosure and add the agreed statement in this section.
\begin{ack}
This research was supported by the Institute of Information \&
Communications Technology Planning \& Evaluation (IITP) grant funded
by the Korean government (MSIT) (RS-2025-02305801, RS-2026-25552484),
by the National Research Foundation of Korea (NRF) funded by the
Korean government (MSIT) (RS-2022-NR068758), and by the InnoCORE
program of the Ministry of Science and ICT (AI Meta-Scientist,
N10260110).
\end{ack}

% \section*{References}

% References follow the acknowledgments in the camera-ready paper. Use unnumbered first-level heading for
% the references. Any choice of citation style is acceptable as long as you are
% consistent. It is permissible to reduce the font size to \verb+small+ (9 point)
% when listing the references.
% Note that the Reference section does not count towards the page limit.
% \medskip

% {
% \small

% [1] Alexander, J.A.\ \& Mozer, M.C.\ (1995) Template-based algorithms for
% connectionist rule extraction. In G.\ Tesauro, D.S.\ Touretzky and T.K.\ Leen
% (eds.), {\it Advances in Neural Information Processing Systems 7},
% pp.\ 609--616. Cambridge, MA: MIT Press.

% [2] Bower, J.M.\ \& Beeman, D.\ (1995) {\it The Book of GENESIS: Exploring
%   Realistic Neural Models with the GEneral NEural SImulation System.}  New York:
% TELOS/Springer--Verlag.

% [3] Hasselmo, M.E., Schnell, E.\ \& Barkai, E.\ (1995) Dynamics of learning and
% recall at excitatory recurrent synapses and cholinergic modulation in rat
% hippocampal region CA3. {\it Journal of Neuroscience} {\bf 15}(7):5249-5262.
% }

\bibliographystyle{plainnat}
\bibliography{reference}
%%%%%%%%%%%%%%%%%%%%%%%%%%%%%%%%%%%%%%%%%%%%%%%%%%%%%%%%%%%%

\appendix

\section{Study Documents}
\label{appendix:study_documents}

\raggedbottom

\subsection*{A.1 Study Documents}

We provide the following documents in the supplementary folder:

% \begin{itemize}
%   \item \textbf{IRB Approval Letter} -- The institutional review board approval letter authorising the data collection studies.
%   \item \textbf{Consent Form (Translated)} -- The IRB-approved informed consent form presented to participants prior to enrollment, including a clause on voluntary public dataset release.
%   \item \textbf{Dataset Documentation File} -- A comprehensive reference document describing dataset structure, file schemas, and usage notes.
% \end{itemize}

\begin{itemize}
  \item \textbf{OT Materials (Original, Korean)} -- The original participant orientation materials administered in Korean during the onboarding session.

  \item \textbf{OT Materials (English Translation)} -- An English translation of the participant orientation materials provided for reviewer accessibility.

  \item \textbf{Informed Consent Form (English Translation)} -- An English translation of the informed consent form shown to participants before enrollment.
\end{itemize}

\subsection*{A.2 Study Demographics}
\label{appendix:study_demographics}

Table~\ref{tab:study_demographics} provides the full per-wave participant and data summary. Counts are reported both for all recruited participants and for the analysis-eligible subset after quality-control (QC) filtering. QC excluded participants with wearable data coverage below 50\%, smartphone data coverage below 50\%, or an ESM response rate below 30\%.
\vspace{-1.0em}
\begin{table}[H]
\centering
\caption{Per-wave participant and data summary.}
\label{tab:study_demographics}
\renewcommand{\arraystretch}{1.2}
\resizebox{\textwidth}{!}{%
\begin{tabular}{lcccc}
\toprule
\textbf{Characteristic} & \textbf{D-1 (2020)} & \textbf{D-2 (2021)} & \textbf{D-3 (2022)} & \textbf{Total / Notes} \\
\midrule
\multicolumn{5}{l}{\textit{Demographics}} \\
\quad Recruited ($N$) & 102 & 112 & 114 & 328 \\
\quad Final sample ($N$) & 92 & 99 & 106 & 297 \\
\quad Gender (M/F) & 64/28 & 75/24 & 67/39 & 206/91 \\
\quad Mean age, SD & 22.9 (3.4) & 24.0 (3.6) & 23.6 (3.5) & — \\
% \quad Psych.\ baselines & PSS-10, GHQ-12, PHQ-9 & PSS-10, GHQ-12, PHQ-9 & PSS-10, GHQ-12 & PHQ-9 absent in D\#3 \\
\midrule
\multicolumn{5}{l}{\textit{Data Collection}} \\
\quad Nominal study window & $\sim$4 weeks & $\sim$4--5 weeks & $\sim$4--5 weeks & participant span varies \\
\quad Mean span (days)/person & 22.51 & 31.67 & 30.18 & realized observation span \\
\quad Mean active days/person & 17.97 & 30.36 & 29.66 & active day = $\geq$1 valid ESM \\
\quad ESM sent freq.\ (per day) & 15 & 12 & 10 & $\downarrow$ intentional reduction \\
\quad Avg.\ ESM responses per active day & 6.70 & 6.99 & 7.03 & computed over active days only \\
\quad Total ESM responses & 10,259 & 21,042 & 21,838 & 53,139 \\
\quad ESM avg/person, SD & 111.51 (51.09) & 212.55 (44.05) & 206.02 (24.72) & — \\
\bottomrule
\end{tabular}}
\end{table}

% \begin{table}[H]
% \centering
% \footnotesize
% \renewcommand{\arraystretch}{1.0}
% \hbadness=10000
% \captionsetup{skip=4pt}
% \caption{Per-wave participant and data summary.}
% \label{tab:A1}
% \begin{tabularx}{\textwidth}{p{4.2cm}XXX}
% \toprule
% \textbf{Property} & \textbf{D-1 (2020)} & \textbf{D-2 (2021)} & \textbf{D-3 (2022)} \\
% \midrule
% Collection period
%   & Feb 7 -- Apr 2, 2020 ($\sim$28 days)
%   & Dec 7, 2020 -- Jan 27, 2021 ($\sim$28 days)
%   & Nov 23, 2021 -- Jan 11, 2022 ($\sim$28 days) \\[4pt]
% Total recruited      & 102 (32 female) & 112 (27 female) & 114 (42 female) \\
% After QC             & 92 (28 female)  & 99 (24 female)  & 106 (39 female) \\
% Mean age, recruited (SD) & 22.9 yr (3.4) & 24.0 yr (3.6) & 23.5 yr (3.4) \\
% \midrule
% Total ESM responses  & 10,774 & 22,232 & 23,280 \\
% ESM responses after QC & 9,914 (92.0\%) & 18,812 (84.6\%) & 20,636 (88.6\%) \\
% Mean ESM / person after QC (SD) & 107.8 (48.7) & 190.0 (39.1) & 194.7 (31.8) \\
% \bottomrule
% \end{tabularx}
% \end{table}

All three cohorts were drawn from the same university student population, resulting in broadly comparable age distributions across waves. The female proportion ranges from 23\% to 37\% across waves after QC. The PSS-10 and GHQ-12 psychological baseline measures are available for all three waves and are the recommended baseline psychological covariates for cross-wave analyses. Nominal study window rows in Table~\ref{tab:study_demographics} report the overall recruitment and data-collection span for each wave.

\subsection*{A.3 Data Collection Infrastructure and Cross-Wave Design Differences}

\subsubsection*{A.3.1 Smartphone Data Collection Application}

Passive sensor data and ESM prompts were collected via ABC Logger, a custom Android application deployed on participants' personal devices. The application runs continuously in the background, logging sensor streams and delivering randomised ESM prompts within a defined daily window. Data were uploaded to a remote server hourly in D-2 and D-3, and via periodic background batches in D-1. Participants were required to keep the application and all wearable devices active for at least 12 hours per day.

Android version requirements differed across waves: Android $\geq$ 7.0 was required in D-1, and Android $\geq$ 8.0 was required in D-2 and D-3. The research team monitored each participant's upload status and ESM compliance every 2--3 days throughout each collection period. Participants who were not correctly uploading data, or who fell below a 50\% ESM response rate, were individually notified and prompted to supplement their responses.

A participant-facing progress counter was introduced in D-2 to support ESM completion. Its effect cannot be isolated from other protocol changes. After the collection period, participants returned all wearable sensors, uninstalled the application, and completed a set of post-study psychological questionnaires.

\subsubsection*{A.3.2 Wearable Devices}

Fitbit Inspire HR was used across all three waves; the Polar H10 chest sensor was also deployed for assigned sub-periods in D-1 and D-2.

\textbf{Fitbit Inspire HR} -- a wrist-worn device collecting heart rate (1~Hz), step count, calorie burn, and distance (all at 1/60~Hz). Used continuously throughout all three waves for all participants.
\vspace{-1.2em}
% \begin{table}[H]
% \centering
% \caption{Wearable device availability per wave.}
% \label{tab:A2}
% \begin{tabular}{llll}
% \hline
% \textbf{Device} & \textbf{Type} & \textbf{Waves} & \textbf{Coverage Notes} \\
% \hline
% Fitbit Inspire HR & Wrist-worn & D-1, D-2, D-3 & Continuous throughout study period \\
% \hline
% \end{tabular}
% \end{table}

\begin{table}[H]
\centering
\footnotesize
\renewcommand{\arraystretch}{1.0}
\hbadness=10000
\captionsetup{skip=4pt}
\caption{Wearable device availability per wave.}
\label{tab:A2}
\begin{tabular}{llll}
\toprule
\textbf{Device} & \textbf{Type} & \textbf{Waves} & \textbf{Coverage Notes} \\
\midrule
Fitbit Inspire HR & Wrist-worn & D-1, D-2, D-3 & Continuous throughout study period \\
\bottomrule
\end{tabular}
\end{table}

\subsubsection*{A.3.3 Cross-Wave Design Differences}

Although the three waves share a common collection backbone, the three waves differ in several collection parameters. Table~\ref{tab:A3} summarises the key protocol parameters per wave, followed by the Key differences across waves.

% \begin{table}[H]
% \centering
% \caption{Cross-wave protocol parameters.}
% \label{tab:A3}
% \begin{tabular}{lccc}
% \hline
% \textbf{Parameter} & \textbf{D-1} & \textbf{D-2} & \textbf{D-3} \\
% \hline
% Android requirement              & $\geq$ 7.0         & $\geq$ 8.0 & $\geq$ 8.0 \\
% ESM prompts / day                & 15 (random)        & 12 (random) & 10 (random) \\
% Data upload cadence              & Periodic batches   & Hourly      & Hourly \\
% Keystroke sensing                & $\times$           & \checkmark  & \checkmark \\
% Participant-facing progress counter & $\times$        & \checkmark  & \checkmark \\
% ESM questionnaire items          & 8                  & 9           & 12 \\
% \hline
% \end{tabular}
% \end{table}

\begin{table}[H]
\centering
\footnotesize
\renewcommand{\arraystretch}{1.0}
\hbadness=10000
\captionsetup{skip=4pt}
\caption{Cross-wave protocol parameters.}
\label{tab:A3}
\begin{tabular}{lccc}
\toprule
\textbf{Parameter} & \textbf{D-1} & \textbf{D-2} & \textbf{D-3} \\
\midrule
Android requirement              & $\geq$ 7.0         & $\geq$ 8.0 & $\geq$ 8.0 \\
ESM prompts / day                & 15 (random)        & 12 (random) & 10 (random) \\
Data upload cadence              & Periodic batches   & Hourly      & Hourly \\
Keystroke sensing                & $\times$           & \checkmark  & \checkmark \\
Participant-facing progress counter & $\times$        & \checkmark  & \checkmark \\
ESM questionnaire items          & 8                  & 9           & 12 \\
\bottomrule
\end{tabular}
\end{table}

\noindent\textbf{Key differences across waves:}

\begin{itemize}
  \item \textbf{Keystroke sensing added in D-2.} KeyEvent data was introduced to capture fine-grained smartphone interaction dynamics -- typing speed, inter-key intervals, and keyboard type -- as potential behavioural indicators of cognitive and emotional state.
  \item \textbf{ESM prompt frequency reduced across waves.} The intentional reduction from 15 $\to$ 12 $\to$ 10 prompts per day reflects lessons learned about participant burden. Fewer prompts were intended to reduce participant burden; observed label volume also depends on participation and observation span.
  \item \textbf{Hourly upload cadence from D-2 onward.} D-1 used periodic background batch uploads, which occasionally produced multi-hour data gaps. Hourly synchronisation in D-2 and D-3 enabled more frequent upload checks; differences in data completeness may have multiple causes.
  \item \textbf{ESM questionnaire expansion.} D-1 used an 8-item instrument. In D-2, the single post-ESM Change item was split into ChangedValence and ChangedArousal. D-3 introduced eight custom, non-standardized affect-intensity items and rescaled Stress and Disturbance from $-3/+3$ to $0/+6$.
\end{itemize}

\noindent\textbf{Android device heterogeneity.}
Participants used their own primary Android smartphones to preserve
naturalistic sensing conditions. Device information was available
for 92/92, 96/99, and 106/106 retained participants in D-1, D-2,
and D-3, respectively. The observed Android versions spanned
8--10 in D-1, 7--10 in D-2, and 8--12 in D-3.
We identified 34 distinct device models in D-1, 36 in D-2,
and 37 normalized device-model families in D-3.
These counts are descriptive and not directly comparable
because model-identification granularity differs across waves.
Device and operating-system heterogeneity may affect sensor
availability, background data collection, and data completeness.
We did not conduct controlled cross-device compatibility testing.

\subsection*{A.4 Study Intrinsic Bias}

We discuss potential sources of intrinsic bias in our datasets that readers and future users should be aware of.

\begin{enumerate}
  \item \textbf{Recruitment and Population Bias.} Participants were recruited via a campus bulletin board at a single South Korean university, representing a self-selected subset of a narrow, young adult student population. Findings may not generalise to other age groups, cultural contexts, or non-student populations.
  \item \textbf{Gender Imbalance.} Male participants outnumber female participants across all three waves in both recruited and retained samples; after QC, the counts are D-1: 64M/28F, D-2: 75M/24F, and D-3: 67M/39F.
  \item \textbf{Android-Only Bias.} The requirement for an Android smartphone excludes all iOS users, which may introduce bias related to platform-specific usage patterns and sensing coverage.
  \item \textbf{Temporal and Contextual Bias.} All three waves were collected under distinct COVID-19 conditions and in winter months, meaning affective baselines, mobility, and social behaviour may not reflect typical everyday life and are not directly comparable across seasons.
  \item \textbf{Compliance and Exclusion Bias.} Quality-control exclusions based on low ESM response rates and sensor coverage thresholds are non-random, likely removing participants with higher affective instability or greater device burden, and thus biasing the dataset toward participants with more stable behavioural patterns.
\end{enumerate}

\vspace{-2em}

\begin{table}[H]
\centering
\renewcommand{\arraystretch}{0.92}
\hbadness=10000
\captionsetup{skip=4pt}
\caption{Post-study psychological baseline questionnaires.}
\label{tab:A4}
\footnotesize

\begin{threeparttable}
\begin{tabularx}{\textwidth}{
    p{1.2cm}
    p{2.6cm}
    p{1.6cm}
    p{2.0cm}
    >{\raggedright\arraybackslash}X
    p{0.8cm}
}
\toprule
\textbf{Scale} &
\textbf{Full Name} &
\textbf{Items} &
\textbf{Scoring Range} &
\textbf{Interpretation} &
\textbf{Waves} \\
\midrule

K-BFI-15
& Korean Big Five Inventory (15-item)
& 15
& Per trait: 3--15
& Measures five personality traits: Openness,
Conscientiousness, Neuroticism, Extraversion, and
Agreeableness. Each trait is scored by summing its
three constituent items. Higher scores indicate
a stronger tendency toward that trait.
& D-1, D-2, D-3 \\

\midrule

PSS-10
& Perceived Stress Scale (10-item)
& 10
& 0--40
& Measures perceived stress over the past month
using five-point Likert items. Scored by summing
all responses. Higher scores indicate greater
perceived stress. Scores $\geq 20$ suggest
elevated stress.
& D-1, D-2, D-3 \\

\midrule

GHQ-12
& General Health Questionnaire (12-item)
& 12
& 0--36
& Measures general psychological distress using
four-point Likert items. Scores of 11--12 are
typical; scores $>15$ indicate distress;
scores $>20$ indicate severe problems.
& D-1, D-2, D-3 \\

\midrule

PHQ-9
& Patient Health Questionnaire (9-item)
& 9
& 0--27
& Measures depression symptom severity.
Severity: 0--4 minimal; 5--9 mild;
10--14 moderate; 15--19 moderately severe;
20--27 severe.
& D-1, D-2 only \\

\midrule

CESD-R
& CES-D Scale -- Revised \newline
\cite{Eaton2004}
& 20
& 0--60
& Measures depression severity over the past
week. Scores $\geq 16$ indicate clinically
significant symptoms.
\textbf{Replaces PHQ-9 from D-3 onward.}
& D-3 only \\

\midrule

SWLS-overall
& Satisfaction with Life Scale
\cite{Diener1985}
& 5
& 5--35
& Global life-satisfaction measure using
seven-point Likert items. Higher scores
indicate greater satisfaction.
& D-1 only \\

\midrule

SWLS-present
& SWLS present-focused item subset
& 3
& 3--21
& Present-focused subset of SWLS items.
Higher scores indicate greater satisfaction
with current life.
& D-1 only \\

\midrule

SWLS-past
& SWLS past-focused item subset
& 2
& 2--14
& Past-focused subset of SWLS items.
Higher scores indicate greater satisfaction
with past life.
& D-1 only \\

\midrule

RSES
& Rosenberg Self-Esteem Scale
\cite{Rosenberg1965}
& 10
& 10--40 \newline (1--4 per item)
& Measures global self-esteem.
Higher scores indicate greater self-esteem.
& D-1 only \\

\midrule

PCQ (PPC-\*)
& Psychological Capital Questionnaire
\cite{Luthans2007}; subscales:
\textbf{self-efficacy}, \textbf{optimism},
\textbf{hope}, \textbf{resiliency}
& 6 each (self-efficacy, optimism, hope);
3 (resiliency)
& Self-efficacy / optimism / hope:
6--30; resiliency: 3--18
& Measures positive psychological resources.
In D-1, stored as
\texttt{PPC-self\_efficacy}, etc.
In D-3, stored without prefix
(\texttt{self-efficacy}, etc.).
& D-1, D-3 only \\

\bottomrule
\end{tabularx}

\begin{tablenotes}
\footnotesize
\item[] \textit{Note.}
Depression was measured with PHQ-9 in D-1/D-2
and replaced by CESD-R in D-3.
SWLS (overall/present/past) and RSES are available
in D-1 only. PCQ subscales
(\texttt{PPC-\*}) are available in D-1 and
reappear in D-3 without the prefix.
\end{tablenotes}

\end{threeparttable}
\end{table}

\vspace{-1.0em}
\subsection*{A.5 Survey Details}

\vspace{-0.5em}

This section describes all surveys administered across the three waves. Surveys fall into two categories: post-study psychological baseline questionnaires collected once per participant at the end of each wave, and ESM label instruments collected repeatedly throughout the study period.

\subsubsection*{A.5.1 Post-Study Psychological Baseline Questionnaires}

% The following questionnaires were administered after the data collection period ended. Responses are stored in \texttt{UserInfo.csv} per wave.

% \input{Tables/appendix_a_poststudy_questionnaire}
\vspace{-0.5em}

The post-study psychological baseline questionnaires are summarised in Table~\ref{tab:A4} above. These were administered after the data collection period ended, and responses are stored in \texttt{UserInfo.csv} per wave.

\subsubsection*{A.5.2 ESM Label Instruments}

\vspace{-0.5em}

Affective and cognitive state labels were collected via ESM prompts delivered through ABC Logger throughout each wave. The ESM questionnaire differs across waves while preserving a shared core label set to support cross-wave benchmarking.

\vspace{-1.2em}
% \begin{table}[H]
% \centering
% \small
% \renewcommand{\arraystretch}{1.4}
% \caption{D-1 ESM questionnaire fields (8 items).}
% \label{tab:A6}
% \begin{tabularx}{\textwidth}{p{2.2cm}p{2.8cm}X}
% \toprule
% \textbf{Field} & \textbf{Scale} & \textbf{Description} \\
% \midrule
% Valence     & $-3$ to $+3$    & Emotion valence: very negative $\to$ very positive \\
% Arousal     & $-3$ to $+3$    & Emotion arousal: very calm $\to$ very excited \\
% Stress      & $-3$ to $+3$    & Stress level: not stressed at all $\to$ very stressed \\
% Duration    & 0--60 min / N/A & Duration of the current emotional state in minutes \\
% Attention   & $-3$ to $+3$    & Attention level: very bored $\to$ very engaged \\
% Disturbance & $-3$ to $+3$    & Task disturbance: did not disturb at all $\to$ very disturbing \\
% Mental      & $-3$ to $+3$    & Mental load of the ongoing task: very low $\to$ very high \\
% Change      & $-3$ to $+3$    & Emotion change while answering: felt more negative $\to$ more positive \\
% \bottomrule
% \end{tabularx}
% \end{table}

\begin{table}[H]
\centering
\footnotesize
\renewcommand{\arraystretch}{1.0}
\hbadness=10000
\captionsetup{skip=4pt}
\caption{D-1 ESM questionnaire fields (8 items).}
\label{tab:A6}
\begin{tabularx}{\textwidth}{p{2.0cm}p{2.2cm}X}
\toprule
\textbf{Field} & \textbf{Scale} & \textbf{Description} \\
\midrule
Valence     & $-3$ to $+3$    & Emotion valence: very negative $\to$ very positive \\
Arousal     & $-3$ to $+3$    & Emotion arousal: very calm $\to$ very excited \\
Stress      & $-3$ to $+3$    & Stress level: not stressed at all $\to$ very stressed \\
Duration    & 0--60 min / N/A & Duration of the current emotional state in minutes \\
Attention   & $-3$ to $+3$    & Attention level: very bored $\to$ very engaged \\
Disturbance & $-3$ to $+3$    & Task disturbance: did not disturb at all $\to$ very disturbing \\
Mental      & $-3$ to $+3$    & Mental load of the ongoing task: very low $\to$ very high \\
Change      & $-3$ to $+3$    & Emotion change while answering: felt more negative $\to$ more positive \\
\bottomrule
\end{tabularx}
\end{table}
\vspace{-2.2em}
% \begin{table}[H]
% \centering
% \small
% \renewcommand{\arraystretch}{1.4}
% \caption{D-2 ESM questionnaire fields (9 items).}
% \label{tab:A7}
% \begin{tabularx}{\textwidth}{p{2.8cm}p{2.8cm}X}
% \toprule
% \textbf{Field} & \textbf{Scale} & \textbf{Description} \\
% \midrule
% Valence        & $-3$ to $+3$    & Emotion valence: very negative $\to$ very positive \\
% Arousal        & $-3$ to $+3$    & Emotion arousal: very calm $\to$ very excited \\
% Stress         & $-3$ to $+3$    & Stress level: not stressed at all $\to$ very stressed \\
% Duration       & 0--60 min / N/A & Duration of the current emotional state in minutes \\
% Attention      & $-3$ to $+3$    & Attention level: very bored $\to$ very engaged \\
% Disturbance    & $-3$ to $+3$    & Task disturbance: did not disturb at all $\to$ very disturbing \\
% Mental         & $-3$ to $+3$    & Mental load of the ongoing task: very low $\to$ very high \\
% ChangedValence & $-3$ to $+3$    & Valence state immediately after answering the ESM. Replaces the single Change item from D-1. \\
% ChangedArousal & $-3$ to $+3$    & Arousal state immediately after answering the ESM. Replaces the single Change item from D-1. \\
% \bottomrule
% \end{tabularx}
% \end{table}

\begin{table}[H]
\centering
\footnotesize
\renewcommand{\arraystretch}{1.0}
\hbadness=10000
\captionsetup{skip=4pt}
\caption{D-2 ESM questionnaire fields (9 items).}
\label{tab:A7}
\begin{tabularx}{\textwidth}{p{2.0cm}p{2.2cm}X}
\toprule
\textbf{Field} & \textbf{Scale} & \textbf{Description} \\
\midrule
Valence        & $-3$ to $+3$    & Emotion valence: very negative $\to$ very positive \\
Arousal        & $-3$ to $+3$    & Emotion arousal: very calm $\to$ very excited \\
Stress         & $-3$ to $+3$    & Stress level: not stressed at all $\to$ very stressed \\
Duration       & 0--60 min / N/A & Duration of the current emotional state in minutes \\
Attention      & $-3$ to $+3$    & Attention level: very bored $\to$ very engaged \\
Disturbance    & $-3$ to $+3$    & Task disturbance: did not disturb at all $\to$ very disturbing \\
Mental         & $-3$ to $+3$    & Mental load of the ongoing task: very low $\to$ very high \\
ChangedValence & $-3$ to $+3$    & Valence state immediately after answering the ESM. Replaces the single Change item from D-1. \\
ChangedArousal & $-3$ to $+3$    & Arousal state immediately after answering the ESM. Replaces the single Change item from D-1. \\
\bottomrule
\end{tabularx}
\end{table}
\vspace{-2.2em}
\begin{table}[H]
\centering
\footnotesize
\renewcommand{\arraystretch}{1.0}
\hbadness=10000
\captionsetup{skip=4pt}
\caption{D-3 ESM questionnaire fields (12 items).}
\label{tab:A8}
\begin{threeparttable}
\begin{tabularx}{\textwidth}{p{2.0cm}p{2.2cm}X}
\toprule
\textbf{Field} & \textbf{Scale} & \textbf{Description} \\
\midrule
\multicolumn{3}{l}{\textit{Custom affect-intensity items (new in D-3)}} \\
\midrule
Happy       & 0 to $+6$ & Intensity of feeling happy \\
Relaxed     & 0 to $+6$ & Intensity of feeling relaxed \\
Cheerful    & 0 to $+6$ & Intensity of feeling cheerful \\
Content     & 0 to $+6$ & Intensity of feeling content \\
Sad         & 0 to $+6$ & Intensity of feeling sad \\
Anxious     & 0 to $+6$ & Intensity of feeling anxious \\
Depressed   & 0 to $+6$ & Intensity of feeling depressed \\
Angry       & 0 to $+6$ & Intensity of feeling angry \\
\midrule
\multicolumn{3}{l}{\textit{Core labels (shared with D-1 and D-2)}} \\
\midrule
Valence     & $-3$ to $+3$ & Emotion valence: very negative $\to$ very positive. Scale unchanged from D-1/D-2. \\
Arousal     & $-3$ to $+3$ & Emotion arousal: very calm $\to$ very excited. Scale unchanged from D-1/D-2. \\
Stress      & 0 to $+6$\tnote{*} & Stress level. Scale changed from $-3/+3$ to $0/+6$ relative to D-1/D-2. \\
Disturbance & 0 to $+6$\tnote{*} & Task disturbance level. Scale changed from $-3/+3$ to $0/+6$ relative to D-1/D-2. \\
\bottomrule
\end{tabularx}
% \begin{tablenotes}
% % \footnote{}
% \item[*] Scale changed from $-3/+3$ in D-1/D-2; linearly rescaled back to $-3/+3$ before binarisation for cross-wave consistency.
% \item[*] \textit{Note on questionnaire evolution.} D-1 established the base 8-item instrument. In D-2, the single Change item was split into ChangedValence and ChangedArousal to better capture the multidimensional nature of emotional shifts. D-3 introduced eight PANAS-style discrete affect-word items and rescaled Stress and Disturbance from $-3/+3$ to $0/+6$. 
% % The four shared core labels -- Valence, Arousal, Stress, and Disturbance -- are present across all three waves and serve as the basis for cross-wave benchmarking.
% \end{tablenotes}
\end{threeparttable}
\end{table}

\subsection*{A.6 Sensor and Modality Details}

All raw sensor data are stored as per-participant UTF-8 encoded CSV files. Timestamps across all files use Unix epoch time in UTC+0 (milliseconds since 1 January 1970). Data are organised into two sources: smartphone streams collected via ABC Logger, and wearable streams from the Fitbit Inspire HR. Smartphone streams are further categorised into system sensing, interaction sensing, and context sensing.

Figure~\ref{fig:sensor_label_matrix} (in the main text) provides a visual summary of sensor and modality availability across the three waves.

\subsubsection*{A.6.1 Smartphone Sensor Files}

% \begin{table}[H]
% \centering
% \footnotesize
% \renewcommand{\arraystretch}{1.2}
% \caption{System sensing files.}
% \label{tab:system_sensing}
% \begin{tabularx}{\textwidth}{p{2.3cm}p{3.0cm}p{1.3cm}>{\raggedright\arraybackslash}X}
% \toprule
% \textbf{File} & \textbf{Description} & \textbf{Waves} & \textbf{Key Fields} \\
% \midrule
% BatteryEvent.csv
%   & Battery level and charging state (event-based)
%   & D-1, D-2, D-3
%   & \texttt{status} (CHARGING / DISCHARGING / FULL / NOT\_CHARGING), \texttt{temperature} (deci-°C), \texttt{level} (\%), \texttt{plugged} (AC / USB / UNDEFINED) \\
% \midrule
% InstalledApp.csv
%   & Installed applications list, queried twice per day
%   & D-1, D-2, D-3
%   & \texttt{name}, \texttt{packageName}, \texttt{isSystemApp}, \texttt{isUpdatedSystemApp}, \texttt{firstInstallTime}, \texttt{lastUpdateTime} \\
% \bottomrule
% \end{tabularx}
% \end{table}

\begin{table}[H]
\centering
\footnotesize
\renewcommand{\arraystretch}{1.0}
\hbadness=10000
\captionsetup{skip=4pt}
\caption{System sensing files.}
\label{tab:system_sensing}
\begin{tabularx}{\textwidth}{p{2.3cm}p{3.0cm}p{1.3cm}>{\raggedright\arraybackslash}X}
\toprule
\textbf{File} & \textbf{Description} & \textbf{Waves} & \textbf{Key Fields} \\
\midrule
BatteryEvent.csv
  & Battery level and charging state (event-based)
  & D-1, D-2, D-3
  & \texttt{status} (CHARGING / DISCHARGING / FULL / NOT\_CHARGING), \texttt{temperature} (deci-°C), \texttt{level} (\%), \texttt{plugged} (AC / USB / UNDEFINED) \\
\midrule
InstalledApp.csv
  & Installed applications list, queried twice per day
  & D-1, D-2, D-3
  & \texttt{name}, \texttt{packageName}, \texttt{isSystemApp}, \texttt{isUpdatedSystemApp}, \texttt{firstInstallTime}, \texttt{lastUpdateTime} \\
\bottomrule
\end{tabularx}
\end{table}
\vspace{-1.0em}

\begin{table}[H]
\centering
\footnotesize
\renewcommand{\arraystretch}{1.0}
\hbadness=10000
\captionsetup{skip=4pt}
\caption{Interaction sensing files.}
\label{tab:interaction_sensing}
\begin{tabularx}{\textwidth}{p{2.7cm}p{3.0cm}p{1.4cm}>{\raggedright\arraybackslash}X}
\toprule
\textbf{File} & \textbf{Description} & \textbf{Waves} & \textbf{Key Fields} \\
\midrule
AppUsageEvent.csv
  & App foreground / background state changes
  & D-1, D-2, D-3
  & \texttt{name}, \texttt{packageName}, \texttt{isSystemApp}, \texttt{type} (ACTIVITY\_PAUSED, ACTIVITY\_RESUMED, SCREEN\_INTERACTIVE, SCREEN\_NON\_INTERACTIVE, \ldots) \\
\midrule
NotificationEvent.csv
  & App notification history
  & D-1, D-2, D-3
  & \texttt{name}, \texttt{packageName}, \texttt{visibility} (PUBLIC / PRIVATE / SECRET / UNDEFINED), \texttt{category} (alarm / call / email / msg / social / \ldots) \\
\midrule
CallEvent.csv
  & Incoming / outgoing / missed call history
  & D-1, D-2, D-3
  & \texttt{type} (INCOMING / OUTGOING / MISSED / UNDEFINED), \texttt{number} (MD5-hashed), \texttt{isPinned}, \texttt{isStarred}, \texttt{duration} (s), \texttt{dataUsage} \\
\midrule
MessageEvent.csv
  & SMS / MMS history
  & D-1, D-2, D-3
  & \texttt{number} (MD5-hashed), \texttt{messageClass} (SMS / MMS / UNDEFINED), \texttt{messageBox} (INBOX / SENT / UNDEFINED) \\
\midrule
DeviceEvent.csv
  & System-level device state changes
  & D-1, D-2, D-3
  & \texttt{type} (SCREEN\_ON / SCREEN\_OFF, BLUETOOTH\_ACTIVATE / DEACTIVATE, POWER\_CONNECTED / DISCONNECTED, RINGER\_MODE\_NORMAL / SILENT / VIBRATE, \ldots) \\
\midrule
KeyEvent.csv
  & Keystroke touch interaction events
  & D-2, D-3 only
  & \texttt{keyboardType} (CHUNJIIN / QWERTY\_KOR / OTHERS), \texttt{prevKeyType} / \texttt{currKeyType} (BACKSPACE / KOR / ENG / NUMBER / SPECIAL), \texttt{distance} (key-grid), \texttt{timeTaken} (ms) \\
\bottomrule
\end{tabularx}
\end{table}

\begin{table}[H]
\centering
\footnotesize
\renewcommand{\arraystretch}{1.0}
\hbadness=10000
\captionsetup{skip=4pt}
\caption{Context sensing files.}
\label{tab:context_sensing}
\begin{threeparttable}
\begin{tabularx}{\textwidth}{p{2.8cm}p{2.5cm}p{1.5cm}p{1.3cm}>{\raggedright\arraybackslash}X}
\toprule
\textbf{File} & \textbf{Description} & \textbf{Sampling Rate} & \textbf{Waves} & \textbf{Key Fields} \\
\midrule
Light.csv
  & Ambient light intensity
  & 5 Hz
  & D-1, D-2, D-3
  & \texttt{brightness} (lux) \\
\midrule
ActivityEvent.csv
  & Physical activity type with confidence
  & Event-based
  & D-1, D-2, D-3
  & \texttt{type} (STILL / WALKING / RUNNING / IN\_VEHICLE / ON\_BICYCLE / TILTING / UNKNOWN), \texttt{confidence} (0--100\%) \\
\midrule
ActivityTransition.csv
  & Activity state transitions
  & Event-based
  & D-1, D-2, D-3
  & \texttt{type}, \texttt{isEntered} (TRUE / FALSE) \\
\midrule
Location.csv
  & GPS location changes
  & Event-based
  & D-1, D-2, D-3
  & \texttt{latitude}, \texttt{longitude}\tnote{$\dagger$}, \texttt{altitude}, \texttt{accuracy} (m), \texttt{speed} (m/s) \\
\midrule
DataTraffic.csv
  & Network data usage
  & Periodic
  & D-1, D-2\tnote{$\ddagger$}, D-3
  & \texttt{fromTime}, \texttt{toTime}, \texttt{rxBytes}, \texttt{txBytes}, \texttt{mobileRxBytes}, \texttt{mobileTxBytes} \\
\midrule
WiFiScan.csv
  & Nearby Wi-Fi access points
  & $\sim$1/min
  & D-1, D-2, D-3
  & \texttt{bssid}\tnote{$\S$} (UUID-replaced), \texttt{frequency} (MHz), \texttt{rssi} (dBm) \\
\midrule
BluetoothScan.csv
  & Nearby Bluetooth devices
  & $\sim$1/min
  & D-1, D-2, D-3
  & \texttt{address}\tnote{$\S$} (UUID-replaced), \texttt{rssi} (dBm) \\
\bottomrule
\end{tabularx}
\begin{tablenotes}
  \small
  \item[$\dagger$] GPS longitude is offset by a fixed random per-participant constant displacement to preserve relative spatial structure without exposing precise locations.
  \item[$\ddagger$] Network traffic data experienced a collection application malfunction in D-2 and should be treated with caution for that wave.
  \item[$\S$] Wi-Fi BSSID and Bluetooth MAC addresses are replaced with UUIDs using the Python \texttt{uuid} library.
\end{tablenotes}
\end{threeparttable}
\end{table}

\subsubsection*{A.6.2 Wearable Sensor Files}

\begin{table}[H]
\centering
\footnotesize
\renewcommand{\arraystretch}{1.0}
\hbadness=10000
\captionsetup{skip=4pt}
\caption{Wearable sensor files.}
\label{tab:A9}
\begin{tabularx}{\textwidth}{p{2.3cm}p{2.2cm}p{1.3cm}p{1.7cm}>{\raggedright\arraybackslash}X}
\toprule
\textbf{File} & \textbf{Sensor} & \textbf{Sampling Rate} & \textbf{Waves} & \textbf{Description} \\
\midrule
Fitbit-HeartRate.csv
  & Fitbit Inspire HR & 1 Hz    & D-1, D-2, D-3 & Heart rate (BPM) \\
\midrule
Fitbit-StepCount.csv
  & Fitbit Inspire HR & 1/60 Hz & D-1, D-2, D-3 & Steps walked per minute \\
\midrule
Fitbit-Calorie.csv
  & Fitbit Inspire HR & 1/60 Hz & D-1, D-2, D-3 & Calories burned per minute; activity level (0--3); METs \\
\midrule
Fitbit-Distance.csv
  & Fitbit Inspire HR & 1/60 Hz & D-1, D-2, D-3 & Distance traveled per minute (meters) \\
\bottomrule
\end{tabularx}
\end{table}

\subsubsection*{A.6.3 Privacy Transformations}

The following privacy-preserving transformations were applied to the raw data before release:

\begin{itemize}
  \item \textbf{GPS location} -- Longitude values received a fixed random per-participant constant displacement. This generates a pseudo-location trace that preserves relative distances and allows location clustering while concealing absolute geographic positions.
  \item \textbf{Call and message contact identifiers} -- Contact numbers in \texttt{CallEvent.csv} and \texttt{MessageEvent.csv} were one-way hashed using MD5. Original digits cannot be recovered, but within-contact consistency (e.g., call frequency counts) is preserved.
  \item \textbf{Wi-Fi and Bluetooth hardware addresses} -- Wi-Fi BSSID values and Bluetooth MAC addresses were replaced with UUIDs using the Python \texttt{uuid} library. This allows co-presence and proximity density estimation without exposing device identities.
\end{itemize}

\subsection*{A.7 Feature Extraction Details}

\subsubsection*{A.7.1 Pipeline Overview}

The raw sensor CSV files described in Section~A.6 are processed into benchmark-ready feature matrices using the preprocessing and feature extraction pipeline described in~\cite{Zhang2024}. The pipeline converts per-participant raw smartphone and wearable sensor streams into fixed-format feature vectors aligned to each ESM prompt. All benchmark features are derived from sensor data available before the corresponding ESM prompt; prior ESM label values are not used as input features in the reported benchmark.

\subsubsection*{A.7.2 Feature Naming Convention}

Every feature column in the benchmark-ready files follows a structured naming convention based on feature group, descriptor, optional categorical value, and optional time-window suffix:

\begin{center}
\texttt{\{FEATURE\_GROUP\}\#\{DESCRIPTOR\}[=\{VALUE\}][\#\{WINDOW\}]}
\end{center}

\noindent In the released feature files, some historical feature names preserve implementation-specific suffixes using either \texttt{\#} or \texttt{\_} before the window name. During cross-wave alignment, these names are canonicalized before constructing the shared feature intersection.

\begin{itemize}
  \item \textbf{FEATURE\_GROUP} -- the feature group identifier, such as \texttt{CALL\_DUR}, \texttt{APP\_CAT}, or \texttt{FitbitHeartrate}.
  \item \textbf{DESCRIPTOR} -- the feature descriptor. Numeric descriptors include \texttt{VAL} (instantaneous value), \texttt{AVG} (mean), \texttt{STD} (standard deviation), \texttt{SKW} (skewness), \texttt{KUR} (kurtosis), \texttt{MED} (median), \texttt{BEP} (binned entropy), \texttt{ASC} (absolute sum of changes), and \texttt{TSC} (time-series complexity). Categorical descriptors include one-hot value indicators and categorical summary descriptors such as entropy, duration-since-change, and relative-support features.
  \item \textbf{=VALUE} (optional) -- appended for categorical features to specify the encoded category. For example, \texttt{APP\_CAT\#VAL=SOCIAL} indicates that the foreground app belongs to the social category.
  \item \textbf{WINDOW} (optional) -- the time window over which the descriptor is computed. The pipeline uses two rolling windows immediately preceding the prompt, \texttt{ImmediatePast\_15} and \texttt{ImmediatePast\_30}, together with six three-hour epoch windows: \texttt{Dawn}, \texttt{Morning}, \texttt{Afternoon}, \texttt{LateAfternoon}, \texttt{Evening}, and \texttt{Night}. These epoch windows are generated for both the current day and the previous day using \texttt{Today} and \texttt{Yesterday} prefixes.
\end{itemize}

% \subsubsection*{A.7.3 Feature Categories and Time Windows}

% The feature groups in Table~\ref{tab:A10} are organised into six sensing categories. For each ESM prompt, the pipeline extracts sensor-derived features from data preceding the prompt. The implemented time windows are: (i) 15-minute and 30-minute rolling windows immediately before the prompt; (ii) six three-hour epoch windows for the current day and the previous day, namely \texttt{Dawn} (06:00--09:00), \texttt{Morning} (09:00--12:00), \texttt{Afternoon} (12:00--15:00), \texttt{LateAfternoon} (15:00--18:00), \texttt{Evening} (18:00--21:00), and \texttt{Night} (21:00--24:00); and (iii) a nightly sleep window derived from screen-event patterns for sleep-related features such as sleep duration, onset, and midpoint.

% Across these windows, the pipeline computes statistical and temporal descriptors including mean, standard deviation, skewness, kurtosis, median, binned entropy, absolute sum of changes, and time-series complexity, together with categorical summaries where applicable. Features derived from prior ESM labels, such as last observed label values, prior HIGH-label likelihoods, or rolling ESM-label means and standard deviations, are not included in the reported benchmark feature matrices.
\subsubsection*{A.7.3 Feature Categories, Prompt Alignment, and Time Windows}

The feature groups in Table~\ref{tab:A10} are organised into six sensing categories. Each feature vector is aligned to an ESM prompt at time $t$, and all benchmark features are extracted only from sensor records observed at or before $t$. The current feature extraction pipeline uses the following active windows.

\begin{itemize}
  \item \textbf{Immediate-past windows.} Two short look-back windows are used: \texttt{ImmediatePast\_15}, covering $[t-15\text{min}, t]$, and \texttt{ImmediatePast\_30}, covering $[t-30\text{min}, t]$. Other short windows such as 5, 10, and 45 minutes are not used in the reported benchmark.

  \item \textbf{Previous-day epoch windows.} Six fixed three-hour epochs from the previous calendar day are used: \texttt{Dawn} (06:00--09:00), \texttt{Morning} (09:00--12:00), \texttt{Afternoon} (12:00--15:00), \texttt{LateAfternoon} (15:00--18:00), \texttt{Evening} (18:00--21:00), and \texttt{Night} (21:00--24:00). These features use the \texttt{Yesterday} prefix.

  \item \textbf{Current-day epoch windows.} The same six three-hour epochs are generated for the current day using the \texttt{Today} prefix. Only epochs whose start time is no later than $t$ are emitted, and the final active epoch is clipped at $t$ so that no post-prompt sensor records are used.

  \item \textbf{Point-in-time features.} Some descriptors do not use an aggregation window. These include \texttt{\#VAL}, the last observed value at or before $t$; \texttt{\#DSC} and \texttt{\#DSC=\{c\}}, the time since the last state change for categorical streams; and calendar descriptors such as \texttt{Time\#DOW}, \texttt{Time\#WKD}, and \texttt{Time\#HRN}.

  \item \textbf{Sleep features.} Sleep-related features, including \texttt{Sleep\#Duration}, \texttt{Sleep\#Onset}, and \texttt{Sleep\#Midpoint}, are derived from screen-event patterns over the overnight window from the previous day 21:00 to the current day 14:00.
\end{itemize}

Across these windows, the pipeline computes statistical and temporal descriptors including mean, standard deviation, skewness, kurtosis, median, binned entropy, absolute sum of changes, and time-series complexity, together with categorical summaries where applicable. Features derived from prior ESM labels, such as last observed label values, prior HIGH-label likelihoods, or rolling ESM-label means and standard deviations, are not included in the reported benchmark feature matrices.

% \begin{table}[H]
% \centering
% \small
% \renewcommand{\arraystretch}{1.4}
% \caption{Feature categories and raw data sources.}
% \label{tab:A11}
% \begin{tabularx}{\textwidth}{p{2.8cm}>{\raggedright\arraybackslash}Xp{3.2cm}}
% \toprule
% \textbf{Category} & \textbf{Raw Data Sources} & \textbf{Waves} \\
% \midrule
% Social interaction
%   & CallEvent.csv, MessageEvent.csv
%   & D-1, D-2, D-3 \\
% \midrule
% Physical activity
%   & ActivityEvent.csv, ActivityTransition.csv, Fitbit-StepCount.csv, Fitbit-Distance.csv, Fitbit-Calorie.csv
%   & D-1, D-2, D-3 \\
% \midrule
% Context and mobility
%   & Location.csv, DataTraffic.csv
%   & D-1, D-2, D-3 (DataTraffic in D-2 treated with caution -- see Table~\ref{tab:A10} footnote $\ddagger$) \\
% \midrule
% Phone usage
%   & AppUsageEvent.csv, InstalledApp.csv, DeviceEvent.csv, BatteryEvent.csv, NotificationEvent.csv, KeyEvent.csv
%   & D-1, D-2, D-3 (KeyEvent: D-2, D-3 only) \\
% \midrule
% Physiological
%   & Fitbit-HeartRate.csv
%   & D-1, D-2, D-3 \\
% \midrule
% Wireless environment
%   & WiFiScan.csv, BluetoothScan.csv
%   & D-1, D-2, D-3 (WiFiScan: D-1, D-2 only) \\
% \bottomrule
% \end{tabularx}
% \end{table}

\begin{table}[H]
\centering
\footnotesize
\renewcommand{\arraystretch}{1.0}
\hbadness=10000
\captionsetup{skip=4pt}
\caption{Feature categories and raw data sources.}
\label{tab:A11}
\begin{tabularx}{\textwidth}{p{2.8cm}>{\raggedright\arraybackslash}Xp{3.2cm}}
\toprule
\textbf{Category} & \textbf{Raw Data Sources} & \textbf{Waves} \\
\midrule
Social interaction
  & CallEvent.csv, MessageEvent.csv
  & D-1, D-2, D-3 \\
\midrule
Physical activity
  & ActivityEvent.csv, ActivityTransition.csv, Fitbit-StepCount.csv, Fitbit-Distance.csv, Fitbit-Calorie.csv
  & D-1, D-2, D-3 \\
\midrule
Context and mobility
  & Location.csv, DataTraffic.csv
  & D-1, D-2, D-3 (DataTraffic in D-2 treated with caution -- see Table~\ref{tab:A10} footnote $\ddagger$) \\
\midrule
Phone usage
  & AppUsageEvent.csv, InstalledApp.csv, DeviceEvent.csv, BatteryEvent.csv, NotificationEvent.csv, KeyEvent.csv
  & D-1, D-2, D-3 (KeyEvent: D-2, D-3 only) \\
\midrule
Physiological
  & Fitbit-HeartRate.csv
  & D-1, D-2, D-3 \\
\midrule
Wireless environment
  & WiFiScan.csv, BluetoothScan.csv
  & D-1, D-2, D-3 (WiFiScan: D-1, D-2 only) \\
\bottomrule
\end{tabularx}
\end{table}

\subsubsection*{A.7.4 Feature Groups and Source Files}
\vspace{-1.2em}

\begin{table}[H]
\centering
\caption{Mapping of feature groups to source CSV files.}
\label{tab:A10}
\footnotesize
\hbadness=10000
\captionsetup{skip=4pt}
\begin{threeparttable}
\begin{tabularx}{\textwidth}{p{2.2cm}p{2.5cm}>{\raggedright\arraybackslash}Xp{2.0cm}}
\toprule
\textbf{CSV Source} & \textbf{Feature Group} & \textbf{Description} & \textbf{Waves} \\
\midrule
\textit{Location.csv}        & LOC\_CLS    & Location class label & D-1, D-2, D-3 \\
                             & LOC\_DST    & Distance between consecutive location readings & D-1, D-2, D-3 \\
                             & LOC\_LABEL  & Semantic location label (home / work / other) & D-1, D-2, D-3 \\
\midrule
\textit{AppUsageEvent.csv}   & APP\_CAT       & App usage category indicator & D-1, D-2, D-3 \\
                             & APP\_DUR\_SOCIAL  & App usage duration: social category & D-1, D-2, D-3 \\
                             & APP\_DUR\_WORK    & App usage duration: work category & D-1, D-2, D-3 \\
                             & APP\_DUR\_INFO    & App usage duration: information category & D-1, D-2, D-3 \\
                             & APP\_DUR\_ENTER   & App usage duration: entertainment category & D-1, D-2, D-3 \\
                             & APP\_DUR\_HEALTH  & App usage duration: health category & D-1, D-2, D-3 \\
                             & APP\_DUR\_SYSTEM  & App usage duration: system category & D-1, D-2, D-3 \\
                             & APP\_DUR\_UNKNOWN & App usage duration: unknown category & D-1, D-2, D-3 \\
\midrule
\textit{InstalledApp.csv}    & INST\_JAC   & Installed apps Jaccard similarity between consecutive readings & D-1, D-2, D-3 \\
\midrule
\textit{ActivityEvent.csv}   & ACE\_WLK    & Activity type confidence: walking & D-1, D-2, D-3 \\
                             & ACE\_RUN    & Activity type confidence: running & D-1, D-2, D-3 \\
                             & ACE\_VHC    & Activity type confidence: in vehicle & D-1, D-2, D-3 \\
                             & ACE\_FOT    & Activity type confidence: on foot & D-1, D-2, D-3 \\
                             & ACE\_BCC    & Activity type confidence: bicycling & D-1, D-2, D-3 \\
                             & ACE\_TLT    & Activity type confidence: tilting & D-1, D-2, D-3 \\
                             & ACE\_UNK    & Activity type confidence: unknown & D-1, D-2, D-3 \\
\midrule
\textit{ActivityTransition.csv} & ACT      & Activity transition event type & D-1, D-2, D-3 \\
\midrule
\textit{Fitbit-HeartRate.csv}   & FitbitHeartrate  & Fitbit heart rate   & D-1, D-2, D-3 \\
\textit{Fitbit-StepCount.csv}   & FitbitStepcount  & Fitbit step count   & D-1, D-2, D-3 \\
\textit{Fitbit-Calorie.csv}     & Fitbitcalorie    & Fitbit calories burned & D-1, D-2, D-3 \\
\textit{Fitbit-Distance.csv}    & Fitbitdistance   & Fitbit distance traveled & D-1, D-2, D-3 \\
\midrule
\textit{CallEvent.csv}       & CALL\_CNT   & Number of calls       & D-1, D-2, D-3 \\
                             & CALL\_DUR   & Total call duration   & D-1, D-2, D-3 \\
\midrule
\textit{MessageEvent.csv}    & MSG\_SNT    & Sent messages         & D-1, D-2, D-3 \\
                             & MSG\_RCV    & Received messages     & D-1, D-2, D-3 \\
                             & MSG\_ALL    & Total messages (sent + received) & D-1, D-2, D-3 \\
\midrule
\textit{BluetoothScan.csv}   & BT\_RSSI      & Bluetooth RSSI signal strength & \textbf{D-1 only} \\
                             & BT\_BondState & Bluetooth device bond state    & \textbf{D-2, D-3 only} \\
                             & BT\_DeviceType & Bluetooth device type         & \textbf{D-2, D-3 only} \\
                             & BT\_classType  & Bluetooth device class type   & \textbf{D-2, D-3 only} \\
\midrule
\textit{DataTraffic.csv}     & DATA\_RCV   & Total data received   & D-1, D-2, D-3\tnote{$\ddagger$} \\
                             & DATA\_SNT   & Total data sent       & D-1, D-2, D-3\tnote{$\ddagger$} \\
                             & DATA\_MRCV  & Mobile data received  & D-1, D-2, D-3\tnote{$\ddagger$} \\
                             & DATA\_MSNT  & Mobile data sent      & D-1, D-2, D-3\tnote{$\ddagger$} \\
% \midrule
% \textit{WiFiScan.csv}        & WIFI\_JAC   & Wi-Fi BSSID Jaccard similarity & \textbf{D-1, D-2 only} \\
%                              & WIFI\_COS   & Wi-Fi RSSI cosine similarity   & \textbf{D-1, D-2 only} \\
%                              & WIFI\_EUC   & Wi-Fi RSSI Euclidean distance  & \textbf{D-1, D-2 only} \\
%                              & WIFI\_MAN   & Wi-Fi RSSI Manhattan distance  & \textbf{D-1, D-2 only} \\
\midrule
\textit{WiFiScan.csv}        & WIFI\_JAC   & Wi-Fi BSSID Jaccard similarity & \textbf{D-1, D-2 only}$^{\dagger}$ \\
                             & WIFI\_COS   & Wi-Fi RSSI cosine similarity   & \textbf{D-1, D-2 only}$^{\dagger}$ \\
                             & WIFI\_EUC   & Wi-Fi RSSI Euclidean distance  & \textbf{D-1, D-2 only}$^{\dagger}$ \\
                             & WIFI\_MAN   & Wi-Fi RSSI Manhattan distance  & \textbf{D-1, D-2 only}$^{\dagger}$ \\
\midrule
\textit{DeviceEvent.csv}     & SCR         & Screen event count          & D-1, D-2, D-3 \\
                             & SCR\_DUR    & Screen-on usage duration    & D-1, D-2, D-3 \\
\midrule
\textit{NotificationEvent.csv} & Notification\_CAT & Notification category indicator  & D-1, D-2, D-3 \\
                             & Notification\_VIS & Notification visibility indicator & D-1, D-2, D-3 \\
\midrule
\textit{KeyEvent.csv}        & keyevent\_CAT  & Key event app category          & \textbf{D-2, D-3 only} \\
                             & keyevent\_DIST & Keystroke distance              & \textbf{D-2, D-3 only} \\
                             & keyevent\_TIME & Keystroke inter-key time delta (ms) & \textbf{D-2, D-3 only} \\
\midrule
\textit{BatteryEvent.csv}    & BAT\_LEV    & Battery level (\%)   & D-1, D-2, D-3 \\
                             & BAT\_TMP    & Battery temperature (°C) & D-1, D-2, D-3 \\
                             & BAT\_PLG    & Battery plug state   & D-1, D-2, D-3 \\
                             & BAT\_STA    & Battery status       & D-1, D-2, D-3 \\
\bottomrule
\end{tabularx}
% \begin{tablenotes}
%   \footnotesize
%   \item[$\ddagger$] DataTraffic-derived features are retained in all three waves' feature matrices, but D-2 DataTraffic streams experienced a collection application malfunction (see Section~A.6.1 footnote $\ddagger$) and D-2 values should be treated with caution.
%   \item[$^{\dagger}$]  \texttt{WiFiScan.csv} exists in D-3, but its schema changed to
%     \texttt{timestamp}, \texttt{address}, \texttt{rssi}, and \texttt{frequency};
%     therefore, the Wi-Fi similarity feature groups listed here are generated only
%     for D-1/D-2.
% \end{tablenotes}
\begin{tablenotes}
  \scriptsize
  \item[$\ddagger$] D-2 DataTraffic is unreliable due to a logger malfunction.
  \item[$\dagger$] D-3 WiFi logs use a different schema; Wi-Fi similarity features are D-1/D-2 only.
\end{tablenotes}
\end{threeparttable}
\end{table}

\subsubsection*{A.7.5 Feature Counts Per Wave}

% \begin{table}[H]
% \centering
% \caption{Feature counts per wave.}
% \label{tab:A12}
% \begin{tabular}{lcc}
% \hline
% \textbf{Dataset} & \textbf{Total feature columns} & \textbf{Columns shared across all three waves} \\
% \hline
% D-1 & 8,037  & 5,017 \\
% D-2 & 10,122 & 5,017 \\
% D-3 & 10,581 & 5,017 \\
% \hline
% \end{tabular}
% \end{table}

\begin{table}[H]
\centering
\footnotesize
\renewcommand{\arraystretch}{1.0}
\hbadness=10000
\captionsetup{skip=4pt}
\caption{Feature counts per wave.}
\label{tab:A12}
\begin{tabular}{lcc}
\toprule
\textbf{Dataset} & \textbf{Total feature columns} & \textbf{Columns shared across all three waves} \\
\midrule
D-1 & 8,037  & 4,890 \\
D-2 & 10,122 & 4,890 \\
D-3 & 10,581 & 4,890 \\
\bottomrule
\end{tabular}
\end{table}

Differences in feature counts across waves arise from sensor availability differences and minor logger-format differences across study waves. In particular, KeyEvent-derived features are available only from D-2 onward, Wi-Fi similarity features are available in D-1/D-2, Bluetooth feature groups differ between D-1 and later waves, and minor changes to the data collection application affected the set of collectible device-event types. For cross-wave evaluation, these wave-specific feature spaces are canonicalized and intersected as described in Appendix~C.1.1.

\subsubsection*{A.7.6 Label Binarisation}

% All ESM labels are first mapped to a common scale before binarisation. For D-3, Stress, Disturbance, and the PANAS-style affective intensity items (Happy, Relaxed, Cheerful, Content, Sad, Anxious, Depressed, and Angry) were originally collected on a 0 to $+6$ scale; these labels are linearly rescaled to $-3$ to $+3$ so that label encodings are comparable with the shared D-1/D-2 scale.

% We construct binary HIGH/LOW benchmark targets by thresholding at the \emph{training-data median} for each label (HIGH if the response is strictly above the threshold, LOW otherwise). The threshold is fitted on labeled training data only; validation/test labels are never used to estimate it. HIGH/LOW therefore represents a benchmark-relative split, not a clinical or absolute affect level.

\noindent Binarisation thresholds:
\begin{itemize}
  \item \textbf{Temporal benchmark.} For each participant, the median is estimated from that participant's chronological training slice and applied unchanged to validation/test responses.
  \item \textbf{Within-wave and cross-wave benchmarks.} Each label's threshold is estimated from the training participants/source wave(s) and applied unchanged to held-out participants or waves.
\end{itemize}

\noindent The resulting binary label columns are used only as prediction targets. They are not used to construct input features in the reported benchmark.

\subsubsection*{A.7.7 Missing Sensor Data Handling}

Missingness arises naturally in passive sensing because streams are event-based, devices may be offline, permissions may be unavailable, and wearable or smartphone uploads may be interrupted. The feature extraction pipeline handles missingness at the stream, window, and feature-matrix levels.

First, raw streams are resampled at sensor-specific rates before feature computation. Duplicate timestamps are resolved by keeping the last observation. Numeric streams are resampled by averaging within each interval, linearly interpolated across gaps, and then stripped of leading or trailing missing values. Categorical streams are resampled using forward fill. If invalid values such as remaining NaN or infinite values are encountered during descriptor extraction, the corresponding window-level feature is treated as missing.

Second, point-in-time features use fixed fallback values when no valid prior observation is available. For \texttt{\#VAL}, the fallback is 0. For duration-since-change descriptors, \texttt{\#DSC} uses a sentinel value of $-1$ when no prior state change can be identified.

Third, after per-window feature extraction, remaining invalid entries are sanitized before constructing the benchmark-ready matrix. Numeric missing or invalid values are set to 0.0, while object or Boolean missing values are set to \texttt{False}. If an entire sensor stream is unavailable for a participant, features depending on that stream are absent for that participant and are filled during the final feature-matrix alignment. These fixed missing-value rules are applied uniformly and do not use ESM label values.

\newpage
\section {Additional Dataset Characterization}
\label{appendix:dataset_characterization_validation}

This appendix provides extended empirical characterization of the three-wave dataset across six dimensions: dataset overview, label richness, data density and participation, sensor coverage, distributional shift across waves, and individual variability. Each subsection is accompanied by representative figures from the full exploratory data analysis.

\subsection*{B.1 The Multi-Year Dataset: Overview}

\begin{figure}[H]
    \centering
    \includegraphics[width=\textwidth]{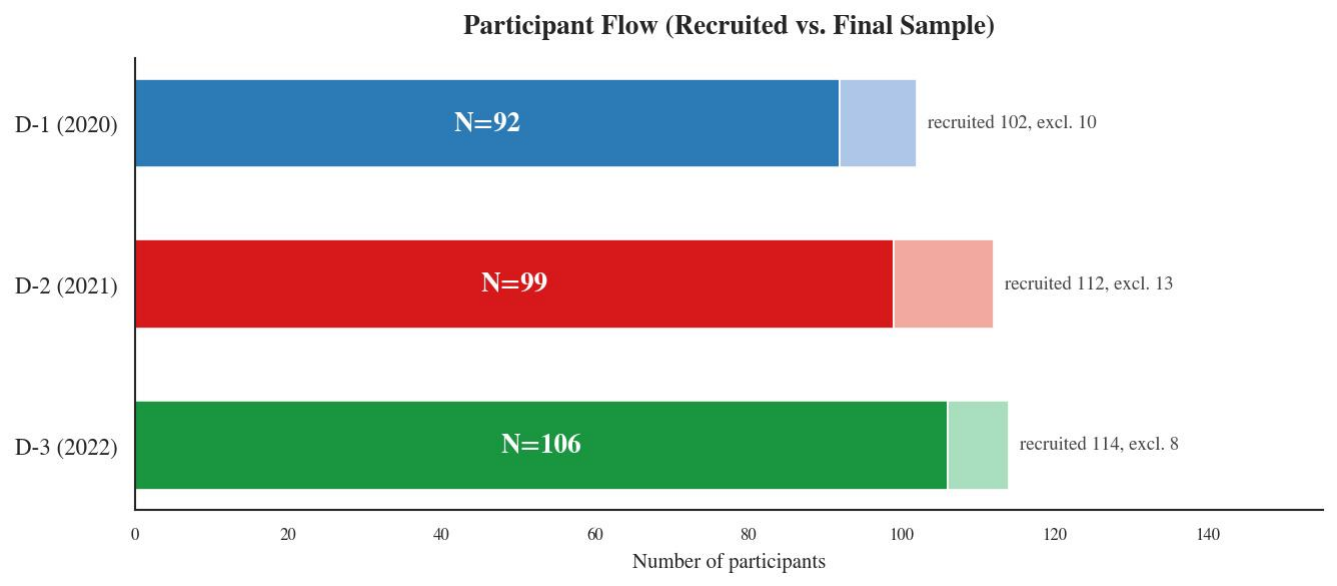}
    \caption{Participant flow from recruitment to final analysis-eligible sample per wave. Solid bars show the final retained sample (D-1: $N = 92$, D-2: $N = 99$, D-3: $N = 106$); lighter extensions show excluded participants. Exclusion rates were 9.8\%, 11.6\%, and 7.0\% for D-1, D-2, and D-3 respectively.}
    \label{fig:participant_flow}
\end{figure}

Figure~\ref{fig:study_design_protocol} demonstrates the study design protocol. Figure~\ref{fig:participant_flow} shows the participant flow from recruitment to the final analysis-eligible sample across all three collection waves. A total of 328 participants were recruited (D-1: 102, D-2: 112, D-3: 114), of whom 297 remained after quality-control (QC) filtering (D-1: $N = 92$, D-2: $N = 99$, D-3: $N = 106$). QC excluded participants with wearable data coverage below 50\%, smartphone data coverage below 50\%, or an ESM response rate below 30\%. Exclusion rates were modest and broadly comparable across waves (D-1: 9.8\%, D-2: 11.6\%, D-3: 7.0\%), indicating stable recruitment and data collection procedures.

\begin{figure}[H]
    \centering
    \includegraphics[width=\textwidth]{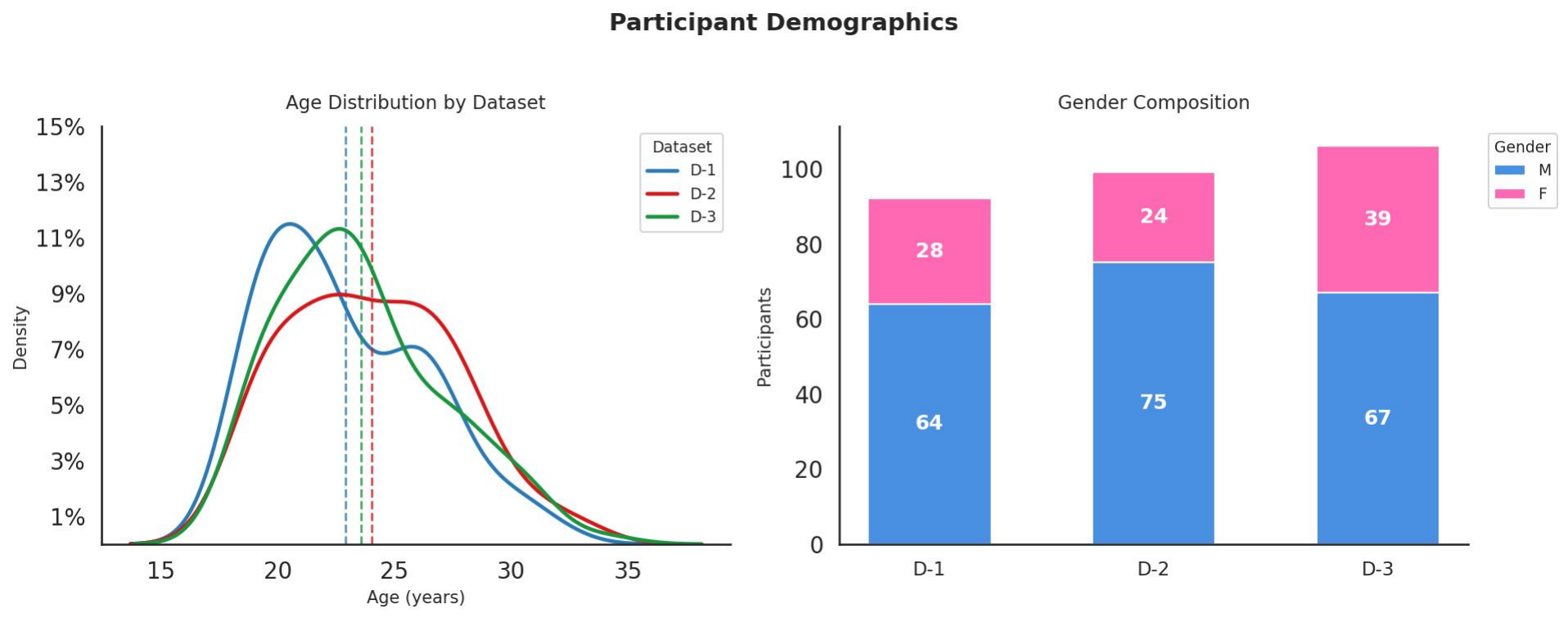}
    \caption{Participant demographics across waves. Left: age distribution per wave (line plot, dashed lines indicate wave means). Right: gender composition per wave (stacked bar). Age distributions are closely matched across waves; the female proportion ranges from 24\% (D-2) to 37\% (D-3).}
    \label{fig:demographics}
\end{figure}

Figure~\ref{fig:demographics} characterises the age and gender composition of the retained samples. All three cohorts were drawn from the same university student population, resulting in highly comparable age distributions across waves (means: D-1 = 22.9 yr, D-2 = 24.0 yr, D-3 = 23.6 yr). The female proportion ranged from 24\% to 37\% across waves, with male participants constituting the majority in all three. The age-by-gender histogram confirms that the gender imbalance is consistent across age bins, rather than being concentrated in any specific age subgroup, suggesting that the imbalance reflects the broader gender composition of the source population rather than differential dropout or self-selection by gender and age jointly.

\subsection*{B.2 Label Richness Across Waves}

\begin{figure}[H]
    \centering
    \includegraphics[width=\textwidth]{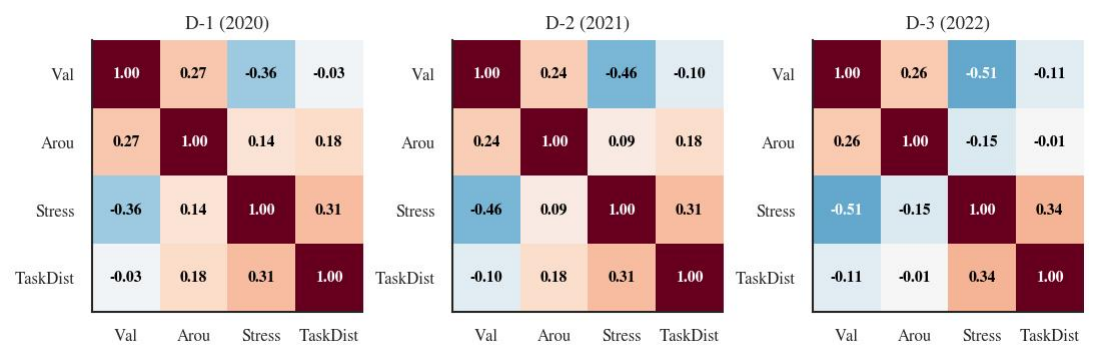}
    \caption{Pairwise Spearman rank correlation matrices for the four shared ESM labels (Valence, Arousal, Stress, Task Disturbance) computed per wave. Red cells indicate positive correlation; blue cells indicate negative correlation. The negative Valence--Stress correlation strengthens monotonically from D-1 to D-3 ($\rho$: $-0.35 \rightarrow -0.45 \rightarrow -0.51$).}
    \label{fig:label_correlation}
\end{figure}

% The dataset employs a wave-specific ESM label instrument with a shared core across the three waves, with a shared core of four affect dimensions -- Valence, Arousal, Stress, and Task Disturbance -- present in all waves and constituting the primary benchmark targets. Figure~\ref{fig:label_correlation} presents the pairwise Spearman rank correlation matrices among these four shared labels computed separately for each wave.
The dataset employs wave-specific ESM instruments with a shared core
of four affect dimensions---Valence, Arousal, Stress, and Task
Disturbance---present across all three waves and constituting the
primary benchmark targets. Figure~\ref{fig:label_correlation}
presents the pairwise Spearman rank correlation matrices among
these four shared labels, computed separately for each wave.

The inter-label structure is broadly consistent across waves and reflects well-established affective relationships. Valence and Stress exhibit a moderate-to-strong negative correlation in all three waves (D-1: $\rho = -0.35$; D-2: $\rho = -0.45$; D-3: $\rho = -0.51$), with the association strengthening incrementally across collection years, although these differences cannot be attributed to a single design change. Stress and Task Disturbance are positively correlated across all waves ($\rho = 0.31, 0.31, 0.35$), reflecting their conceptual proximity as negative cognitive-load states. Valence and Arousal are weakly positively correlated ($\rho = 0.24$--$0.28$), consistent with their orthogonality in the circumplex model of affect. Notably, the Arousal--Stress correlation shifts from weak positive in D-1 ($\rho = 0.15$) to near-zero or slightly negative in D-2 and D-3 ($\rho = 0.08$, $-0.15$), a descriptive difference that cannot be attributed specifically to response-scale or questionnaire changes.

\begin{figure}[H]
    \centering
    \includegraphics[width=\linewidth]{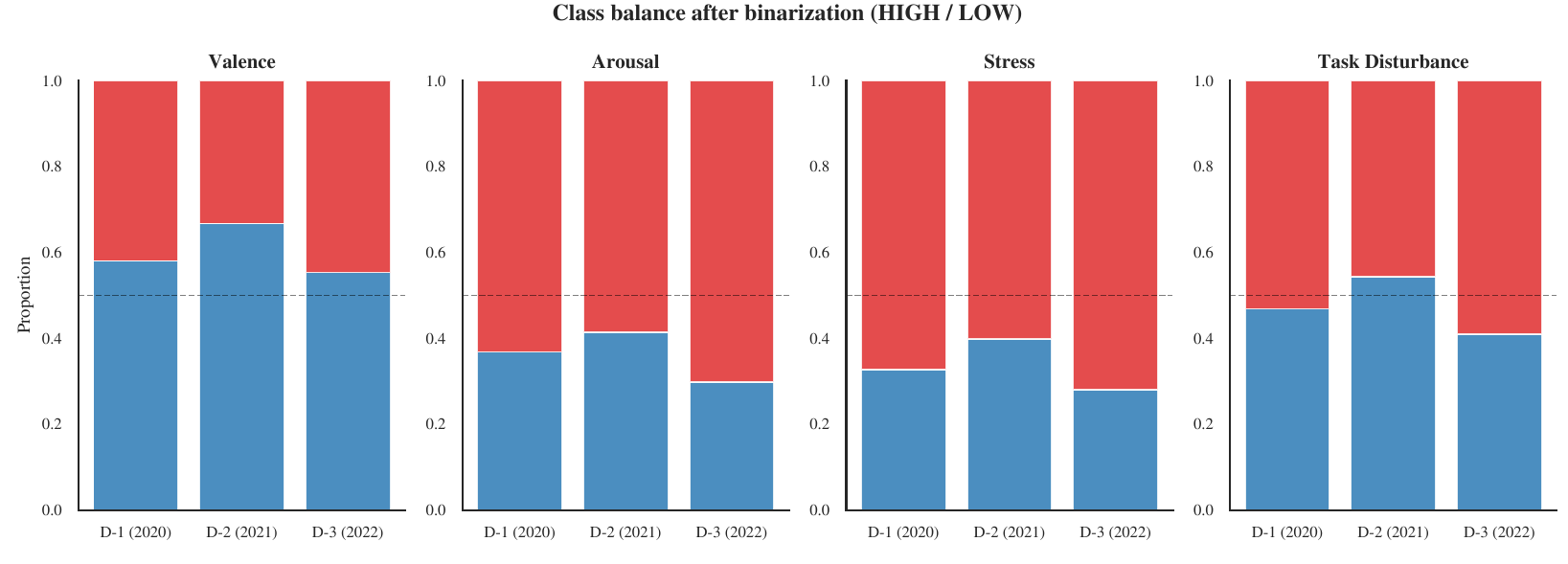}
    % \caption{Class balance after user-level median binarization for the four shared labels (Valence, Arousal, Stress, Task Disturbance) across D-1, D-2, and D-3. Each bar shows the proportion of HIGH and LOW responses per wave.}
    \caption{Exploratory class distributions for the four shared ESM
labels (Valence, Arousal, Stress, and Task Disturbance) across
D-1, D-2, and D-3. Each bar shows the proportion of HIGH and
LOW responses within a wave.}
    \label{fig:label_balance}
\end{figure}

% Figure~\ref{fig:label_balance} complements the inter-label correlation view by showing the class balance resulting from user-level median binarization applied to all benchmark targets. For each of the four shared labels, the proportion of HIGH and LOW responses is shown separately per wave. The resulting class distributions remain reasonably balanced across labels and waves, supporting AUROC-based evaluation while preserving the interpretation of HIGH/LOW as deviations above or below each participant's typical affective level.
Figure~\ref{fig:label_balance} complements the inter-label
correlation analysis by illustrating the empirical class
distributions of the four shared labels across the three waves.
The observed HIGH/LOW proportions provide descriptive information
about class balance and potential differences across waves.
These exploratory distributions should be distinguished from the
benchmark-specific target construction protocol described in
Section~4.3 and Appendix~A.7.6.

\subsection*{B.3 Data Density and Participation}

\begin{figure}[H]
    \centering
    \includegraphics[width=0.55\textwidth]{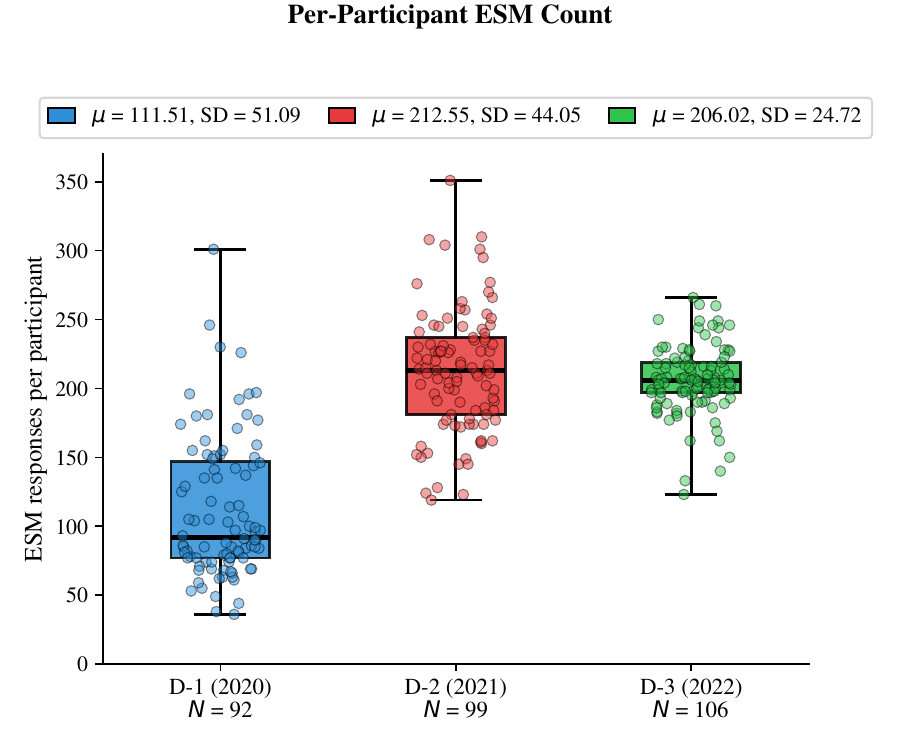}
    % \caption{Per-participant total ESM response count distributions per wave (D-1: $\mu = 108$, SD $= 49$; D-2: $\mu = 190$, SD $= 39$; D-3: $\mu = 195$, SD $= 32$). Box shows IQR; individual points show per-participant values. D-2 and D-3 exhibit substantially higher and more consistent label volumes than D-1, reflecting wave-specific differences in prompt frequency and participation patterns.}
    \caption{Per-participant total ESM response count distributions across the three collection waves (D-1: $\mu=111.51$, SD $\approx51$; D-2: $\mu=212.55$, SD $\approx44$; D-3: $\mu=206.02$, SD $\approx25$). Boxes represent the interquartile range (IQR), and individual points represent retained participants. All statistics are calculated using the same post-QC participant cohorts as Table 2.}
    \label{fig:esm_response_counts}
\end{figure}

% Figure~\ref{fig:esm_response_counts} shows the distribution of per-participant total ESM response counts across all three waves. The substantially wider and lower-centred distribution for D-1 (mean $\mu = 112$, SD $= 51$) compared to D-2 ($\mu = 213$, SD $= 44$) and D-3 ($\mu = 206$, SD $= 25$) reflects the higher ESM prompt frequency in D-1 (15/day) paired with lower per-prompt compliance, alongside the shorter effective participation windows of several D-1 participants. D-2 and D-3 show tight, well-concentrated response count distributions, with most participants contributing between 150 and 260 labelled ESM events. The marked reduction in variance from D-1 to D-3 reflects the combined effect of the participant-facing progress counter introduced in D-2, the intentional reduction in daily prompt frequency, and improved compliance monitoring protocols. Together, the D-2 and D-3 distributions confirm that the dataset provides dense, individual-level label sequences suitable for within-person temporal modelling.
Figure~\ref{fig:esm_response_counts} illustrates the distribution of ESM response counts per retained participant across the three collection waves. D-1 exhibits fewer responses per participant on average ($\mu=111.51$) and greater variability (SD $\approx51$) than D-2 ($\mu=212.55$, SD $\approx44$) and D-3 ($\mu=206.02$, SD $\approx25$). Despite its higher nominal prompting frequency (15 prompts/day), D-1 yielded fewer responses per participant, suggesting differences in participant compliance and effective participation duration. D-2 and D-3 show more concentrated response-count distributions, with the lowest variability observed in D-3. These descriptive differences may reflect changes in prompting schedules, study design, and participant engagement across waves.

\subsection*{B.4 Sensor Data Characterization}

\begin{figure}[H]
    \centering
    \includegraphics[width=\textwidth]{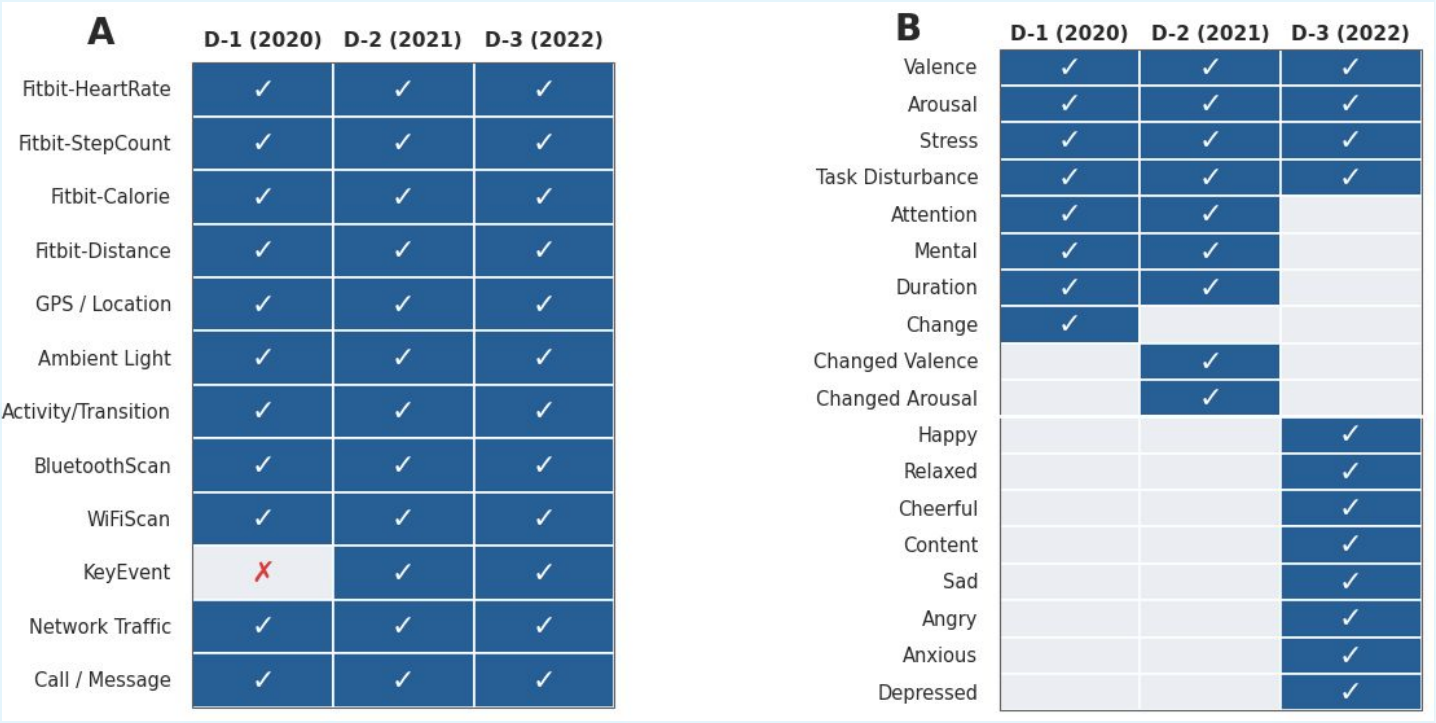}
    \caption{Sensor and label modality coverage matrices. Panel A: sensor stream availability per wave. The KeyEvent stream was added from D-2 onwards. Panel B: ESM label coverage per wave. The four shared core labels (Valence, Arousal, Stress, Task Disturbance) are available in all three waves; wave-specific labels are shown in the stepped staircase pattern.}
    \label{fig:sensor_label_coverage}
\end{figure}

Figure~\ref{fig:sensor_label_coverage} presents the cross-wave sensor and label modality coverage as a combined two-panel matrix. Panel A enumerates all sensor streams collected by the ABC Logger smartphone application and the wearable devices, indicating their presence or absence per wave. Panel B provides the corresponding label coverage matrix, identifying which ESM label dimensions were administered in each wave.

Panel A confirms that the core smartphone sensing suite -- comprising GPS/Location, Bluetooth and WiFi scans, app usage, screen events, notifications, battery events, accelerometer, ambient light, activity transitions, network traffic, and call/message metadata -- was active across all three waves. Keystroke sensing (KeyEvent) was absent in D-1 and introduced from D-2 onwards. The Fitbit wrist wearable streams (heart rate, step count, calorie, distance) were available continuously throughout all three waves.

Panel B illustrates the cross-wave differences in the ESM label schema across waves. The four core dimensions (Valence, Arousal, Stress, Task Disturbance) are shared across D-1, D-2, and D-3. Attention, Mental effort, Activity Duration, and a single Change item were present in D-1 and D-2; the Change item was split into ChangedValence and ChangedArousal in D-2. D-3 introduced eight custom D-3 affect-intensity items (Happy, Relaxed, Cheerful, Content, Sad, Angry, Anxious, Depressed), substantially enriching the affective granularity of that wave. The four shared labels are therefore the recommended benchmark targets for cross-wave evaluation, while wave-specific labels enable single-wave analyses with richer affective resolution.

\subsection*{B.5 Cross-Wave Differences}

\begin{figure}[tb]
\centerline{\includegraphics[width=0.7\columnwidth]{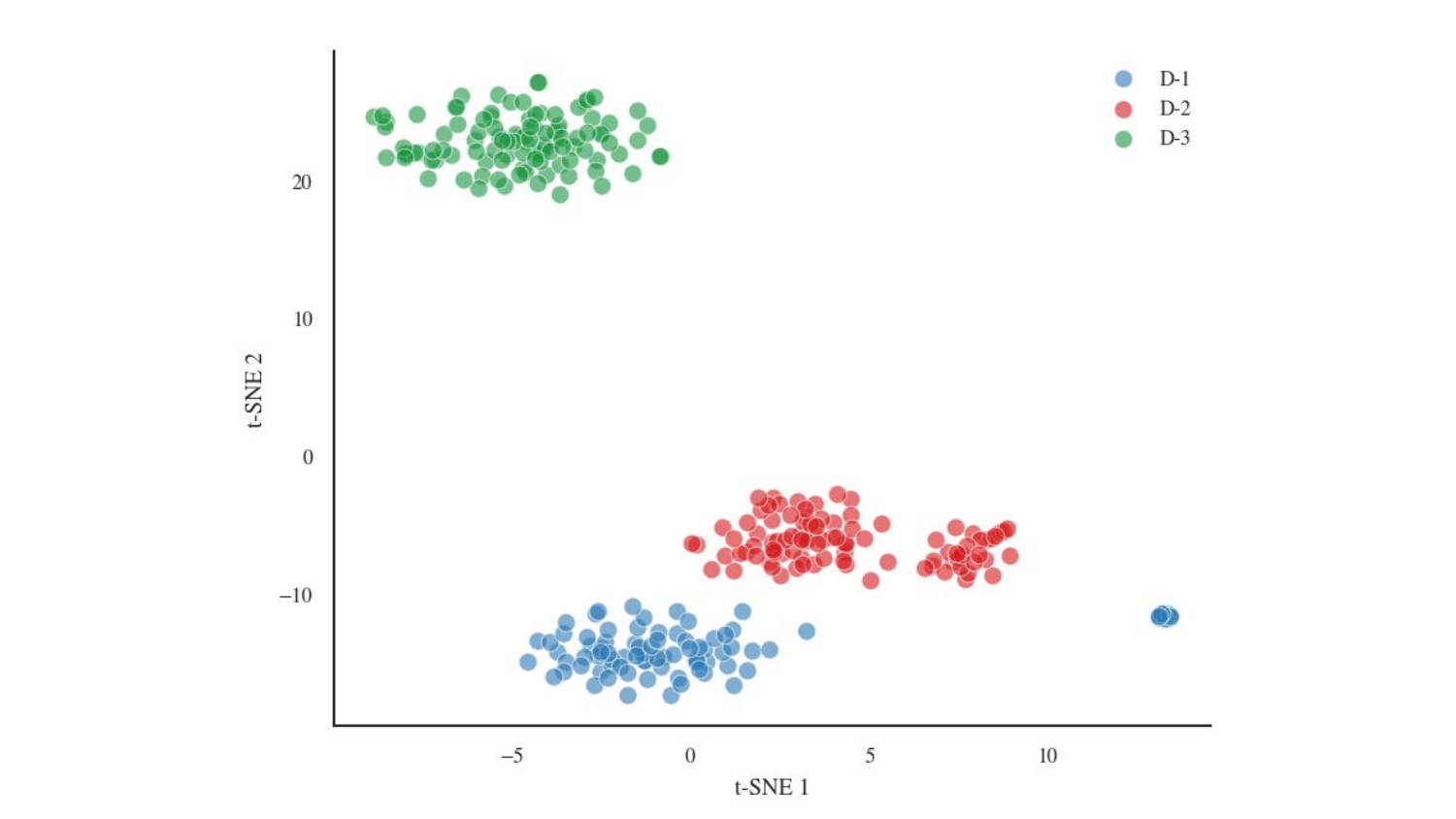}}
\caption{t-SNE visualization of participant-level sensor representations across the three waves. Each point represents one participant, obtained by averaging that participant's preprocessed sensor features over the available observation period.}
\label{fig:TSNE_cross_wave}
\end{figure}

Across D-1, D-2, and D-3, these differences arise from participant composition, wave-specific study design choices, sensing coverage, and label design, making cross-wave generalization a meaningful setting rather than an artificial split.
While they share enough common structure to support unified evaluation, they also differ in ways that make cross-wave evaluation nontrivial. 
Some of these differences follow directly from wave-specific study design choices: D-1 establishes the base protocol, D-2 refines questionnaire structure and increases label volume, and D-3 expands the semantic label layer. Figure~\ref{fig:TSNE_cross_wave} provides a complementary empirical view in feature space. For this visualization, each point is a participant-level aggregate representation obtained by averaging that participant's preprocessed sensor features over the available observation period. The resulting embeddings show visible wave-associated structure rather than a single undifferentiated cloud, indicating that cross-wave differences are reflected not only in protocol design but also in the observed feature space. At the same time, the overlap between waves suggests realistic structured heterogeneity rather than trivial separability.

\begin{table}[t]
\centering
\caption{
Cross-wave sensing-feature distribution shifts measured
using standardized one-dimensional Wasserstein distances.
For each feature, we compute pairwise Wasserstein distances
between the empirical observation distributions and normalize
by the feature's pooled three-wave standard deviation.
Values represent the median standardized distance
within each sensing bundle.
These are descriptive, observation-weighted estimates.
}
\label{tab:crosswave_wasserstein}
\small
\begin{tabular}{lcccc}
\toprule
\textbf{Sensing Bundle}
& \textbf{\# Features}
& \textbf{D-1/D-2}
& \textbf{D-1/D-3}
& \textbf{D-2/D-3} \\
\midrule
Phone use
& 1,531 & 0.18 & 0.29 & 0.20 \\

Mobility/activity
& 1,036 & 0.16 & 0.25 & 0.30 \\

Social/connectivity
& 691 & 0.20 & \textbf{0.43} & 0.23 \\

Wearable physiology
& 454 & 0.21 & 0.37 & \textbf{0.47} \\

Other device state
& 664 & 0.14 & 0.26 & 0.17 \\

\midrule
\textbf{Total}
& \textbf{4,376} & -- & -- & -- \\
\bottomrule
\end{tabular}
\end{table}

\noindent\textbf{Sensing-bundle distribution shifts.}
Table~\ref{tab:crosswave_wasserstein} summarizes
cross-wave covariate differences across five sensing bundles.
Among the observation-weighted comparisons,
social/connectivity features exhibit the largest median
shift between D-1 and D-3 (0.43), whereas wearable
physiology exhibits the largest shift between D-2
and D-3 (0.47).
These descriptive results indicate that cross-wave
heterogeneity is distributed across multiple sensing
modalities rather than concentrated in a single bundle.
However, because the analysis uses observation-level
distributions, participants contributing more observations
receive greater weight.

This subsection also analyzes cross-wave distributional shifts through three complementary lenses: covariate, concept, and conditional shift.

\begin{figure}[tb]
    \centering
    \includegraphics[width=\textwidth]{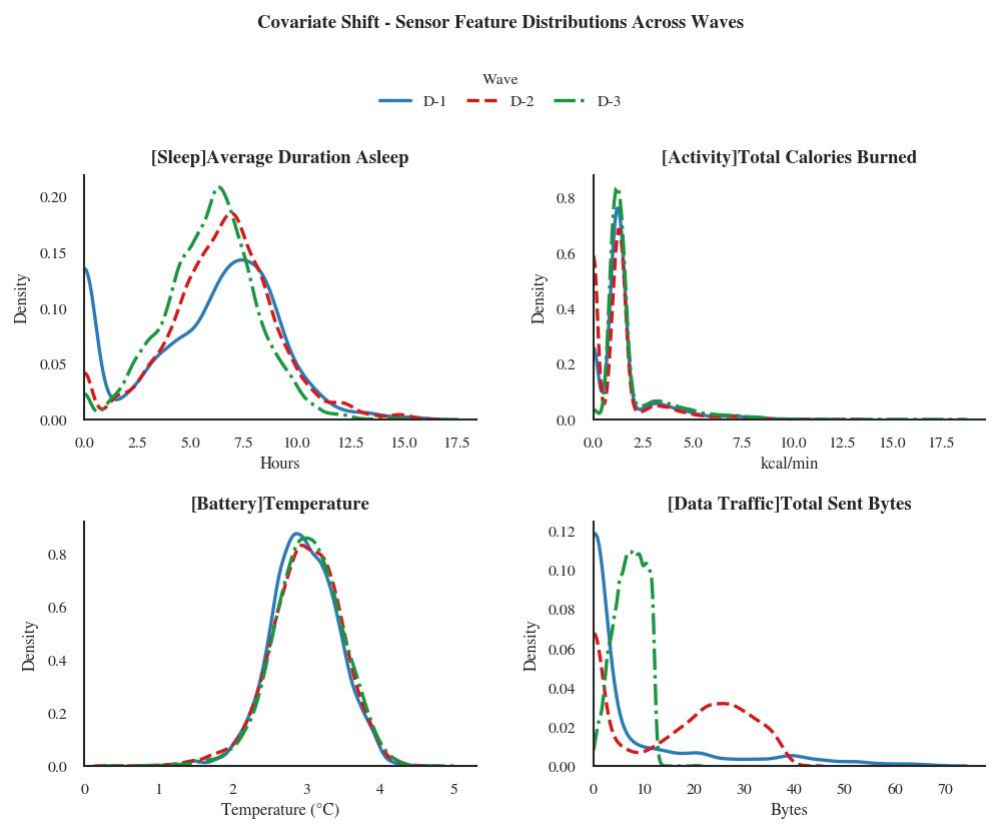}
    \caption{Covariate shift $P(X)$: kernel density estimates of four representative sensor features across the three waves (D-1 blue solid, D-2 red dashed, D-3 green dash-dot). Battery temperature and calorie features are stable across waves; sleep duration and data traffic exhibit clear wave-level distributional differences.}
    \label{fig:covariate_shift}
\end{figure}

Figure~\ref{fig:covariate_shift} shows covariate shift -- $P(X)$ -- by overlaying the distributions of representative sensor features across the three waves. Four features spanning distinct sensing categories are shown: sleep duration (wearable/activity), total calories burned (physical activity), battery temperature (phone system), and total sent bytes (data traffic). Battery temperature and calorie distributions are closely aligned across waves, indicating stable physiological and device characteristics. In contrast, sleep duration and data traffic show pronounced wave-level shifts: D-1 participants exhibit a broader, flatter sleep distribution compared to D-2 and D-3, and data traffic volumes in D-1 are substantially higher and more variable, likely reflecting differences in COVID-19 restriction phases and associated changes in mobility and online activity patterns across the three collection years.

\vspace{2.0em}

\begin{figure}[H]
    \centering
    \includegraphics[width=\textwidth]{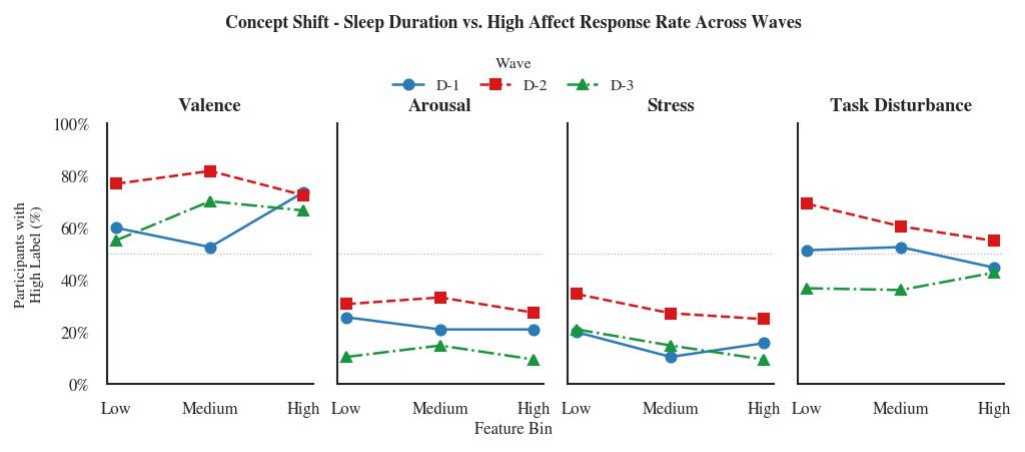}
    \caption{Concept shift $P(Y \mid X)$: proportion of participants reporting HIGH label as a function of binned sleep duration, shown separately for Valence, Arousal, Stress, and Task Disturbance. Each line is one wave. Substantial wave-level divergence indicates that the feature-to-label mapping is not stable across collection years.}
    \label{fig:concept_shift}
\end{figure}

Figure~\ref{fig:concept_shift} illustrates concept shift -- $P(Y \mid X)$ -- using sleep duration as the focal feature. For each of the four shared affect labels, participants are binned into low, medium, and high sleep duration groups, and the proportion reporting a HIGH affect label is plotted per bin and per wave. If the feature-label relationship were stable across waves, the three wave-level curves would overlap. Instead, the curves diverge substantially: for Valence, D-2 participants in every sleep duration bin report HIGH valence at rates 20--25 percentage points above D-3; for Arousal and Stress, D-3 participants consistently show lower HIGH rates regardless of sleep duration bin. This demonstrates that the mapping from sleep behaviour to affective state shifts across waves, meaning that a model trained on one wave cannot rely on the same feature-label associations in another.

\begin{figure}[H]
    \centering
    \includegraphics[width=\textwidth]{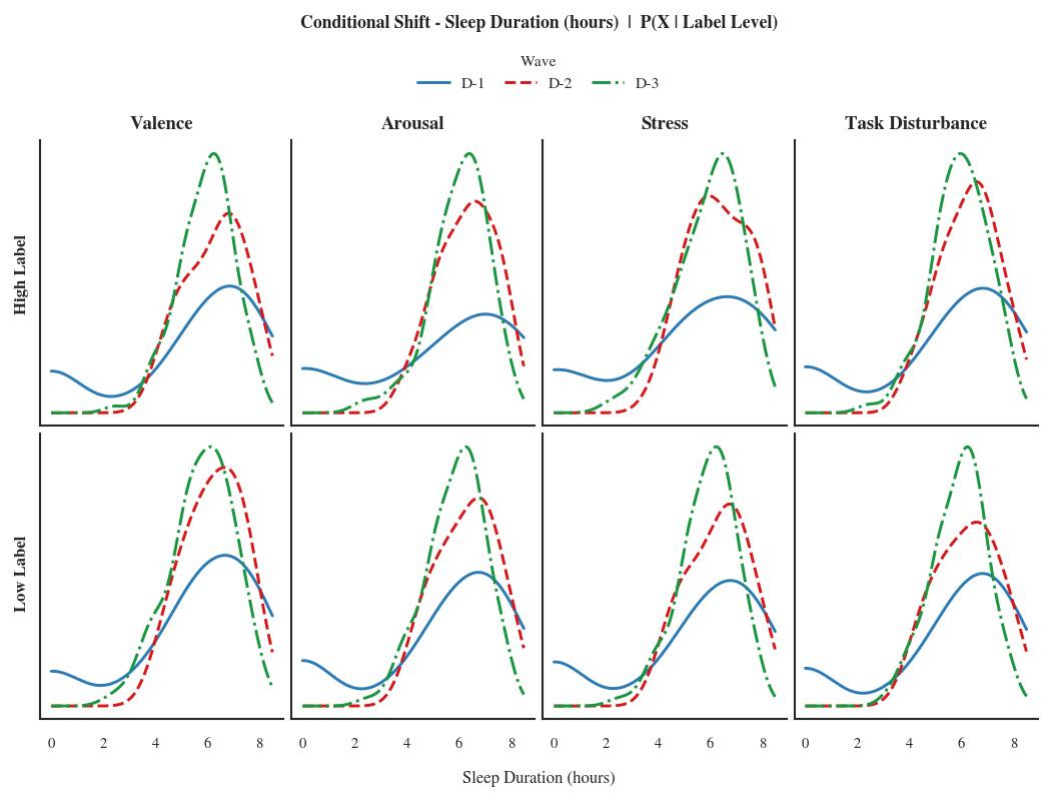}
    \caption{Conditional shift $P(X \mid Y)$: distribution of sleep duration conditioned on label level (HIGH top row, LOW bottom row) for each of the four shared labels, across three waves. Wave-level offsets in both rows demonstrate that sensor distributions shift across collection years independently of label level, violating the $P(X \mid Y)$ stability assumption required for cross-wave generalization.}
    \label{fig:conditional_shift}
\end{figure}

Figure~\ref{fig:conditional_shift} presents conditional shift -- $P(X \mid Y)$ -- using sleep duration as the focal feature, showing how its distribution differs between HIGH and LOW label groups within each wave. The $2 \times 4$ grid of kernel density estimates (rows: HIGH vs.\ LOW label; columns: Valence, Arousal, Stress, Task Disturbance) reveals two distinct patterns. First, within a given wave and label level, D-2 and D-3 participants show sharper, more concentrated sleep distributions than D-1 participants, whose distributions are flatter and broader. Second, within each wave, the HIGH and LOW label groups show only modest distributional separation, consistent with sleep duration being a weak-to-moderate predictor of momentary affect. Crucially, the wave-level offset in both the HIGH and LOW rows demonstrates that conditional shift operates independently of label level: participants from different waves exhibit different sensor distributions even when conditioned on the same affective state, a pattern that directly challenges the assumption of $P(X \mid Y)$ stability required for naive cross-wave generalization.

\subsection*{B.6 Cross-User Differences Within Each Wave}

\begin{figure}[tb]
\centerline{\includegraphics[width=\columnwidth]{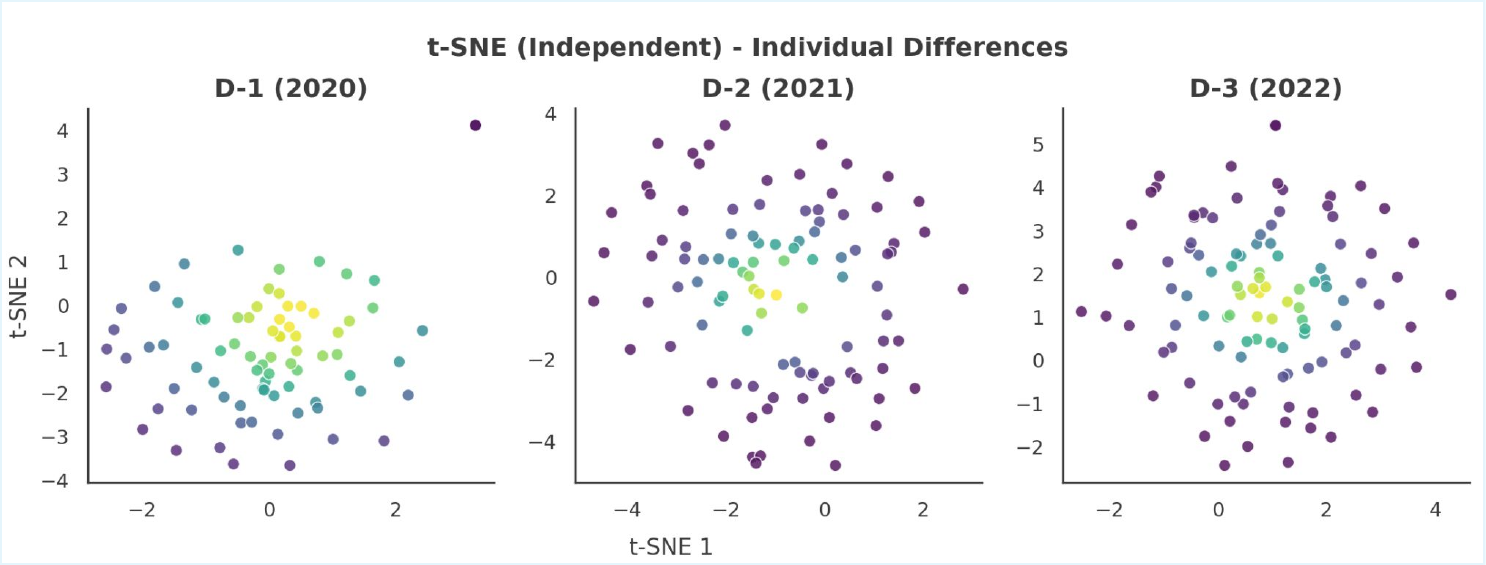}}
\caption{t-SNE visualization of participant-level sensor representations within each wave. Colors encode the participant's distance to the wave-specific centroid in the original feature space, with brighter colors indicating more central participants and darker colors indicating more peripheral ones}
\label{fig:TSNE_cross_user}
\end{figure}

Figure~\ref{fig:TSNE_cross_user} provides an exploratory view of participant-level variability within each wave. It uses the same participant-level aggregate representation as Figure~\ref{fig:TSNE_cross_wave}, but computes and visualizes the embeddings separately for D-1, D-2, and D-3. Within each wave, participants are spread from more central to more peripheral regions of the feature space rather than collapsing into a compact cluster. This pattern is visible across all three waves, indicating that within-wave variability is broad rather than being driven only by a few outliers.

These results show that substantial individual variability remains even within a relatively controlled cohort. The resource is therefore useful not only for descriptive analysis, but also for studying how affect models behave under realistic participant-level heterogeneity.

\begin{figure}[H]
    \centering
    \includegraphics[width=\textwidth]{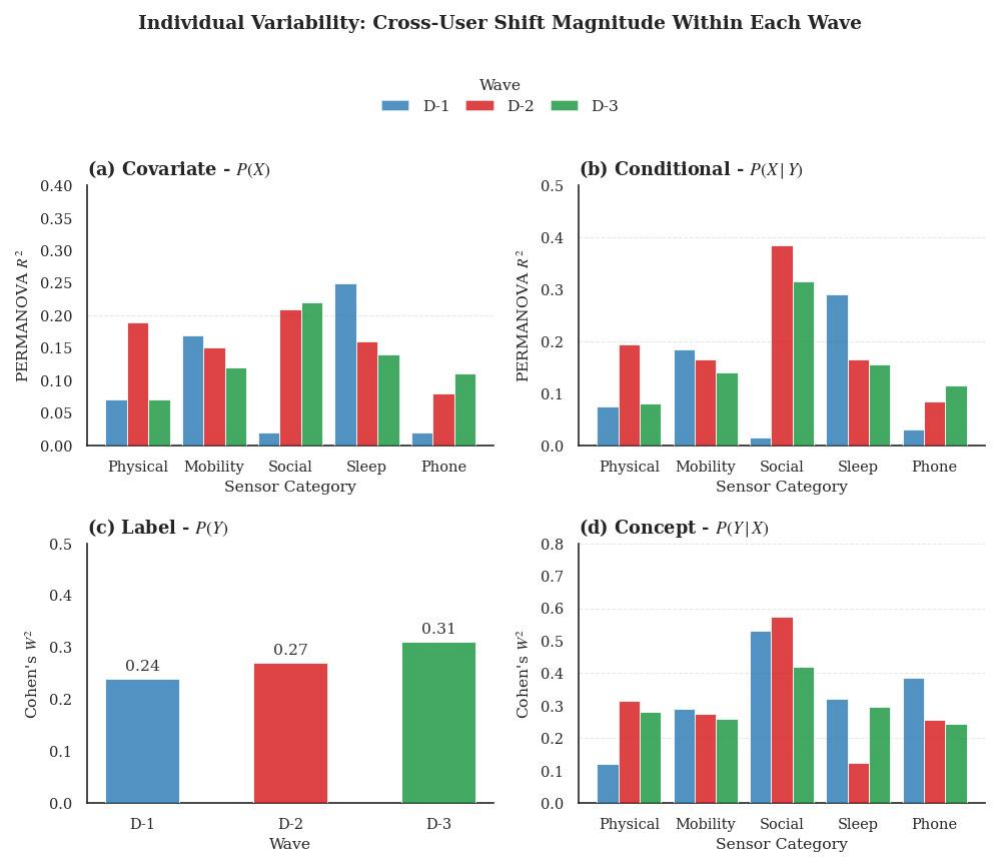}
    \caption{Cross-user shift magnitude within each wave, decomposed by shift type and sensor category. (a) Covariate shift $P(X)$: PERMANOVA $R^2$ of sensor feature distributions explained by participant identity. (b) Conditional shift $P(X \mid Y)$: PERMANOVA $R^2$ of conditional feature distributions. (c) Label shift $P(Y)$: Cohen's $W^2$ of label distributions per wave (0.24 / 0.27 / 0.31 for D-1 / D-2 / D-3). (d) Concept shift $P(Y \mid X)$: Cohen's $W^2$ of feature-to-label mapping variability across individuals. Social features exhibit the highest concept shift; Sleep features exhibit the highest covariate shift.}
    \label{fig:cross_user_shift}
\end{figure}

Figure~\ref{fig:cross_user_shift} quantifies the magnitude of cross-user distributional shift within each wave using a four-panel analysis of shift type and sensor category. Panels (a) and (b) report PERMANOVA $R^2$ values measuring the fraction of variance in sensor feature distributions $P(X)$ and conditional sensor distributions $P(X \mid Y)$, respectively, attributable to participant identity. Panel (c) reports Cohen's $W^2$ for the label distribution $P(Y)$ per wave. Panel (d) reports Cohen's $W^2$ for the concept distribution $P(Y \mid X)$, capturing how strongly the feature-to-label mapping varies across individuals.

Across all four shift types and sensor categories, participant identity accounts for a non-trivial and consistent fraction of total variance. For covariate shift $P(X)$, sleep-related features exhibit the highest cross-user $R^2$ (up to 0.25 in D-1), followed by Social interaction and Mobility features. The Social category shows the strongest individual-level concept shift $P(Y \mid X)$ (Cohen's $W^2 \approx 0.53$--$0.58$ across waves), indicating that social behavioural patterns have the most idiosyncratic relationship with affective states. Label shift $P(Y)$ is also substantial and increases monotonically across waves (Cohen's $W^2 = 0.24, 0.27, 0.31$ for D-1, D-2, D-3 respectively), confirming that individuals differ systematically in their affective baselines and that this heterogeneity grows with the enrichment of the label schema. Taken together, these results establish that user identity is a dominant source of variance at every level of the feature-label pipeline, providing strong quantitative justification for personalised modelling approaches.

\begin{figure}[h]
    \centering
    \includegraphics[width=\linewidth]{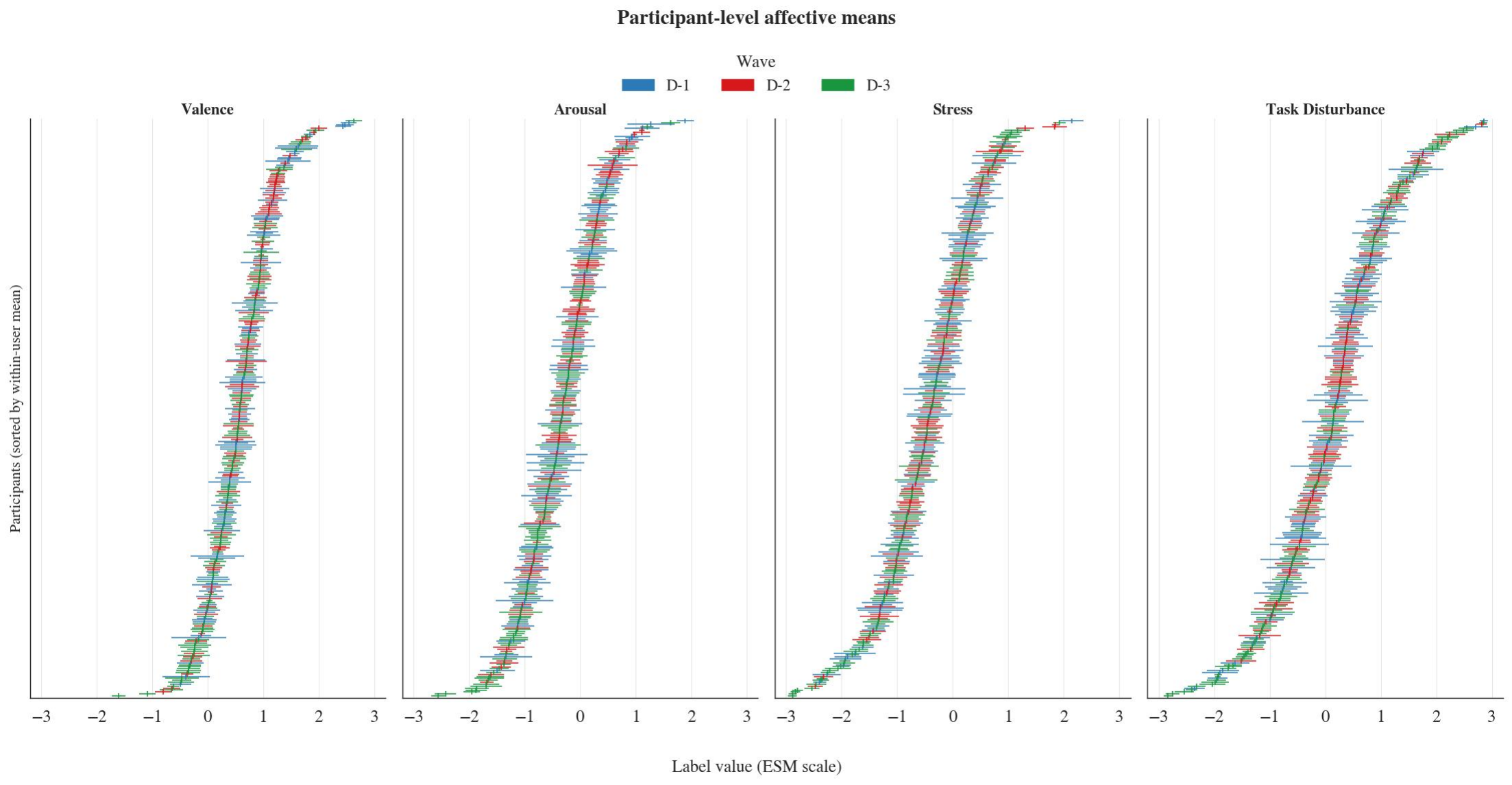}
    \caption{Participant-level affective means for the four shared labels 
    (Valence, Arousal, Stress, Task Disturbance) across all three waves. 
    Each horizontal bar represents one participant's within-user mean $\pm$ 
    one standard deviation, sorted by within-user mean value.}
    \label{fig:participant_label_means}
\end{figure}

Figure~\ref{fig:participant_label_means} provides a complementary participant-level view of this heterogeneity, focusing directly on the label space rather than the feature space. For each of the four shared labels, each participant is represented by their within-user mean and within-user standard deviation, with participants sorted by their mean value. The wide vertical spread across all four panels confirms that individuals differ substantially in their affective baselines. Critically, the interleaving of wave colors throughout the sorted distribution indicates that individual-level spread is not simply a wave-level artifact -- participants from D-1, D-2, and D-3 appear across the full range of affective means, consistent with the quantitative shift analysis in Figure~\ref{fig:cross_user_shift} showing that user identity accounts for a substantial fraction of label variance within each wave.

\clearpage

\section{Benchmark Details and Full Results}
\label{appendix:benchmark}

\raggedbottom

\newenvironment{appendixdesc}{%
  \begin{list}{\textbullet}{%
    \setlength{\leftmargin}{1.5em}%
    \setlength{\itemindent}{0pt}%
    \setlength{\labelsep}{0.45em}%
    \setlength{\labelwidth}{0.8em}%
    \setlength{\topsep}{0.25\baselineskip}%
    \setlength{\partopsep}{0pt}%
    \setlength{\itemsep}{0.25\baselineskip}%
    \setlength{\parsep}{0pt}}%
}{\end{list}}
\newcommand{\appendixitem}[2]{\item \textbf{#1} #2}
% --- Utilities for compact AUROC-only appendix tables ---
% Many appendix tables store metrics as Acc/F1/AUROC. The macros below render only AUROC
% (the third component) without rewriting table contents.
\makeatletter
% AUROC extractor for cells formatted as Acc/F1/AUROC.
\def\auconly#1{\expandafter\auconly@i#1\relax}
\def\auconly@i\textbf#1\relax{\expandafter\auconly@ii#1\relax}
\def\auconly@i#1\relax{\auconly@ii#1\relax}
\def\auconly@ii#1/#2/#3\relax{#3}
\newcolumntype{A}{>{\collectcell\auconly}c<{\endcollectcell}}

% AUROC extractor for cells formatted as Acc/F1/AUROC (± fold std), e.g., 0.1/0.2/0.3 ($\pm$0.01).
\def\auconlypm#1{\expandafter\auconlypm@i#1\relax}
\def\auconlypm@i\textbf#1\relax{\expandafter\auconlypm@ii#1\relax}
\def\auconlypm@i--\relax{--}
\def\auconlypm@i#1\relax{\auconlypm@ii#1\relax}
\def\auconlypm@ii#1/#2/#3 (#4\relax{#3}
\newcolumntype{P}{>{\collectcell\auconlypm}c<{\endcollectcell}}
\makeatother

\noindent\textbf{C.1 Additional Benchmark Evaluation Protocol}

\medskip
\noindent\textbf{C.1.1 Feature Alignment and Preprocessing}

To ensure evaluation integrity, all preprocessing statistics are fit on the training split only and then applied unchanged to validation and test.

For the \textbf{cross-wave benchmark}, we additionally align feature spaces across waves.

\textbf{Shared feature space.} Raw feature sets vary by wave (Appendix~A.7.5, Table~\ref{tab:A12}) due to (i) sensor-availability differences (e.g., \texttt{KeyEvent}-derived features from D-2 onward), (ii) logger-format drift that encodes the same underlying feature under slightly different column names, and (iii) minor case/naming inconsistencies. We construct a shared cross-wave feature space with two steps.
\begin{itemize}
    \item \textbf{Alias canonicalization.} We map each raw feature name to a canonical form by normalizing case and tokens (e.g., \texttt{PIF\#AGE} vs.\ \texttt{PIF\#age}), resolving a small set of categorical aliases (e.g., \texttt{UNKNOWN} vs.\ \texttt{UNDEFINED}), and removing logger artefacts. Leakage-prone columns (timestamps, participant identifiers, labels) are removed before canonicalization.
    \item \textbf{Feature intersection.} We use a fixed three-way intersection of canonicalized feature sets, $\mathcal{F}=\mathcal{F}_{\text{D-1}}\cap\mathcal{F}_{\text{D-2}}\cap\mathcal{F}_{\text{D-3}}$, yielding a common feature space of 4{,}890 features. This triple intersection is used for every cross-wave direction (both one-to-one and two-to-one). Pairwise intersections are recorded for auditing but are not used for training or evaluation.
\end{itemize}

% Missing values are handled upstream in the feature-extraction pipeline (Zhang et al., 2024); the benchmark assumes NaN-free feature tables.

\medskip
\noindent\textbf{C.1.2 Metrics and Model Selection Strategy}

\textbf{Evaluation metrics and uncertainty.} AUROC is the primary metric, and within-wave tables report the mean of the five fold-level test AUROCs. Uncertainty is quantified with a participant-level paired bootstrap. For each evaluation direction (12 within-wave and 36 cross-wave), test participants are resampled with replacement ($B=2{,}000$), all ESM rows of each sampled participant are retained, and AUROC is recomputed on the pooled predictions. Intervals are 95\% percentile intervals. One participant resample is drawn per test wave and shared by every label, training configuration, and model evaluated on that wave. Paired differences ($\Delta$AUROC) therefore use identical resamples, and per-label and pooled contrasts average the per-direction differences within each replicate before taking percentiles. A difference is significant when its paired interval excludes zero, and two-sided bootstrap $p$-values are Holm-adjusted within the comparison families stated with each result. Performance is called above chance only when the AUROC interval excludes 0.5 which by itself does not establish practical utility. Family-best representatives are chosen post hoc by test AUROC, per label and wave within waves and per label across waves. Calibration is summarized by the Brier score and the 10-bin expected calibration error (ECE). These analyses use re-executed runs that saved per-sample predictions (same splits, seed, and tuned hyperparameters), so point values can differ slightly from the tabulated fold means.

\textbf{Validation and model selection.} Hyperparameter tuning and checkpoint selection use only setting-defined validation data.
\begin{appendixdesc}
\appendixitem{Within-wave benchmark:}{We use stratified group 5-fold cross-validation, stratifying by the binary label while grouping by user identity so that label balance is preserved across folds and each user appears in exactly one fold.}
\appendixitem{Cross-wave benchmark:}{In the two-to-one setting, we align and concatenate the two source waves into a single source pool, then hold out a stratified 20\% source-only validation slice. The target wave is used only for final evaluation and does not contribute any selection signal.}
\end{appendixdesc}

\medskip
\noindent\textbf{C.1.3 Normalization and outlier handling}
We apply the same high-level preprocessing steps across settings, with setting-specific normalization. All normalization statistics are fitted only on the training split defined by each benchmark setting and are then applied unchanged to the corresponding validation and test splits. No validation/test feature statistics or affect labels are used when fitting normalization parameters.

\begin{appendixdesc}
% \appendixitem{Temporal/within-wave normalization:}{Per-user $z$-score normalization using that user's training split statistics, applied to all splits for the same user. Features with (near-)zero variance are left effectively unscaled to avoid numerical issues.} 
% \appendixitem{Cross-wave normalization:}{Pool-level $z$-score normalization using the concatenated source training pool only.}
\appendixitem{Temporal benchmark normalization:}{For each user, z-score statistics are computed from that user's temporal training slice and then applied to that user's validation and test slices.}
\appendixitem{Within-wave/Cross-wave benchmark normalization:}{For each fold or transfer direction, feature-wise mean and standard deviation are fitted on the labeled source/training features only and then applied unchanged to validation and test features, including unseen users and the held-out wave.}
\appendixitem{Outlier clipping:}{We clip feature values using a training-fitted threshold based on the $99.9\%$ quantile of $|z|$, with a minimum clipping threshold of 10, applied to all splits. This prevents a small number of extreme feature values from dominating optimization (especially for neural models) while keeping the transformation monotone and training-only to avoid leakage.}
\end{appendixdesc}

\medskip
\subsection*{C.2 Models and training setup}
\subsubsection*{C.2.1 Model implementations and architecture details}
This subsection summarizes model implementations and how domain labels are instantiated. DG methods follow DomainBed-style protocols~\cite{Gulrajani2021}. In the within-wave benchmark, each participant is treated as a domain. In the cross-wave benchmark, the train/test split is defined by wave, but DG domains are still defined at the participant level: each source participant is assigned a unique domain label, identified by its source wave and user ID. This means that both single-source and multi-source cross-wave training contain multiple source domains, corresponding to the users available in the source wave(s).
UDA methods are implemented using the \textbf{Transfer-Learning-Library (TLL)}~\cite{TLL2020} under \emph{transductive} unsupervised domain adaptation (labeled source, unlabeled target). In the within-wave benchmark, the unlabeled target set is the held-out user fold; in the cross-wave benchmark, it is the target wave. Target labels are never used for training or selection. All UDA methods receive the same unlabeled target set under the same backbone and training loop, so differences reflect the adaptation objective rather than privileged access to target information. Baselines and other tabular models are integrated from reference libraries or official repositories and wrapped in a unified pipeline.

The complete model inventory is provided in Table~21, and the hyperparameter search spaces are detailed in Tables~22--24.

\medskip
\noindent\textbf{C.2.1.1 Baseline models}

\modeldesc{XGBoost~\cite{Chen2016}}{XGBoost is the main gradient-boosted tree baseline in the benchmark. It is included because boosted trees remain a strong reference point for dense tabular sensing features with heterogeneous scales and nonlinear interactions.}

\modeldesc{LightGBM~\cite{Ke2017}}{LightGBM is a histogram-based gradient-boosted tree method that complements XGBoost with a second strong tree baseline. It is included because fast boosted-tree methods are often competitive even against more specialized tabular neural models.}

\modeldesc{MLP}{We use a standard multilayer perceptron as the simplest neural baseline for fixed tabular inputs. It provides the reference feedforward model against which the tabular neural, DG, and UDA methods can be compared.}

\modeldesc{ResNet~\cite{He2016}}{ResNet is the residual-feedforward neural baseline. It tests whether residual blocks improve optimization and performance over a plain MLP on the same feature tables.}

\medskip
\noindent\textbf{C.2.1.2 Tabular neural networks}

\modeldesc{TabNet~\cite{Arik2021}}{TabNet is an attentive tabular architecture that learns feature-selection masks over multiple decision steps. It is included as a tabular model that performs adaptive feature selection rather than uniform full-table processing.}

\modeldesc{SAINT~\cite{Somepalli2021}}{SAINT is a transformer-style tabular model with richer attention structure than a plain feedforward network. It is included as a higher-capacity tabular transformer baseline.}

\modeldesc{TabTransformer~\cite{Huang2020}}{TabTransformer contextualizes tabular features through transformer layers applied to embedded inputs. It is included because transformer-style contextualization can capture cross-feature dependencies that are difficult to model with a plain MLP.}

\modeldesc{FTTransformer~\cite{Gorishniy2021}}{FTTransformer is a lighter feature-token transformer for tabular data. It provides a strong transformer baseline with a simpler tokenization scheme than heavier row-column attention variants.}

\modeldesc{DCN~\cite{Wang2021}}{DCN combines explicit cross layers with a deep nonlinear tower. It is included because the cross network can model low-order feature interactions directly.}

\medskip
\noindent\textbf{C.2.1.3 Domain generalization models}

Current benchmark runs evaluate DG models with a shared MLP backbone so that differences primarily reflect the DG objective rather than backbone changes.

\modeldesc{Invariant Risk Minimization (IRM)~\cite{Arjovsky2019}}{IRM encourages the learned representation to support a stable classifier across source domains. It serves as the canonical invariance-based DG baseline.}

\modeldesc{Risk Extrapolation (VREx)~\cite{Krueger2021}}{VREx penalizes variance in risk across source domains. It is included as a close alternative to IRM based on risk equalization rather than invariant-classifier constraints.}

\modeldesc{GroupDRO~\cite{Sagawa2020}}{GroupDRO optimizes the worst-group loss during training, treating domains as groups. It serves as the robust optimization baseline for heterogeneous source domains.}

\modeldesc{MixStyle~\cite{Zhou2021}}{MixStyle perturbs and mixes feature statistics across training domains. It is included as a feature-statistic augmentation baseline for DG.}

\modeldesc{MLDG~\cite{Li2018}}{MLDG is a meta-learning approach to domain generalization. It simulates train-domain and meta-test-domain splits during optimization so that updates favor held-out-domain performance.}

\modeldesc{MASF~\cite{Dou2019}}{MASF combines task generalization with feature-space alignment and class-level metric learning. It is included as a stronger DG baseline that explicitly targets transferable class structure.}

\modeldesc{Fish~\cite{Shi2022}}{Fish performs gradient-matching style meta-updates across domains. It is included as a lightweight DG alternative based on parameter updates that remain consistent across source domains.}

\modeldesc{CSD~\cite{Piratla2020}}{CSD, or Common-Specific Decomposition, separates shared and domain-specific components in the learned representation. It is included as a structured DG model for testing whether explicit decomposition helps under wave shift.}

\modeldesc{SagNet~\cite{Nam2021}}{SagNet, or Style Agnostic Network, tries to separate content and style during training so that the classifier becomes less sensitive to nuisance style variation. It is included as a DG model that explicitly targets style robustness.}

\medskip
\noindent\textbf{C.2.1.4 Domain adaptation models}

Current benchmark runs evaluate UDA models with a shared MLP backbone so that differences primarily reflect the adaptation algorithm rather than backbone changes.

\modeldesc{Domain-Adversarial Neural Networks (DANN)~\cite{Ganin2016}}{DANN uses adversarial training with a domain discriminator so that the representation becomes less informative about the domain label while remaining predictive for the task label. It is the standard adversarial UDA baseline.}

\modeldesc{Conditional Domain Adversarial Network (CDAN)~\cite{Long2018}}{CDAN extends DANN by conditioning the adversarial alignment on task predictions as well as learned features. It is included because conditional alignment can be stronger than marginal adversarial alignment.}

\modeldesc{Deep Adaptation Network (DAN)~\cite{Long2015}}{DAN aligns source and target features by minimizing a multi-kernel maximum mean discrepancy objective. It serves as a non-adversarial UDA baseline based on distribution matching in feature space.}

\modeldesc{DeepCORAL~\cite{Sun2016}}{DeepCORAL aligns second-order statistics between source and target feature activations. It is included as a covariance-matching alternative to adversarial and MMD-based transfer methods.}

\modeldesc{Minimum Class Confusion (MCC)~\cite{Jin2020}}{MCC is a non-adversarial UDA method that reduces class confusion on target predictions. It serves as a strong target-side regularization baseline.}

\modeldesc{Adversarial Discriminative Domain Adaptation (ADDA)~\cite{Tzeng2017}}{ADDA adapts a target encoder toward a fixed source hypothesis through adversarial training. It is included as a classical two-stage UDA method.}

\modeldesc{Maximum Classifier Discrepancy (MCD)~\cite{Saito2018}}{MCD uses two classifiers and explicitly maximizes and then minimizes their prediction discrepancy on target samples. It is included because classifier disagreement provides an alternative transfer signal.}

\modeldesc{Joint Adaptation Network (JAN)~\cite{Long2017}}{JAN aligns joint distributions of features and predictions through a JMMD objective. It is included as a label-aware distribution-matching UDA method.}

\modeldesc{Source Hypothesis Transfer (SHOT)~\cite{Liang2020}}{SHOT is a source-free UDA method that adapts a source hypothesis to unlabeled target data without revisiting source samples. It is included because source-free transfer is relevant in practical adaptation settings.}

\modeldesc{Class-Balanced Self-Training (CBST)~\cite{Zou2018}}{CBST uses iterative pseudo-labeling with class-balanced selection rules on target data. It is included as a self-training transfer baseline.}

\modeldesc{Cross-Domain Gradient Discrepancy Minimization (CGDM)~\cite{Du2021}}{CGDM is a discrepancy-based UDA method ported into the repository with a dedicated MLP-style feature extractor and dual classifiers. It provides an additional transfer baseline beyond the more standard adversarial and moment-matching methods.}

\noindent\begin{minipage}{\linewidth}
\centering
\small
\setlength{\tabcolsep}{8pt} % 열 간격을 조금 넓혀서 시각적 여백 확보
\renewcommand{\arraystretch}{1.2}
\captionof{table}{Benchmark model inventory across baseline, tabular neural, DG, and DA groups.}
\label{tab:appendix_model_inventory}
% 너비를 0.85\linewidth로 제한하여 너무 꽉 차지 않게 설정
\begin{tabular*}{0.85\linewidth}{@{\extracolsep{\fill}} p{0.25\linewidth} p{0.55\linewidth} @{}}
\toprule
Group & Model \\
\midrule
Baseline & XGB, LGB, MLP, ResNet \\
Tabular neural & TabNet, SAINT, TabTransformer, FTTransformer, DCN \\
DG & IRM, VREx, GroupDRO, MixStyle, MLDG, MASF, Fish, CSD, SagNet \\
UDA & DANN, CDAN, DAN, DeepCORAL, MCC, ADDA, MCD, JAN, SHOT, CBST, CGDM \\
\bottomrule
\end{tabular*}
\end{minipage}\par

\medskip
\noindent\textbf{C.2.2 Benchmark-Specific Modifications and External Baselines}

This subsection documents specific architectural adaptations and the integration of baselines residing outside the standard DomainBed and TLL frameworks.

\begin{appendixdesc}
\appendixitem{Tabular Models:}{\emph{TabNet} (via \textbf{pytorch-tabnet}), and \emph{SAINT, TabTransformer, FTTransformer} (via \textbf{pytorch-widedeep}~\cite{Zaurin2023}).}
\appendixitem{Deep CTR \& Domain Adaptation:}{\emph{DCN} is integrated from \textbf{deepctr-torch}~\cite{Shen2020}, while \emph{MASF} and \emph{CGDM} are ported from their original author-released implementations.}
\appendixitem{Advanced UDA Techniques:}{\emph{DeepCORAL}\footnote{\url{https://github.com/SSARCandy/DeepCORAL}}, \emph{ADDA}\footnote{\url{https://github.com/jvanvugt/pytorch-domain-adaptation}}, \emph{MCD}\footnote{\url{https://github.com/mil-tokyo/MCD_DA}}, \emph{SHOT}\footnote{\url{https://github.com/tim-learn/SHOT}}, and \emph{CBST}\footnote{\url{https://github.com/yzou2/CBST}} were adapted from official public repositories to ensure performance fidelity.}
\appendixitem{Efficient attention for high-dimensional tabular inputs:}{Tabular Transformers tokenise each input feature as a separate attention token and therefore incur a self-attention cost that scales quadratically in the number of features. Our feature tables have several thousand columns per wave (Appendix~A.7.5, Table~\ref{tab:A12}), which makes standard softmax self-attention memory-bound on our GPUs even at modest batch sizes. We therefore replace the attention kernel of \textbf{TabTransformer}, \textbf{SAINT}, and \textbf{FTTransformer} with a linear-attention variant~\cite{Katharopoulos2020}. Concretely, we approximate $\mathrm{softmax}(QK^{\top})V$ by $\phi(Q)\left(\phi(K)^{\top}V\right)$ for a shared feature map $\phi(\cdot)$, reducing the attention cost from $\mathcal{O}(n^{2}d)$ to $\mathcal{O}(nd^{2})$ for $n$ feature tokens. TabTransformer uses the linear-attention path exposed by its reference library (\textbf{pytorch-widedeep}); SAINT and FTTransformer are patched to route their encoder blocks through the same operator while preserving all other architectural choices. The substitution is local to the attention kernel only, so the family comparison between tabular Transformers, DG, and UDA models remains capacity-matched under a shared training protocol.}
\end{appendixdesc}

\medskip
\noindent\textbf{C.2.3 Unified Training Protocol}

This subsection lists the values that realise the cross-method comparability controls described in Section~\ref{sec:comparability}. Every baseline, DG, and UDA method is driven by the same training loop, early-stopping contract, and model-selection rule; the only per-model differences are the HPO search spaces reported in Tables~\ref{tab:appendix_hyperparameter_grids_part1},~\ref{tab:appendix_hyperparameter_grids_part2}, and~\ref{tab:appendix_hyperparameter_grids_part3}.

\begin{appendixdesc}
\appendixitem{Optimisation and Loss:}{All neural models are trained with AdamW~\cite{Loshchilov2019} under a cross-entropy classification loss for binary targets, with the learning rate and weight decay drawn from the HPO grid. Tree-based baselines (XGBoost, LightGBM) use their native boosted-tree objective with AUC as the evaluation metric. No explicit class-imbalance correction is applied at the loss level (\texttt{class\_weight} or \texttt{pos\_weight} are not used), so improvements on imbalanced targets are attributable to the objective or architecture rather than to a reweighted loss.}

\appendixitem{Epoch Budget and Early Stopping:}{Every model is trained under a uniform maximum budget of $50$ epochs. A patience-based early-stopping rule monitors validation AUROC with patience $= 20$, minimum improvement $\Delta = 10^{-4}$, and best-weight restoration enabled: when validation AUROC fails to improve by more than $\Delta$ for $20$ consecutive epochs the run terminates and the best-AUROC checkpoint is reloaded before final evaluation. Tree-based baselines use the equivalent built-in \texttt{early\_stopping\_rounds} $= 20$ on the same validation split.}

\appendixitem{Batch Size:}{All models (baselines, DG, and UDA) are trained with a fixed batch size of $512$ applied uniformly across all settings, except where this was not possible due to memory limitations (see code for details). Batch size is not treated as a tunable hyperparameter in this release: it was fixed up-front to keep the per-step optimisation budget consistent across families, since a per-model sweep over batch size would have required GPU memory and run-time budget beyond what was available for the transformer and UDA variants at the larger end of the grid.}

\appendixitem{Hardware:}{Neural network models (MLP, ResNet, all tabular Transformers, and all DG and UDA methods) were trained on a single NVIDIA RTX PRO 5000 Blackwell GPU with 48~GB VRAM. Gradient boosted tree baselines (XGBoost, LightGBM) were trained on CPU rather than GPU, since the per tree fit does not benefit from accelerator parallelism at our feature dimensionality.}

\appendixitem{Validation, Checkpoint, and HPO:}{Model selection is performed exclusively on the validation split defined by each benchmark's protocol (temporal validation slice, within-wave fold-level validation partition, cross-wave source-only validation split). The selected checkpoint is the epoch with the highest validation AUROC observed during training, restored from the best-weight snapshot maintained by the early-stopping callback. HPO itself uses the same validation AUROC as its objective, with $30$ Optuna~\cite{optuna_2019} trials per (benchmark, label, dataset/direction, model) configuration.}

\appendixitem{Reproducibility:}{Each run seeds \nolinkurl{numpy.random}, \nolinkurl{torch.manual_seed}, and \nolinkurl{torch.cuda.manual_seed_all} at the start. Primary benchmark runs use seed $42$. Split construction, HPO sampling, model initialisation, and data-loader shuffling are all seeded, so rerunning on identical hardware reproduces the reported metrics within cuDNN-level floating-point tolerance.}
\end{appendixdesc}

\medskip
\noindent\textbf{C.2.4 Hyperparameter Optimization Search Spaces}

The comprehensive hyperparameter search spaces for all models are detailed in Tables~22--24. Our grid design follows two goals: (i) fidelity to the original implementations and (ii) a comparable optimization budget across models.

\begin{appendixdesc}
\appendixitem{Adaptation for tabular data:}{For DG and UDA methods with limited prior use on high dimensional longitudinal sensing tables, we set each search range starting from the corresponding reference benchmark, namely DomainBed~\cite{Gulrajani2021} for DG methods and the Transfer-Learning-Library~\cite{TLL2020} for UDA methods, and then narrow or extend the ranges of individual hyperparameters to suit our feature dimensionality and per fold sample size.}
\appendixitem{Alignment with prior work:}{For remaining methods, we ensure that our search spaces cover the hyperparameter ranges reported in the original papers or reference benchmarks that introduced or evaluated the method.}
\end{appendixdesc}

\clearpage
\thispagestyle{plain}

\begin{table}[p]
\centering
\footnotesize
\setlength{\tabcolsep}{5pt}
\renewcommand{\arraystretch}{1.15}
\caption{Hyperparameter grids for baseline and tabular neural methods (continued in Table~\ref{tab:appendix_hyperparameter_grids_part2} and Table~\ref{tab:appendix_hyperparameter_grids_part3}).}
\label{tab:appendix_hyperparameter_grids_part1}
\begin{tabularx}{\linewidth}{@{}>{\raggedright\arraybackslash}p{0.14\linewidth}>{\raggedright\arraybackslash}p{0.13\linewidth}>{\raggedright\arraybackslash}p{0.23\linewidth}X@{}}
\toprule
Model & Applicability & Hyperparameter & Values \\
\midrule
\multicolumn{4}{c}{\textbf{Baseline Methods}} \\
\midrule
MLP / ResNet & shared & learning rate & LogUniform$(10^{-5}, 10^{-2})$ \\
 &  & weight decay & LogUniform$(10^{-6}, 10^{-3})$ \\
 &  & hidden dim & Integer$[64, 512]$ \\
 &  & dropout & Uniform$[0.0, 0.5]$ \\
 & MLP & num.\ layers & Integer$[1, 8]$ \\
 & ResNet & num.\ blocks & Integer$[1, 8]$ \\
\midrule
XGB & method-specific & learning rate & LogUniform$(10^{-5}, 1)$ \\
 &  & max depth & Integer$[3, 10]$ \\
 &  & min child weight & LogUniform$(10^{-8}, 10^{5})$ \\
 &  & subsample & Uniform$[0.5, 1.0]$ \\
 &  & column subsample (tree / level) & Uniform$[0.5, 1.0]$ \\
 &  & gamma & \{0\} or LogUniform$(10^{-8}, 10^{2})$ \\
 &  & $\lambda$ & \{0\} or LogUniform$(10^{-8}, 10^{2})$ \\
 &  & $\alpha$ & \{0\} or LogUniform$(10^{-8}, 10^{2})$ \\
 &  & num.\ estimators & Integer$[50, 500]$ \\
\midrule
LGB & method-specific & learning rate & LogUniform$(10^{-3}, 1)$ \\
 &  & num.\ leaves & Integer$[8, 128]$ \\
 &  & num.\ estimators & Integer$[50, 500]$ \\
 &  & min child samples & Integer$[5, 100]$ \\
 &  & subsample & Uniform$[0.5, 1.0]$ \\
 &  & column subsample (tree) & Uniform$[0.5, 1.0]$ \\
\midrule
\multicolumn{4}{c}{\textbf{Tabular Neural Methods}} \\
\midrule
TabNet & method-specific & learning rate & Uniform$[10^{-3}, 5 \times 10^{-3}]$ \\
 &  & weight decay & LogUniform$(10^{-5}, 5 \times 10^{-4})$ \\
 &  & $n_d$, $n_a$ & \{8, 16\} \\
 &  & $n_\text{steps}$ & Integer$[3, 4]$ \\
 &  & $\gamma$ & Uniform$[1.1, 1.5]$ \\
 &  & $\lambda_\text{sparse}$ & LogUniform$(10^{-5}, 10^{-3})$ \\
\midrule
SAINT / TabTransformer & shared & learning rate & LogUniform$(2 \times 10^{-4}, 8 \times 10^{-4})$ \\
 &  & weight decay & LogUniform$(10^{-5}, 5 \times 10^{-4})$ \\
 &  & input dim & \{8, 16\} \\
 &  & $n_\text{heads}$ & \{2, 4\} \\
 &  & $n_\text{blocks}$ & Integer$[1, 2]$ \\
 &  & attn dropout & Uniform$[0.0, 0.2]$ \\
 &  & ff dropout & Uniform$[0.0, 0.2]$ \\
\midrule
FTTransformer & method-specific & learning rate & LogUniform$(5 \times 10^{-5}, 5 \times 10^{-4})$ \\
 &  & weight decay & LogUniform$(10^{-5}, 5 \times 10^{-4})$ \\
 &  & input dim & \{16, 32\} \\
 &  & $n_\text{blocks}$ & Integer$[1, 2]$ \\
 &  & attn dropout & Uniform$[0.0, 0.2]$ \\
 &  & ff dropout & Uniform$[0.0, 0.2]$ \\
 &  & residual dropout & $\{0\} \cup$ Uniform$[0.0, 0.1]$ \\
 &  & ff factor & Uniform$[1.0, 2.0]$ \\
\midrule
DCN & method-specific & learning rate & LogUniform$(5 \times 10^{-5}, 5 \times 10^{-3})$ \\
 &  & weight decay & LogUniform$(10^{-5}, 5 \times 10^{-4})$ \\
 &  & $n_\text{hidden}$ layers & Integer$[1, 2]$ \\
 &  & layer size & Integer$[64, 128]$ \\
 &  & $n_\text{cross}$ layers & Integer$[1, 2]$ \\
 &  & hidden dropout & Uniform$[0.0, 0.2]$ \\
 &  & cross dropout & $\{0\} \cup$ Uniform$[0.0, 0.2]$ \\
\bottomrule
\end{tabularx}
\end{table}

\clearpage

\clearpage
\thispagestyle{plain}

\begin{table}[p]
\centering
\footnotesize
\setlength{\tabcolsep}{5pt}
\renewcommand{\arraystretch}{1.15}
\caption{Hyperparameter grids for domain generalization (DG) methods (continued from Table~\ref{tab:appendix_hyperparameter_grids_part1}; continues in Table~\ref{tab:appendix_hyperparameter_grids_part3}).}
\label{tab:appendix_hyperparameter_grids_part2}
\begin{tabularx}{\linewidth}{@{}>{\raggedright\arraybackslash}p{0.14\linewidth}>{\raggedright\arraybackslash}p{0.13\linewidth}>{\raggedright\arraybackslash}p{0.23\linewidth}X@{}}
\toprule
Model & Applicability & Hyperparameter & Values \\
\midrule
\multicolumn{4}{c}{\textbf{Domain Generalization Methods}} \\
\midrule
DG (shared) & all DG methods & learning rate & LogUniform$(3 \times 10^{-5}, 3 \times 10^{-4})$ \\
 &  & weight decay & LogUniform$(10^{-6}, 10^{-2})$ \\
 &  & dropout & \{0.0, 0.1, 0.5\} \\
 &  & hidden dim & Integer$[64, 512]$ \\
 &  & layers & Integer$[1, 8]$ \\
\midrule
IRM & method-specific & penalty $\lambda$ & LogUniform$(10^{-1}, 10^{5})$ \\
 &  & anneal & $10^u,\; u \sim \mathrm{Uniform}[0,4]$ \\
\midrule
VREx & method-specific & penalty $\lambda$ & LogUniform$(10^{-1}, 10^{5})$ \\
 &  & anneal & $10^u,\; u \sim \mathrm{Uniform}[0,4]$ \\
\midrule
GroupDRO & method-specific & step size $\eta$ & LogUniform$(10^{-3}, 10^{-1})$ \\
\midrule
MixStyle & method-specific & $\alpha$ & Uniform$[0.1, 0.5]$ \\
 &  & $p$ & Uniform$[0.1, 0.9]$ \\
 &  & mix mode & \{random, crossdomain\} \\
\midrule
MLDG & method-specific & meta step $\beta$ & LogUniform$(10^{-1}, 10)$ \\
 &  & $n_\text{meta-test}$ & \{1, 2\} \\
\midrule
MASF & method-specific & inner lr & LogUniform$(10^{-5}, 10^{-2})$ \\
 &  & metric lr & LogUniform$(10^{-5}, 10^{-2})$ \\
 &  & metric weight & LogUniform$(10^{-4}, 10^{-2})$ \\
 &  & margin & Uniform$[0.2, 2.0]$ \\
 &  & temperature & Uniform$[1.0, 5.0]$ \\
 &  & metric dim & \{64, 128, 256\} \\
\midrule
Fish & method-specific & meta lr & \{0.05, 0.1, 0.5\} \\
\midrule
CSD & method-specific & $\lambda$ & Uniform$[0.1, 5.0]$ \\
 &  & $k$ & \{2, 3\} \\
\midrule
SagNet & method-specific & adv weight & LogUniform$(10^{-2}, 10)$ \\
\bottomrule
\end{tabularx}
\end{table}

\clearpage

\clearpage
\thispagestyle{plain}

\vspace*{\fill}
\noindent\begin{minipage}{\linewidth}
\centering
\footnotesize
\setlength{\tabcolsep}{5pt}
\renewcommand{\arraystretch}{1.12}
\captionof{table}{Hyperparameter grids for unsupervised domain adaptation (UDA) methods (continued from Table~\ref{tab:appendix_hyperparameter_grids_part1} and Table~\ref{tab:appendix_hyperparameter_grids_part2}).}
\label{tab:appendix_hyperparameter_grids_part3}
\begin{tabularx}{\linewidth}{@{}>{\raggedright\arraybackslash}p{0.14\linewidth}>{\raggedright\arraybackslash}p{0.13\linewidth}>{\raggedright\arraybackslash}p{0.23\linewidth}X@{}}
\toprule
Model & Applicability & Hyperparameter & Values \\
\midrule
\multicolumn{4}{c}{\textbf{Domain Adaptation Methods}} \\
\midrule
UDA (shared) & all UDA methods & learning rate & LogUniform$(10^{-5}, 10^{-2})$ \\
 &  & weight decay & LogUniform$(10^{-6}, 10^{-3})$ \\
 &  & hidden dim & Integer$[64, 512]$ \\
 &  & layers & Integer$[1, 8]$ \\
\midrule
DANN & method-specific & discriminator lr & LogUniform$(10^{-5}, 10^{-2})$ \\
 &  & disc width & $2^u,\; u \sim \mathrm{Uniform}[6,10]$ \\
 &  & trade-off & LogUniform$(10^{-2}, 10^{2})$ \\
\midrule
CDAN & method-specific & discriminator lr & LogUniform$(10^{-5}, 10^{-2})$ \\
 &  & disc width & $2^u,\; u \sim \mathrm{Uniform}[6,10]$ \\
 &  & trade-off & LogUniform$(10^{-2}, 10^{2})$ \\
 &  & entropy & \{T, F\} \\
 &  & randomized & \{F, T\} \\
 &  & randomized dim & \{256, 512, 1024\} \\
\midrule
DAN & method-specific & trade-off & LogUniform$[0.1, 10.0]$ \\
 &  & kernels & \{3, 5, 7\} \\
 &  & linear & \{F, T\} \\
\midrule
DeepCORAL & method-specific & $\gamma$ & LogUniform$[0.1, 10.0]$ \\
\midrule
MCC & method-specific & temperature & Uniform$[1.0, 5.0]$ \\
 &  & trade-off & LogUniform$(10^{-2}, 10^{1})$ \\
\midrule
JAN & method-specific & JMMD $\lambda$ & Uniform$[0.1, 10.0]$ \\
 &  & kernels & \{3, 5, 7\} \\
 &  & linear & \{F, T\} \\
\midrule
ADDA & method-specific & discriminator lr & LogUniform$(10^{-5}, 10^{-2})$ \\
 &  & disc width & $2^u,\; u \sim \mathrm{Uniform}[6,10]$ \\
\midrule
MCD & method-specific & $k$ (\#discrepancy steps) & Integer$[1, 4]$ \\
 &  & trade-off & Uniform$[0.5, 2.0]$ \\
\midrule
SHOT & method-specific & cls par $\beta_\text{cls}$ & Uniform$[0.0, 1.0]$ \\
 &  & ent par $\beta_\text{ent}$ & Uniform$[0.5, 2.0]$ \\
 &  & label-update interval & Integer$[5, 20]$ \\
 &  & adapt epochs & Integer$[10, 50]$ \\
\midrule
CBST & method-specific & pretrain epochs & Integer$[10, 30]$ \\
 &  & max self-training iter & \{3, 5, 7\} \\
 &  & init portion & Uniform$[0.1, 0.3]$ \\
 &  & portion step & Uniform$[0.05, 0.2]$ \\
 &  & max portion & Uniform$[0.6, 0.9]$ \\
 &  & retrain epochs & \{3, 5, 10\} \\
\midrule
CGDM & method-specific & weight decay & LogUniform$(10^{-5}, 10^{-3})$ \\
\bottomrule
\end{tabularx}
\end{minipage}
\vspace*{\fill}

\clearpage

\medskip
\subsection*{C.3 Additional Results}

Interval-based summaries use the participant-level bootstrap of Appendix~C.1.2. Across 755 contrasts between a DG or UDA model and the strongest supervised model of the same direction (113 within-wave and 642 cross-wave), 69.0\% of the paired 95\% intervals contain zero, 221 favour the supervised model, and 13 favour the adaptation method. AUROC intervals lie above 0.5 for 123 of 186 within-wave and 469 of 897 cross-wave model cells.

\subsubsection*{C.3.1 Within-User Temporal Prediction Results}

The resulting AUROCs quantify the predictive signal carried by the features themselves. Across labels, gradient-boosted trees (XGBoost, LightGBM) lead in most cells of Table~\ref{tab:appendix_tier_a_full_results}, with tabular neural models (SAINT, FTTransformer, TabTransformer) competitive on a handful of (label, dataset) pairs. The predictability of D-3-only affect labels varies significantly between classes, with the best AUROCs from $0.548$ (\emph{Depressed}) to $0.673$ (\emph{Angry}), indicating that the difficulty per-label varies substantially even within a common affect family. Across waves, arousal and valence improve monotonically from D-1 to D-3 ($0.579 \to 0.610 \to 0.647$ and $0.575 \to 0.594 \to 0.635$), while disturbance remains near chance throughout. Stress, in contrast, is roughly stable across waves ($0.636 \to 0.618 \to 0.646$ across D-1, D-2, and D-3). Table~\ref{tab:appendix_tier_a_full_results}(b) reports reference baselines separately for passive sensing and for the same user's earlier labels. Every label-history baseline except the last label for arousal is significantly higher than current-prompt XGBoost, whereas the GRU over earlier prompts is lower on every label (significantly except for valence).

\begin{table}[!htbp]
\centering
\scriptsize
\setlength{\tabcolsep}{4pt}
\renewcommand{\arraystretch}{1.05}
\caption{Full within-user temporal results. (a) Tuned passive-sensing models. Cells show test AUROC (60/20/20 chronological split per user, concatenated across users), and bold marks the best model per row. (b) Reference baselines. Cells show mean test AUROC over the three waves. Labels are binarized at each user's training-portion median, so (b) is directly comparable with the stress rows of (a) only. $^{\dagger}$95\% CI includes 0.5 in at least one wave.}
\label{tab:appendix_tier_a_full_results}
\textit{(a) Tuned passive-sensing models}\par\smallskip
\begin{tabular}{@{}llccccccccc@{}}
\toprule
Label & Dataset & XGB & LGB & MLP & ResNet & TabNet & DCN & SAINT & TabTr & FTT \\
\midrule
\multicolumn{11}{l}{\textbf{Shared labels}} \\
\midrule
arousal & D-1 & \textbf{0.579} & 0.564 & 0.522 & 0.520 & 0.505 & 0.510 & 0.502 & 0.503 & 0.517 \\
 & D-2 & 0.600 & \textbf{0.610} & 0.538 & 0.559 & 0.518 & 0.538 & 0.529 & 0.561 & 0.530 \\
 & D-3 & \textbf{0.647} & 0.645 & 0.561 & 0.542 & 0.540 & 0.536 & 0.592 & 0.562 & 0.558 \\
\addlinespace[0.2em]
disturbance & D-1 & 0.535 & 0.501 & 0.501 & 0.503 & 0.509 & 0.516 & \textbf{0.550} & 0.528 & 0.518 \\
 & D-2 & \textbf{0.587} & 0.575 & 0.494 & 0.511 & 0.481 & 0.518 & 0.560 & 0.550 & 0.532 \\
 & D-3 & 0.476 & 0.472 & \textbf{0.577} & 0.566 & 0.495 & 0.511 & 0.557 & 0.551 & 0.525 \\
\addlinespace[0.2em]
stress & D-1 & 0.618 & \textbf{0.636} & 0.559 & 0.555 & 0.500 & 0.535 & 0.502 & 0.558 & 0.513 \\
 & D-2 & \textbf{0.618} & 0.616 & 0.591 & 0.592 & 0.575 & 0.559 & 0.566 & 0.578 & 0.549 \\
 & D-3 & \textbf{0.646} & 0.619 & 0.562 & 0.586 & 0.544 & 0.572 & 0.544 & 0.579 & 0.539 \\
\addlinespace[0.2em]
valence & D-1 & \textbf{0.575} & 0.558 & 0.560 & 0.530 & 0.507 & 0.541 & 0.560 & 0.554 & 0.537 \\
 & D-2 & 0.592 & 0.592 & 0.556 & 0.554 & 0.503 & 0.521 & 0.571 & \textbf{0.594} & 0.528 \\
 & D-3 & \textbf{0.635} & 0.625 & 0.527 & 0.536 & 0.488 & 0.510 & 0.593 & 0.574 & 0.564 \\
\midrule
\multicolumn{11}{l}{\textbf{D-3 rich labels}} \\
\midrule
angry & D-3 & \textbf{0.673} & 0.633 & 0.537 & 0.544 & 0.492 & 0.502 & 0.533 & 0.520 & 0.515 \\
anxious & D-3 & 0.483 & 0.515 & 0.560 & 0.548 & 0.516 & 0.511 & \textbf{0.575} & 0.556 & 0.540 \\
cheerful & D-3 & 0.574 & \textbf{0.599} & 0.571 & 0.533 & 0.525 & 0.526 & 0.559 & 0.564 & 0.532 \\
content & D-3 & \textbf{0.601} & 0.578 & 0.545 & 0.513 & 0.511 & 0.527 & 0.562 & 0.567 & 0.515 \\
depressed & D-3 & 0.474 & 0.502 & 0.519 & 0.541 & 0.518 & 0.499 & 0.522 & 0.476 & \textbf{0.548} \\
happy & D-3 & 0.562 & \textbf{0.593} & 0.576 & 0.563 & 0.502 & 0.564 & 0.550 & 0.555 & 0.537 \\
relaxed & D-3 & \textbf{0.585} & 0.584 & 0.553 & 0.550 & 0.505 & 0.513 & 0.581 & 0.570 & 0.517 \\
sad & D-3 & 0.563 & \textbf{0.607} & 0.558 & 0.541 & 0.505 & 0.513 & 0.501 & 0.565 & 0.559 \\
\bottomrule
\end{tabular}
\par\smallskip
\textit{(b) Reference baselines by information setting}\par\smallskip
\begin{tabular}{@{}lcccc@{}}
\toprule
Baseline & arousal & disturbance & stress & valence \\
\midrule
\multicolumn{5}{l}{\textbf{Passive sensing only}} \\
Population prior & 0.500 & 0.500 & 0.500 & 0.500 \\
Circadian logistic regression & 0.539$^{\dagger}$ & 0.512$^{\dagger}$ & 0.555$^{\dagger}$ & 0.559 \\
GRU, current + 7 preceding prompts & 0.525$^{\dagger}$ & 0.564$^{\dagger}$ & 0.564$^{\dagger}$ & 0.593 \\
XGBoost, current prompt (default hyperparameters) & 0.635 & 0.616 & 0.608 & 0.630$^{\dagger}$ \\
\midrule
\multicolumn{5}{l}{\textbf{Earlier ESM labels of the same user (separate information setting)}} \\
Last label & 0.625 & 0.647 & 0.656 & 0.673 \\
Mean of all earlier labels & 0.680 & 0.716 & 0.703 & 0.696 \\
Mean of last three labels & 0.666 & 0.708 & 0.711 & 0.728 \\
Previous-24\,h mean & 0.688 & 0.733 & 0.723 & 0.746 \\
\bottomrule
\end{tabular}
\end{table}

\FloatBarrier

\medskip
\subsubsection*{C.3.2 Within-Wave Cross-User Generalization Results}

\begin{figure}[!htbp]
    \centering
    \includegraphics[width=0.65\linewidth]{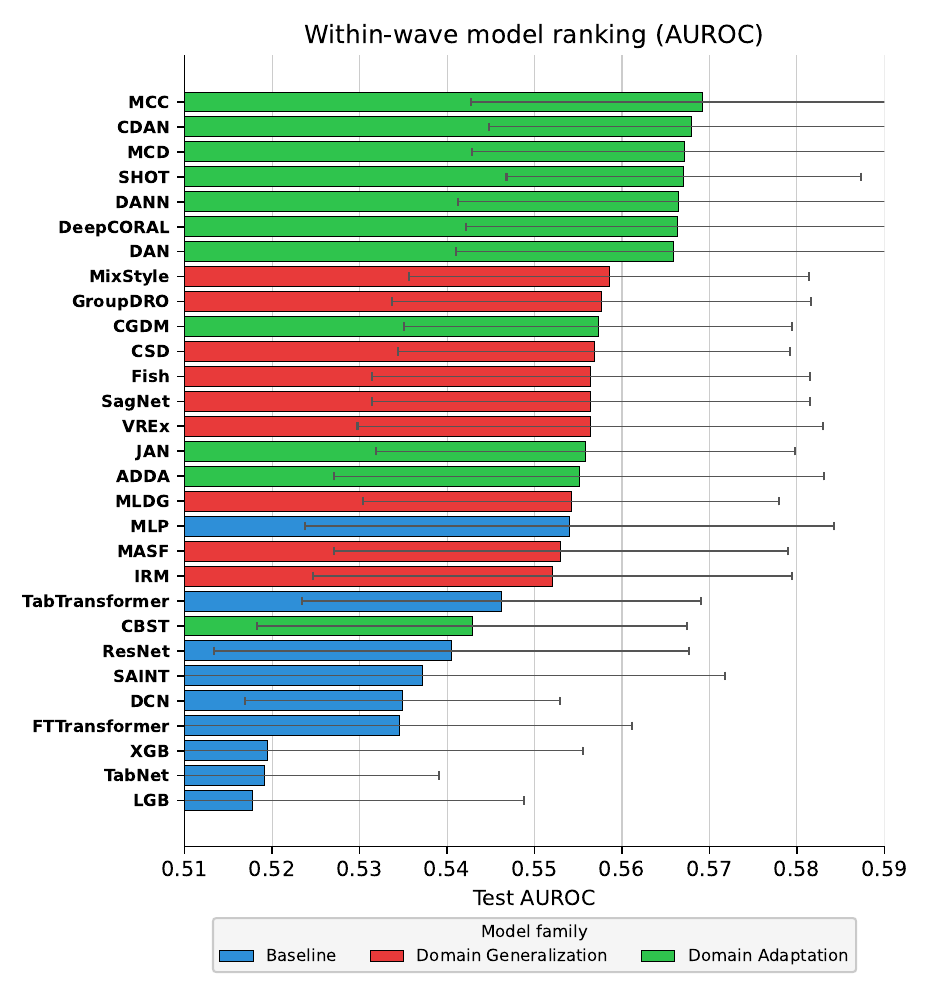}
    \caption{Within-wave model ranking by mean test AUROC. Each bar is the mean of a model's test AUROC across all within-wave (label, dataset) rows; error bars show the standard deviation across those rows. Bars are shaded by family.}
    \label{fig:appendix_c_model_ranking}
\end{figure}

We use a stratified 5-fold group split that holds out users, and report each model's mean test AUROC together with its fold-level standard deviation. The mean shows average generalization, and the spread shows sensitivity to which users are held out. Figure~\ref{fig:appendix_c_model_ranking} ranks all 29 models according to the mean within-wave AUROC for the exact mean values; refer to Table~\ref{tab:appendix_c_model_ranking}.

\FloatBarrier

\medskip
\noindent\textbf{Family comparison.} On fold-mean AUROC, the best UDA model has the highest value in 11 of 12 shared-label cells and in seven of the eight D-3-only labels (Table~\ref{tab:appendix_tier_b_family_comparison}a), but these are point differences. In paired contrasts (Table~\ref{tab:appendix_tier_b_family_comparison}c), family-best DG and UDA exceed the MLP backbone by $+0.0177$ $[0.0094, 0.0252]$ and $+0.0124$ $[0.0053, 0.0197]$, whereas neither family significantly exceeds the strongest supervised model after Holm adjustment (DG $+0.0088$ with $p_{\text{Holm}} = 0.120$, UDA $+0.0035$), and DG and UDA do not differ ($+0.0053$ $[-0.0015, 0.0113]$). No individual DG or UDA method differs from MLP after Holm adjustment. For arousal, disturbance, and valence the re-executed pool has 11 models, so representatives in (c) can differ from (a). CBST is the one UDA method at the bottom of the ranking, plausibly because its iterative pseudo-labelling needs more target data than each fold provides.

Detailed per-direction paired bootstrap comparisons
are reported in
Table~\ref{tab:appendix_tier_b_per_direction_contrasts}.

\begin{table}[t]
\centering\footnotesize
\caption{Within-wave model ranking by mean test AUROC (companion to Figure~\ref{fig:appendix_c_model_ranking}). Each model's score is the mean of its per-row HPO-best test AUROC across all 20 within-wave (label, dataset) rows; $\pm$ is the standard deviation across those rows.}
\label{tab:appendix_c_model_ranking}
\setlength{\tabcolsep}{6pt}
\begin{tabular}{rllc}
\toprule
Rank & Model & Family & Test AUROC \\
\midrule
1 & MCC & DA & 0.569 $\pm$ 0.026 \\
2 & CDAN & DA & 0.568 $\pm$ 0.023 \\
3 & MCD & DA & 0.567 $\pm$ 0.024 \\
4 & SHOT & DA & 0.567 $\pm$ 0.020 \\
5 & DANN & DA & 0.566 $\pm$ 0.025 \\
6 & DeepCORAL & DA & 0.566 $\pm$ 0.024 \\
7 & DAN & DA & 0.566 $\pm$ 0.025 \\
8 & MixStyle & DG & 0.559 $\pm$ 0.023 \\
9 & GroupDRO & DG & 0.558 $\pm$ 0.024 \\
10 & CGDM & DA & 0.557 $\pm$ 0.022 \\
11 & CSD & DG & 0.557 $\pm$ 0.022 \\
12 & Fish & DG & 0.556 $\pm$ 0.025 \\
13 & SagNet & DG & 0.556 $\pm$ 0.025 \\
14 & VREx & DG & 0.556 $\pm$ 0.027 \\
15 & JAN & DA & 0.556 $\pm$ 0.024 \\
16 & ADDA & DA & 0.555 $\pm$ 0.028 \\
17 & MLDG & DG & 0.554 $\pm$ 0.024 \\
18 & MLP & Base & 0.554 $\pm$ 0.030 \\
19 & MASF & DG & 0.553 $\pm$ 0.026 \\
20 & IRM & DG & 0.552 $\pm$ 0.027 \\
21 & TabTransformer & Base & 0.546 $\pm$ 0.023 \\
22 & CBST & DA & 0.543 $\pm$ 0.025 \\
23 & ResNet & Base & 0.541 $\pm$ 0.027 \\
24 & SAINT & Base & 0.537 $\pm$ 0.035 \\
25 & DCN & Base & 0.535 $\pm$ 0.018 \\
26 & FTTransformer & Base & 0.535 $\pm$ 0.027 \\
27 & XGB & Base & 0.519 $\pm$ 0.036 \\
28 & TabNet & Base & 0.519 $\pm$ 0.020 \\
29 & LGB & Base & 0.518 $\pm$ 0.031 \\
\bottomrule
\end{tabular}
\end{table}

\begin{table}[!htbp]
\centering
\scriptsize
\setlength{\tabcolsep}{4pt}
\renewcommand{\arraystretch}{1.05}
\caption{Within-wave cross-user results for baseline and tabular models. Cells show test AUROC (mean over 5 stratified user folds). Bold marks the best model per row.}
\label{tab:appendix_tier_b_base_dg_results}
\begin{tabular}{@{}llccccccccc@{}}
\toprule
Label & Dataset & XGB & LGB & MLP & ResNet & TabNet & DCN & SAINT & TabTr & FTT \\
\midrule
\multicolumn{11}{l}{\textbf{Shared labels}} \\
\midrule
arousal & D-1 & 0.545 & 0.554 & 0.537 & 0.542 & 0.536 & 0.527 & 0.556 & \textbf{0.557} & 0.534 \\
 & D-2 & 0.535 & 0.540 & 0.542 & 0.552 & 0.542 & 0.534 & \textbf{0.562} & 0.555 & 0.517 \\
 & D-3 & 0.580 & 0.590 & \textbf{0.610} & 0.605 & 0.579 & 0.571 & 0.583 & 0.599 & 0.592 \\
\addlinespace[0.2em]
disturbance & D-1 & 0.495 & 0.517 & 0.537 & 0.506 & 0.506 & 0.525 & 0.521 & \textbf{0.542} & 0.519 \\
 & D-2 & 0.494 & 0.504 & 0.560 & 0.542 & 0.533 & 0.532 & \textbf{0.565} & 0.564 & 0.536 \\
 & D-3 & 0.501 & 0.493 & \textbf{0.523} & 0.515 & 0.494 & 0.520 & 0.510 & 0.520 & 0.502 \\
\addlinespace[0.2em]
stress & D-1 & 0.495 & 0.515 & 0.505 & 0.515 & 0.498 & 0.519 & 0.507 & 0.509 & \textbf{0.521} \\
 & D-2 & 0.531 & 0.500 & \textbf{0.546} & 0.537 & 0.500 & 0.539 & 0.530 & 0.535 & 0.539 \\
 & D-3 & 0.538 & 0.523 & \textbf{0.580} & 0.563 & 0.521 & 0.544 & 0.536 & 0.556 & 0.553 \\
\addlinespace[0.2em]
valence & D-1 & 0.488 & 0.498 & \textbf{0.558} & 0.538 & 0.529 & 0.531 & 0.530 & 0.549 & 0.522 \\
 & D-2 & 0.537 & 0.516 & \textbf{0.545} & 0.536 & 0.504 & 0.523 & 0.507 & 0.532 & 0.520 \\
 & D-3 & 0.556 & 0.551 & \textbf{0.576} & 0.555 & 0.530 & 0.538 & 0.566 & 0.571 & 0.527 \\
\midrule
\multicolumn{11}{l}{\textbf{D-3 rich labels}} \\
\midrule
angry & D-3 & 0.474 & 0.473 & \textbf{0.523} & 0.518 & 0.504 & 0.519 & 0.497 & 0.511 & 0.510 \\
anxious &  & 0.524 & 0.553 & \textbf{0.554} & 0.520 & 0.500 & 0.524 & 0.519 & 0.529 & 0.531 \\
cheerful &  & 0.553 & 0.525 & \textbf{0.609} & 0.568 & 0.530 & 0.568 & 0.578 & 0.573 & 0.576 \\
content &  & 0.533 & 0.540 & \textbf{0.594} & 0.577 & 0.521 & 0.568 & 0.583 & 0.561 & 0.578 \\
depressed &  & 0.441 & 0.462 & 0.518 & 0.516 & 0.508 & 0.514 & \textbf{0.532} & 0.530 & 0.504 \\
happy &  & 0.502 & 0.477 & \textbf{0.584} & 0.566 & 0.520 & 0.551 & 0.566 & 0.563 & 0.568 \\
relaxed &  & 0.484 & 0.512 & \textbf{0.556} & 0.548 & 0.508 & 0.539 & 0.550 & 0.544 & 0.542 \\
sad &  & \textbf{0.581} & 0.511 & 0.522 & 0.493 & 0.519 & 0.510 & 0.446 & 0.524 & 0.501 \\
\bottomrule
\end{tabular}
\end{table}

\begin{table}[!htbp]
\centering
\scriptsize
\setlength{\tabcolsep}{4pt}
\renewcommand{\arraystretch}{1.05}
\caption{Within-wave cross-user results for domain generalization (DG) models. Cells show test AUROC; bold marks the best model per row.}
\label{tab:appendix_tier_b_dg_results}
\begin{tabular}{@{}llccccccccc@{}}
\toprule
Label & Dataset & IRM & VREx & GDRO & MixStyle & MLDG & MASF & Fish & CSD & SagNet \\
\midrule
\multicolumn{11}{l}{\textbf{Shared labels}} \\
\midrule
arousal & D-1 & 0.529 & 0.546 & 0.544 & 0.538 & 0.527 & \textbf{0.546} & 0.543 & 0.540 & 0.521 \\
 & D-2 & 0.567 & 0.562 & \textbf{0.567} & 0.563 & 0.556 & 0.562 & 0.565 & 0.566 & 0.556 \\
 & D-3 & 0.606 & \textbf{0.609} & 0.603 & 0.596 & 0.598 & 0.600 & 0.606 & 0.604 & 0.605 \\
\addlinespace[0.2em]
disturbance & D-1 & 0.535 & 0.533 & 0.550 & 0.545 & 0.542 & 0.527 & 0.539 & \textbf{0.550} & 0.540 \\
 & D-2 & 0.558 & 0.559 & 0.559 & \textbf{0.564} & 0.560 & 0.552 & 0.559 & 0.557 & 0.561 \\
 & D-3 & 0.523 & 0.537 & \textbf{0.537} & 0.536 & 0.532 & 0.532 & 0.534 & 0.535 & 0.535 \\
\addlinespace[0.2em]
stress & D-1 & 0.531 & 0.514 & 0.532 & 0.534 & \textbf{0.536} & 0.520 & 0.524 & 0.531 & 0.534 \\
 & D-2 & 0.536 & 0.554 & 0.559 & 0.557 & 0.554 & 0.555 & 0.553 & 0.555 & \textbf{0.562} \\
 & D-3 & 0.571 & 0.571 & 0.570 & 0.569 & 0.567 & \textbf{0.575} & 0.568 & 0.568 & 0.564 \\
\addlinespace[0.2em]
valence & D-1 & 0.542 & 0.547 & \textbf{0.551} & 0.543 & 0.538 & 0.545 & 0.550 & 0.551 & 0.536 \\
 & D-2 & 0.539 & \textbf{0.554} & 0.538 & 0.543 & 0.528 & 0.525 & 0.538 & 0.552 & 0.537 \\
 & D-3 & 0.562 & 0.574 & 0.574 & 0.577 & 0.571 & 0.559 & 0.572 & 0.567 & \textbf{0.579} \\
\midrule
\multicolumn{11}{l}{\textbf{D-3 rich labels}} \\
\midrule
angry & D-3 & 0.519 & 0.524 & 0.522 & \textbf{0.526} & 0.519 & 0.523 & 0.521 & 0.523 & 0.525 \\
anxious &  & 0.553 & 0.551 & 0.555 & \textbf{0.558} & 0.555 & 0.556 & 0.549 & 0.545 & 0.548 \\
cheerful &  & 0.594 & \textbf{0.607} & 0.605 & 0.605 & 0.601 & 0.598 & 0.606 & 0.604 & 0.603 \\
content &  & 0.588 & \textbf{0.589} & 0.576 & 0.584 & 0.585 & 0.582 & 0.583 & 0.580 & 0.587 \\
depressed &  & 0.532 & 0.540 & 0.525 & 0.540 & 0.537 & 0.527 & 0.530 & 0.538 & \textbf{0.544} \\
happy &  & 0.589 & 0.583 & 0.592 & 0.592 & 0.582 & \textbf{0.594} & 0.589 & 0.582 & 0.592 \\
relaxed &  & 0.560 & 0.558 & 0.555 & \textbf{0.562} & 0.553 & 0.556 & 0.559 & 0.555 & 0.553 \\
sad &  & 0.507 & 0.516 & 0.538 & 0.537 & 0.541 & 0.526 & 0.541 & 0.535 & \textbf{0.548} \\
\bottomrule
\end{tabular}
\end{table}

\clearpage
\begin{table}[!htbp]
\centering
\scriptsize
\setlength{\tabcolsep}{4pt}
\renewcommand{\arraystretch}{1.05}
\caption{Within-wave cross-user results for unsupervised domain adaptation (UDA) models. Cells show test AUROC; bold marks the best model per row.}
\label{tab:appendix_tier_b_da_only_results}
\resizebox{\linewidth}{!}{%
\begin{tabular}{@{}llccccccccccc@{}}
\toprule
Label & Dataset & DANN & CDAN & DAN & CORAL & MCC & ADDA & MCD & JAN & SHOT & CBST & CGDM \\
\midrule
\multicolumn{13}{l}{\textbf{Shared labels}} \\
\midrule
arousal & D-1 & 0.570 & 0.561 & 0.531 & 0.569 & 0.566 & 0.552 & \textbf{0.576} & 0.562 & 0.566 & 0.534 & 0.566 \\
 & D-2 & 0.568 & 0.572 & 0.573 & 0.571 & 0.573 & 0.563 & 0.568 & 0.552 & \textbf{0.573} & 0.540 & 0.561 \\
 & D-3 & \textbf{0.611} & 0.607 & 0.608 & 0.601 & 0.609 & 0.606 & 0.608 & 0.592 & 0.608 & 0.562 & 0.602 \\
\addlinespace[0.2em]
disturbance & D-1 & 0.553 & 0.559 & 0.563 & 0.558 & 0.551 & 0.543 & \textbf{0.566} & 0.554 & 0.559 & 0.524 & 0.548 \\
 & D-2 & 0.558 & 0.566 & 0.563 & 0.566 & 0.564 & 0.555 & 0.561 & 0.560 & \textbf{0.567} & 0.538 & 0.545 \\
 & D-3 & 0.538 & 0.544 & 0.546 & 0.534 & \textbf{0.555} & 0.525 & 0.537 & 0.518 & 0.550 & 0.508 & 0.526 \\
\addlinespace[0.2em]
stress & D-1 & 0.503 & 0.513 & 0.516 & 0.512 & 0.503 & 0.508 & 0.510 & 0.508 & \textbf{0.538} & 0.505 & 0.506 \\
 & D-2 & 0.548 & 0.550 & 0.549 & 0.541 & 0.548 & \textbf{0.555} & 0.544 & 0.516 & 0.552 & 0.550 & 0.548 \\
 & D-3 & 0.579 & 0.579 & \textbf{0.581} & 0.575 & 0.572 & 0.564 & 0.573 & 0.551 & 0.573 & 0.575 & 0.557 \\
\addlinespace[0.2em]
valence & D-1 & 0.570 & \textbf{0.570} & 0.563 & 0.564 & 0.567 & 0.539 & 0.565 & 0.555 & 0.565 & 0.532 & 0.553 \\
 & D-2 & 0.573 & 0.573 & 0.570 & 0.566 & \textbf{0.575} & 0.573 & 0.573 & 0.567 & 0.563 & 0.523 & 0.552 \\
 & D-3 & 0.581 & 0.575 & 0.578 & 0.579 & \textbf{0.586} & 0.532 & 0.586 & 0.563 & 0.577 & 0.536 & 0.573 \\
\midrule
\multicolumn{13}{l}{\textbf{D-3 rich labels}} \\
\midrule
angry & D-3 & 0.539 & \textbf{0.544} & 0.543 & 0.539 & 0.533 & 0.522 & 0.542 & 0.536 & 0.542 & 0.512 & 0.540 \\
anxious &  & 0.560 & 0.562 & 0.548 & \textbf{0.565} & 0.559 & 0.559 & 0.557 & 0.558 & 0.556 & 0.536 & 0.554 \\
cheerful &  & 0.599 & 0.603 & 0.611 & 0.609 & \textbf{0.613} & 0.601 & 0.607 & 0.590 & 0.608 & 0.584 & 0.591 \\
content &  & 0.592 & \textbf{0.600} & 0.593 & 0.593 & 0.598 & 0.593 & 0.598 & 0.574 & 0.587 & 0.594 & 0.578 \\
depressed &  & \textbf{0.557} & 0.553 & 0.556 & 0.553 & 0.556 & 0.529 & 0.557 & 0.543 & 0.547 & 0.537 & 0.549 \\
happy &  & 0.604 & 0.601 & 0.600 & 0.604 & \textbf{0.609} & 0.599 & 0.596 & 0.597 & 0.597 & 0.571 & 0.587 \\
relaxed &  & 0.569 & 0.571 & 0.572 & 0.570 & \textbf{0.574} & 0.547 & 0.567 & 0.569 & 0.566 & 0.557 & 0.564 \\
sad &  & 0.556 & 0.556 & 0.552 & 0.558 & \textbf{0.571} & 0.535 & 0.555 & 0.550 & 0.548 & 0.539 & 0.545 \\
\bottomrule
\end{tabular}
}
\end{table}

\setcounter{table}{29}
\begin{table}[!htbp]
\centering
\footnotesize
\setlength{\tabcolsep}{4pt}
\renewcommand{\arraystretch}{1.05}
\caption{Within-wave family-best comparison. (a) Best model per family, \textit{Model: AUROC ($\pm$ fold std)}. Bold marks the highest value, not a significant difference. (b) MLP backbone. Prev. is the share of HIGH labels, and AUROC is shown with its participant-level 95\% CI in brackets, followed by the Brier score and ECE. $^{\dagger}$CI includes 0.5. (c) Paired $\Delta$AUROC of family-best models on re-executed runs with 95\% CIs in brackets, per label (three waves) and pooled over 12 directions. Sup includes MLP. Bold marks a CI that excludes 0, $^{*}$marks significance after Holm adjustment over the 24 per-label contrasts, and $p_{\text{Holm}}$ is computed over the 12 principal pooled contrasts.}
\label{tab:appendix_tier_b_family_comparison}
\textit{(a) Family-best models, fold-mean AUROC}\par\smallskip
\begin{tabular}{@{}llccc@{}}
\toprule
Label & Dataset & Baseline & DG & UDA \\
\midrule
\multicolumn{5}{l}{\textbf{Shared labels}} \\
\midrule
arousal & D-1 & TabTr: 0.557 ($\pm$0.016) & MASF: 0.546 ($\pm$0.013) & \textbf{MCD: 0.576 ($\pm$0.013)} \\
 & D-2 & SAINT: 0.562 ($\pm$0.025) & GDRO: 0.567 ($\pm$0.016) & \textbf{SHOT: 0.573 ($\pm$0.015)} \\
 & D-3 & MLP: 0.610 ($\pm$0.007) & VREx: 0.609 ($\pm$0.005) & \textbf{DANN: 0.611 ($\pm$0.003)} \\
\addlinespace[0.2em]
disturbance & D-1 & TabTr: 0.542 ($\pm$0.011) & CSD: 0.550 ($\pm$0.018) & \textbf{MCD: 0.566 ($\pm$0.029)} \\
 & D-2 & SAINT: 0.565 ($\pm$0.014) & MixStyle: 0.564 ($\pm$0.011) & \textbf{SHOT: 0.567 ($\pm$0.008)} \\
 & D-3 & MLP: 0.523 ($\pm$0.011) & GDRO: 0.537 ($\pm$0.013) & \textbf{MCC: 0.555 ($\pm$0.023)} \\
\addlinespace[0.2em]
stress & D-1 & FTT: 0.521 ($\pm$0.020) & MLDG: 0.536 ($\pm$0.008) & \textbf{SHOT: 0.538 ($\pm$0.017)} \\
 & D-2 & MLP: 0.546 ($\pm$0.015) & \textbf{SagNet: 0.562 ($\pm$0.016)} & ADDA: 0.555 ($\pm$0.019) \\
 & D-3 & MLP: 0.580 ($\pm$0.016) & MASF: 0.575 ($\pm$0.014) & \textbf{DAN: 0.581 ($\pm$0.018)} \\
\addlinespace[0.2em]
valence & D-1 & MLP: 0.558 ($\pm$0.019) & GDRO: 0.551 ($\pm$0.031) & \textbf{CDAN: 0.570 ($\pm$0.015)} \\
 & D-2 & MLP: 0.545 ($\pm$0.013) & VREx: 0.554 ($\pm$0.023) & \textbf{MCC: 0.575 ($\pm$0.028)} \\
 & D-3 & MLP: 0.576 ($\pm$0.019) & SagNet: 0.579 ($\pm$0.021) & \textbf{MCC: 0.586 ($\pm$0.014)} \\
\midrule
\multicolumn{5}{l}{\textbf{D-3 rich labels}} \\
\midrule
angry & D-3 & MLP: 0.523 ($\pm$0.019) & MixStyle: 0.526 ($\pm$0.022) & \textbf{CDAN: 0.544 ($\pm$0.026)} \\
anxious &  & MLP: 0.554 ($\pm$0.016) & MixStyle: 0.558 ($\pm$0.020) & \textbf{CORAL: 0.565 ($\pm$0.011)} \\
cheerful &  & MLP: 0.609 ($\pm$0.025) & VREx: 0.607 ($\pm$0.025) & \textbf{MCC: 0.613 ($\pm$0.032)} \\
content &  & MLP: 0.594 ($\pm$0.012) & VREx: 0.589 ($\pm$0.013) & \textbf{CDAN: 0.600 ($\pm$0.013)} \\
depressed &  & SAINT: 0.532 ($\pm$0.043) & SagNet: 0.544 ($\pm$0.024) & \textbf{DANN: 0.557 ($\pm$0.010)} \\
happy &  & MLP: 0.584 ($\pm$0.023) & MASF: 0.594 ($\pm$0.013) & \textbf{MCC: 0.609 ($\pm$0.012)} \\
relaxed &  & MLP: 0.556 ($\pm$0.013) & MixStyle: 0.562 ($\pm$0.013) & \textbf{MCC: 0.574 ($\pm$0.026)} \\
sad &  & \textbf{XGB: 0.581 ($\pm$0.059)} & SagNet: 0.548 ($\pm$0.053) & MCC: 0.571 ($\pm$0.024) \\
\bottomrule
\end{tabular}
\par\smallskip
\textit{(b) MLP backbone per (label, dataset)}\par\smallskip
{\renewcommand{\arraystretch}{0.9}
\begin{tabular}{@{}llrrcccc@{}}
\toprule
Label & Dataset & Users & Instances & Prev. & AUROC & Brier & ECE \\
\midrule
arousal & D-1 & 92 & 10,075 & 0.340 & 0.549 [0.529, 0.568] & 0.225 & 0.033 \\
 & D-2 & 99 & 20,831 & 0.313 & 0.544 [0.529, 0.559] & 0.261 & 0.174 \\
 & D-3 & 106 & 21,619 & 0.330 & 0.605 [0.590, 0.620] & 0.214 & 0.013 \\
\addlinespace[0.2em]
disturbance & D-1 & 92 & 10,075 & 0.315 & 0.528 [0.503, 0.550] & 0.222 & 0.060 \\
 & D-2 & 99 & 20,831 & 0.283 & 0.555 [0.537, 0.572] & 0.203 & 0.031 \\
 & D-3 & 106 & 21,619 & 0.301 & 0.490$^{\dagger}$ [0.447, 0.530] & 0.211 & 0.025 \\
\addlinespace[0.2em]
stress & D-1 & 92 & 10,075 & 0.457 & 0.484$^{\dagger}$ [0.452, 0.520] & 0.249 & 0.002 \\
 & D-2 & 99 & 20,831 & 0.466 & 0.509$^{\dagger}$ [0.479, 0.539] & 0.251 & 0.016 \\
 & D-3 & 106 & 21,619 & 0.447 & 0.579 [0.565, 0.593] & 0.246 & 0.045 \\
\addlinespace[0.2em]
valence & D-1 & 92 & 10,075 & 0.261 & 0.558 [0.537, 0.576] & 0.209 & 0.098 \\
 & D-2 & 99 & 20,831 & 0.226 & 0.546 [0.529, 0.563] & 0.202 & 0.114 \\
 & D-3 & 106 & 21,619 & 0.268 & 0.554 [0.536, 0.572] & 0.210 & 0.098 \\
\bottomrule
\end{tabular}}
\par\smallskip
\textit{(c) Paired contrasts}\par\smallskip
{\scriptsize\setlength{\tabcolsep}{3pt}\renewcommand{\arraystretch}{0.9}
\begin{tabular}{@{}lcccccc@{}}
\toprule
Contrast & arousal & disturbance & stress & valence & Pooled (12) & $p_{\text{Holm}}$ \\
\midrule
Best DG $-$ MLP & +.0065 & \textbf{+.0193} & \textbf{+.0329}$^{*}$ & +.0123 & \textbf{+.0177} & 0.012 \\
 & [$-$.0032, +.0160] & \textbf{[+.0020, +.0356]} & \textbf{[+.0175, +.0482]} & [$-$.0039, +.0268] & \textbf{[+.0094, +.0252]} & \\
\addlinespace[0.15em]
Best UDA $-$ MLP & +.0051 & \textbf{+.0176} & \textbf{+.0214} & +.0057 & \textbf{+.0124} & 0.021 \\
 & [$-$.0053, +.0156] & \textbf{[+.0021, +.0331]} & \textbf{[+.0045, +.0376]} & [$-$.0072, +.0193] & \textbf{[+.0053, +.0197]} & \\
\addlinespace[0.15em]
Best Sup $-$ MLP & +.0023 & +.0030 & \textbf{+.0285}$^{*}$ & +.0019 & \textbf{+.0089} & 0.018 \\
 & [$-$.0050, +.0100] & [$-$.0088, +.0154] & \textbf{[+.0115, +.0457]} & [$-$.0078, +.0119] & \textbf{[+.0027, +.0153]} & \\
\addlinespace[0.15em]
Best DG $-$ best Sup & +.0042 & +.0163 & +.0043 & +.0104 & \textbf{+.0088} & 0.120 \\
 & [$-$.0062, +.0140] & [$-$.0002, +.0321] & [$-$.0097, +.0180] & [$-$.0074, +.0268] & \textbf{[+.0016, +.0159]} & \\
\addlinespace[0.15em]
Best UDA $-$ best Sup & +.0028 & \textbf{+.0146} & $-$.0071 & +.0038 & +.0035 & 0.566 \\
 & [$-$.0090, +.0146] & \textbf{[+.0014, +.0278]} & [$-$.0152, +.0007] & [$-$.0106, +.0190] & [$-$.0027, +.0102] & \\
\addlinespace[0.15em]
Best DG $-$ best UDA & +.0014 & +.0017 & +.0114 & +.0067 & +.0053 & 0.390 \\
 & [$-$.0102, +.0122] & [$-$.0074, +.0103] & [$-$.0034, +.0258] & [$-$.0077, +.0194] & [$-$.0015, +.0113] & \\
\bottomrule
\end{tabular}}
\end{table}

% Camera-ready: within-wave paired bootstrap comparisons
% Source: recovered server joint-bootstrap results (B=2,000).
\begin{table}[!htbp]
\centering
\scriptsize
\setlength{\tabcolsep}{2.5pt}
\renewcommand{\arraystretch}{0.9}
\caption{Within-wave per-direction paired family contrasts on re-executed runs. Each cell is $\Delta$AUROC (first row) with its participant-level 95\% CI (second row) from the joint bootstrap of Appendix~C.1.2. Family representatives are chosen per label and wave by pooled test AUROC, as in Table~\ref{tab:appendix_tier_b_family_comparison}(c). Sup includes MLP. Bold marks a CI that excludes 0. These per-direction intervals are not multiplicity adjusted. For arousal, disturbance and valence the re-executed pool has 11 models.}
\label{tab:appendix_tier_b_per_direction_contrasts}
\resizebox{\linewidth}{!}{%
\begin{tabular}{@{}llcccccc@{}}
\toprule
Label & Dataset & Best DG $-$ MLP & Best UDA $-$ MLP & Best DG $-$ best Sup & Best UDA $-$ best Sup & Best DG $-$ best UDA & Best Sup $-$ MLP \\
\midrule
arousal & D-1 & $-$0.0052 & +0.0040 & $-$0.0122 & $-$0.0030 & $-$0.0092 & +0.0070 \\
 & & [$-$0.0273, +0.0164] & [$-$0.0211, +0.0308] & [$-$0.0363, +0.0119] & [$-$0.0340, +0.0289] & [$-$0.0384, +0.0186] & [$-$0.0150, +0.0300] \\
 & D-2 & \textbf{+0.0226} & +0.0137 & \textbf{+0.0226} & +0.0137 & +0.0088 & \textit{Sup = MLP} \\
 & & \textbf{[+0.0056, +0.0395]} & [$-$0.0013, +0.0292] & \textbf{[+0.0056, +0.0395]} & [$-$0.0013, +0.0292] & [$-$0.0067, +0.0228] &  \\
 & D-3 & +0.0023 & $-$0.0024 & +0.0023 & $-$0.0024 & +0.0046 & \textit{Sup = MLP} \\
 & & [$-$0.0052, +0.0099] & [$-$0.0083, +0.0040] & [$-$0.0052, +0.0099] & [$-$0.0083, +0.0040] & [$-$0.0036, +0.0133] &  \\
\addlinespace[0.2em]
disturbance & D-1 & +0.0153 & +0.0088 & +0.0153 & +0.0088 & +0.0065 & \textit{Sup = MLP} \\
 & & [$-$0.0113, +0.0439] & [$-$0.0096, +0.0291] & [$-$0.0113, +0.0439] & [$-$0.0096, +0.0291] & [$-$0.0149, +0.0270] &  \\
 & D-2 & $-$0.0032 & +0.0004 & $-$0.0032 & +0.0004 & $-$0.0037 & \textit{Sup = MLP} \\
 & & [$-$0.0161, +0.0095] & [$-$0.0090, +0.0104] & [$-$0.0161, +0.0095] & [$-$0.0090, +0.0104] & [$-$0.0116, +0.0044] &  \\
 & D-3 & \textbf{+0.0458} & \textbf{+0.0435} & +0.0368 & \textbf{+0.0345} & +0.0023 & +0.0090 \\
 & & \textbf{[+0.0060, +0.0849]} & \textbf{[+0.0021, +0.0841]} & [$-$0.0006, +0.0738] & \textbf{[+0.0025, +0.0673]} & [$-$0.0112, +0.0151] & [$-$0.0263, +0.0462] \\
\addlinespace[0.2em]
stress & D-1 & \textbf{+0.0440} & +0.0333 & $-$0.0072 & \textbf{$-$0.0179} & +0.0107 & \textbf{+0.0512} \\
 & & \textbf{[+0.0089, +0.0756]} & [$-$0.0001, +0.0707] & [$-$0.0447, +0.0316] & \textbf{[$-$0.0362, $-$0.0002]} & [$-$0.0302, +0.0487] & \textbf{[+0.0135, +0.0909]} \\
 & D-2 & \textbf{+0.0519} & \textbf{+0.0343} & \textbf{+0.0176} & $-$0.0000 & \textbf{+0.0176} & \textbf{+0.0343} \\
 & & \textbf{[+0.0200, +0.0841]} & \textbf{[+0.0015, +0.0664]} & \textbf{[+0.0037, +0.0312]} & [$-$0.0108, +0.0108] & \textbf{[+0.0050, +0.0302]} & \textbf{[+0.0030, +0.0673]} \\
 & D-3 & +0.0027 & $-$0.0033 & +0.0027 & $-$0.0033 & +0.0060 & \textit{Sup = MLP} \\
 & & [$-$0.0065, +0.0111] & [$-$0.0145, +0.0079] & [$-$0.0065, +0.0111] & [$-$0.0145, +0.0079] & [$-$0.0077, +0.0186] &  \\
\addlinespace[0.2em]
valence & D-1 & +0.0018 & +0.0002 & +0.0018 & +0.0002 & +0.0016 & \textit{Sup = MLP} \\
 & & [$-$0.0364, +0.0360] & [$-$0.0207, +0.0236] & [$-$0.0364, +0.0360] & [$-$0.0207, +0.0236] & [$-$0.0295, +0.0270] &  \\
 & D-2 & +0.0102 & +0.0147 & +0.0102 & +0.0147 & $-$0.0045 & \textit{Sup = MLP} \\
 & & [$-$0.0089, +0.0280] & [$-$0.0124, +0.0422] & [$-$0.0089, +0.0280] & [$-$0.0124, +0.0422] & [$-$0.0298, +0.0186] &  \\
 & D-3 & \textbf{+0.0249} & +0.0021 & +0.0192 & $-$0.0036 & \textbf{+0.0228} & +0.0057 \\
 & & \textbf{[+0.0034, +0.0457]} & [$-$0.0150, +0.0201] & [$-$0.0127, +0.0509] & [$-$0.0315, +0.0251] & \textbf{[+0.0041, +0.0412]} & [$-$0.0235, +0.0357] \\
\bottomrule
\end{tabular}}
\end{table}

\clearpage

\medskip
\subsubsection*{C.3.3 Cross-Wave Generalization Results}

Tables~\ref{tab:appendix_tier_c_baseline_tabular_results}--\ref{tab:appendix_tier_c_da_results} report per-direction results for baseline, DG, and UDA models across the six single-source and three multi-source directions. Family-level rankings are substantially reordered. DG achieves the best mean under single-source transfer ($0.542 \pm 0.017$), the baseline family leads under multi-source pooling ($0.551 \pm 0.015$) and UDA trails in both ($0.531$ single, $0.532$ multi). Multi-source pooling yields modest, label-dependent gains over single-source transfer ($+0.004$ AUROC for arousal, $+0.006$ for valence, $+0.005$ for disturbance, and $+0.005$ for stress), and most of this improvement accrues to baseline models rather than to methods that explicitly model domain structure. Direction asymmetries are most significant for arousal involving D-3, with $\text{D-2}\!\to\!\text{D-3}$ exceeding $\text{D-3}\!\to\!\text{D-2}$ by $0.035$ AUROC and $\text{D-1}\!\to\!\text{D-3}$ exceeding $\text{D-3}\!\to\!\text{D-1}$ by $0.023$, whereas the remaining shared labels are markedly more symmetric (asymmetry below $0.02$ AUROC for valence and stress and below $0.01$ for disturbance).

\noindent\textbf{Comparison with the MLP backbone and with supervised models.} On re-executed runs, all nine DG objectives have positive mean AUROC differences from MLP over the label and direction cells ($+0.0032$ to $+0.0073$). Their intervals exclude zero for eight of them, and four (CSD, MLDG, IRM, and MixStyle) remain significant after Holm adjustment over the DG and UDA methods. UDA is less reliable. Only DAN has an interval above zero ($+0.0065$, $p_{\text{Holm}} = 0.052$), and DANN, MCC, and MCD are significantly below MLP after adjustment. With one family representative per label (Appendix~C.1.2), family-best DG and UDA exceed MLP by $+0.0078$ $[0.0044, 0.0113]$ and $+0.0077$ $[0.0036, 0.0121]$ but lie $0.0065$ below the strongest supervised model, XGBoost or LightGBM (intervals $[-0.0122, -0.0009]$ and $[-0.0125, -0.0009]$, not significant after Holm adjustment). The strongest supervised model exceeds MLP by $+0.0143$ $[0.0085, 0.0204]$, about twice the adaptation gains. Under transfer, MLP is also poorly calibrated (Brier $0.247$ to $0.383$ and ECE $0.220$ to $0.355$ across labels) compared with XGBoost (Brier $0.191$ to $0.251$ and ECE $0.041$ to $0.047$).

\noindent\textbf{Recurring participants.} Participant codes are assigned per wave, so equal codes in different waves do not identify the same person. The 14 verified recurring individuals are identified from the enrolment crosswalk; unconfirmed potential links are treated as distinct participants. Removing from the training waves every participant linked to the target wave (5 to 11 per direction, with test sets unchanged) changes mean AUROC by $-0.0022$ (stress), $+0.0003$ (valence), $+0.0001$ (disturbance), and $-0.0020$ (arousal), and none of the 103 model and label differences is significant after Holm adjustment. Separately, removing training participants whose wave-local Pcode strings happen to match target Pcode strings (not identity matches), an intentionally over-conservative training-size control (63 to 98 per direction), changes these means by $-0.0047$, $-0.0010$, $-0.0069$, and $+0.0007$. Recurrence therefore does not materially drive cross-wave performance.

Detailed per-direction paired bootstrap comparisons
are reported in
Table~\ref{tab:appendix_tier_c_per_direction_contrasts}.

\begin{table}[!htbp]
\centering
\scriptsize
\setlength{\tabcolsep}{4pt}
\renewcommand{\arraystretch}{1.05}
\caption{Full cross-wave AUROC results for baseline and tabular models. Best per row in \textbf{bold}.}
\label{tab:appendix_tier_c_baseline_tabular_results}
\begin{tabular}{@{}llccccccccc@{}}
\toprule
Label & Train $\rightarrow$ Test & XGB & LGB & MLP & ResNet & TabNet & DCN & SAINT & TabTr & FTT \\
\midrule
arousal & D-1 $\rightarrow$ D-2 & 0.535 & 0.529 & 0.498 & 0.494 & 0.500 & 0.531 & \textbf{0.538} & 0.510 & 0.538 \\
 & D-1 $\rightarrow$ D-3 & 0.545 & \textbf{0.572} & 0.520 & 0.535 & 0.499 & 0.540 & 0.545 & 0.515 & 0.545 \\
 & D-2 $\rightarrow$ D-1 & 0.538 & 0.528 & 0.512 & 0.534 & 0.510 & 0.525 & \textbf{0.553} & 0.517 & 0.553 \\
 & D-2 $\rightarrow$ D-3 & \textbf{0.582} & 0.565 & 0.539 & 0.554 & 0.507 & 0.566 & 0.580 & 0.535 & 0.580 \\
 & D-3 $\rightarrow$ D-1 & 0.548 & 0.541 & 0.545 & 0.537 & 0.515 & 0.538 & \textbf{0.552} & 0.533 & 0.552 \\
 & D-3 $\rightarrow$ D-2 & 0.529 & 0.541 & 0.527 & 0.514 & 0.510 & \textbf{0.543} & 0.533 & 0.527 & 0.533 \\
 & D-1+D-2 $\rightarrow$ D-3 & \textbf{0.578} & 0.577 & 0.537 & 0.552 & 0.504 & 0.551 & 0.562 & 0.536 & 0.562 \\
 & D-1+D-3 $\rightarrow$ D-2 & 0.569 & \textbf{0.571} & 0.529 & 0.541 & 0.523 & 0.526 & 0.538 & 0.537 & 0.538 \\
 & D-2+D-3 $\rightarrow$ D-1 & 0.532 & \textbf{0.542} & 0.527 & 0.531 & 0.525 & 0.523 & 0.539 & 0.520 & 0.539 \\
\addlinespace[0.2em]
disturbance & D-1 $\rightarrow$ D-2 & 0.518 & 0.512 & 0.552 & 0.527 & 0.499 & 0.526 & 0.540 & \textbf{0.553} & 0.540 \\
 & D-1 $\rightarrow$ D-3 & 0.515 & 0.517 & 0.525 & 0.502 & 0.501 & \textbf{0.541} & 0.517 & 0.513 & 0.517 \\
 & D-2 $\rightarrow$ D-1 & 0.528 & 0.529 & 0.536 & \textbf{0.560} & 0.500 & 0.527 & 0.552 & 0.547 & 0.552 \\
 & D-2 $\rightarrow$ D-3 & 0.515 & 0.498 & 0.530 & 0.525 & 0.501 & \textbf{0.542} & 0.532 & 0.520 & 0.532 \\
 & D-3 $\rightarrow$ D-1 & \textbf{0.541} & 0.523 & 0.513 & 0.508 & 0.490 & 0.527 & 0.508 & 0.516 & 0.508 \\
 & D-3 $\rightarrow$ D-2 & \textbf{0.538} & 0.533 & 0.506 & 0.519 & 0.498 & 0.526 & 0.512 & 0.504 & 0.512 \\
 & D-1+D-2 $\rightarrow$ D-3 & 0.537 & 0.518 & 0.525 & 0.523 & 0.497 & \textbf{0.542} & 0.520 & 0.524 & 0.520 \\
 & D-1+D-3 $\rightarrow$ D-2 & \textbf{0.565} & 0.550 & 0.536 & 0.536 & 0.510 & 0.527 & 0.540 & 0.529 & 0.540 \\
 & D-2+D-3 $\rightarrow$ D-1 & 0.559 & \textbf{0.561} & 0.536 & 0.534 & 0.507 & 0.527 & 0.534 & 0.534 & 0.534 \\
\addlinespace[0.2em]
stress & D-1 $\rightarrow$ D-2 & 0.525 & 0.524 & 0.533 & 0.517 & 0.497 & 0.502 & 0.494 & \textbf{0.540} & 0.522 \\
 & D-1 $\rightarrow$ D-3 & 0.511 & 0.500 & 0.507 & 0.502 & 0.497 & 0.489 & 0.493 & \textbf{0.525} & 0.518 \\
 & D-2 $\rightarrow$ D-1 & \textbf{0.532} & 0.526 & 0.531 & 0.532 & 0.511 & 0.513 & 0.506 & 0.500 & 0.523 \\
 & D-2 $\rightarrow$ D-3 & 0.524 & \textbf{0.534} & 0.504 & 0.502 & 0.505 & 0.527 & 0.492 & 0.509 & 0.532 \\
 & D-3 $\rightarrow$ D-1 & 0.504 & \textbf{0.530} & 0.518 & 0.515 & 0.502 & 0.505 & 0.499 & 0.492 & 0.508 \\
 & D-3 $\rightarrow$ D-2 & 0.519 & 0.524 & \textbf{0.542} & 0.540 & 0.503 & 0.509 & 0.522 & 0.526 & 0.518 \\
 & D-1+D-2 $\rightarrow$ D-3 & \textbf{0.522} & 0.520 & 0.512 & 0.512 & 0.491 & 0.522 & 0.491 & 0.493 & 0.508 \\
 & D-1+D-3 $\rightarrow$ D-2 & \textbf{0.532} & 0.531 & 0.518 & 0.513 & 0.505 & 0.523 & 0.512 & 0.516 & 0.513 \\
 & D-2+D-3 $\rightarrow$ D-1 & 0.528 & 0.535 & \textbf{0.536} & 0.521 & 0.507 & 0.519 & 0.509 & 0.493 & 0.498 \\
\addlinespace[0.2em]
valence & D-1 $\rightarrow$ D-2 & 0.521 & 0.514 & 0.512 & 0.504 & 0.505 & 0.507 & \textbf{0.534} & 0.515 & 0.534 \\
 & D-1 $\rightarrow$ D-3 & \textbf{0.551} & 0.533 & 0.530 & 0.531 & 0.499 & 0.541 & 0.523 & 0.536 & 0.523 \\
 & D-2 $\rightarrow$ D-1 & 0.530 & 0.509 & 0.514 & 0.516 & 0.506 & 0.518 & \textbf{0.541} & 0.512 & 0.541 \\
 & D-2 $\rightarrow$ D-3 & \textbf{0.542} & 0.529 & 0.535 & 0.519 & 0.505 & 0.541 & 0.528 & 0.526 & 0.528 \\
 & D-3 $\rightarrow$ D-1 & 0.527 & \textbf{0.561} & 0.522 & 0.516 & 0.508 & 0.520 & 0.525 & 0.534 & 0.525 \\
 & D-3 $\rightarrow$ D-2 & 0.528 & 0.536 & 0.536 & 0.540 & 0.507 & 0.507 & 0.545 & \textbf{0.546} & 0.545 \\
 & D-1+D-2 $\rightarrow$ D-3 & 0.543 & \textbf{0.562} & 0.527 & 0.531 & 0.499 & 0.540 & 0.532 & 0.524 & 0.532 \\
 & D-1+D-3 $\rightarrow$ D-2 & 0.553 & \textbf{0.556} & 0.521 & 0.527 & 0.512 & 0.507 & 0.537 & 0.520 & 0.537 \\
 & D-2+D-3 $\rightarrow$ D-1 & 0.535 & \textbf{0.559} & 0.514 & 0.523 & 0.528 & 0.520 & 0.546 & 0.520 & 0.546 \\
\bottomrule
\end{tabular}
\end{table}

\begin{table}[!htbp]
\centering
\scriptsize
\setlength{\tabcolsep}{4pt}
\renewcommand{\arraystretch}{1.05}
\caption{Full cross-wave AUROC results for DG models. Best per row in \textbf{bold}.}
\label{tab:appendix_tier_c_dg_results}
\begin{tabular}{@{}llccccccccc@{}}
\toprule
Label & Train $\rightarrow$ Test & IRM & VREx & GDRO & MixStyle & MLDG & MASF & Fish & CSD & SagNet \\
\midrule
arousal & D-1 $\rightarrow$ D-2 & \textbf{0.565} & 0.544 & 0.554 & 0.553 & 0.559 & 0.560 & 0.556 & 0.547 & 0.533 \\
 & D-1 $\rightarrow$ D-3 & 0.582 & 0.532 & 0.564 & 0.561 & \textbf{0.583} & 0.561 & 0.574 & 0.558 & 0.557 \\
 & D-2 $\rightarrow$ D-1 & 0.544 & 0.546 & 0.542 & 0.543 & 0.533 & 0.531 & 0.544 & \textbf{0.548} & 0.540 \\
 & D-2 $\rightarrow$ D-3 & 0.595 & \textbf{0.595} & 0.590 & 0.591 & 0.570 & 0.585 & 0.593 & 0.593 & 0.581 \\
 & D-3 $\rightarrow$ D-1 & 0.550 & 0.550 & 0.550 & 0.552 & 0.553 & 0.559 & 0.554 & \textbf{0.560} & 0.547 \\
 & D-3 $\rightarrow$ D-2 & 0.556 & 0.556 & 0.555 & \textbf{0.560} & 0.557 & 0.552 & 0.558 & 0.559 & 0.550 \\
 & D-1+D-2 $\rightarrow$ D-3 & 0.590 & 0.589 & 0.568 & 0.584 & 0.575 & 0.586 & 0.587 & \textbf{0.591} & 0.558 \\
 & D-1+D-3 $\rightarrow$ D-2 & 0.552 & \textbf{0.555} & 0.545 & 0.553 & 0.545 & 0.535 & 0.553 & 0.552 & 0.546 \\
 & D-2+D-3 $\rightarrow$ D-1 & 0.542 & 0.541 & 0.547 & 0.531 & \textbf{0.552} & 0.544 & 0.544 & 0.533 & 0.532 \\
\addlinespace[0.2em]
disturbance & D-1 $\rightarrow$ D-2 & 0.545 & 0.549 & 0.552 & 0.555 & 0.553 & 0.546 & 0.551 & 0.548 & \textbf{0.556} \\
 & D-1 $\rightarrow$ D-3 & 0.535 & 0.527 & 0.529 & 0.529 & 0.532 & 0.528 & 0.525 & \textbf{0.537} & 0.530 \\
 & D-2 $\rightarrow$ D-1 & 0.545 & 0.545 & 0.541 & 0.546 & 0.541 & 0.536 & 0.546 & \textbf{0.549} & 0.524 \\
 & D-2 $\rightarrow$ D-3 & 0.524 & 0.524 & 0.530 & 0.529 & 0.532 & 0.513 & 0.530 & \textbf{0.535} & 0.507 \\
 & D-3 $\rightarrow$ D-1 & 0.535 & 0.534 & 0.529 & 0.532 & 0.531 & 0.533 & 0.529 & \textbf{0.536} & 0.515 \\
 & D-3 $\rightarrow$ D-2 & 0.529 & 0.529 & 0.522 & \textbf{0.533} & 0.523 & 0.517 & 0.522 & 0.525 & 0.512 \\
 & D-1+D-2 $\rightarrow$ D-3 & 0.524 & 0.525 & 0.518 & \textbf{0.525} & 0.519 & 0.514 & 0.513 & 0.521 & 0.499 \\
 & D-1+D-3 $\rightarrow$ D-2 & 0.543 & 0.543 & 0.539 & 0.542 & 0.540 & 0.544 & 0.540 & \textbf{0.553} & 0.535 \\
 & D-2+D-3 $\rightarrow$ D-1 & \textbf{0.540} & 0.524 & 0.525 & 0.527 & 0.536 & 0.532 & 0.525 & 0.525 & 0.534 \\
\addlinespace[0.2em]
stress & D-1 $\rightarrow$ D-2 & 0.535 & 0.529 & \textbf{0.538} & 0.526 & 0.531 & 0.531 & 0.527 & 0.520 & 0.521 \\
 & D-1 $\rightarrow$ D-3 & 0.503 & \textbf{0.511} & 0.509 & 0.504 & 0.507 & 0.508 & 0.502 & 0.503 & 0.499 \\
 & D-2 $\rightarrow$ D-1 & 0.527 & 0.527 & 0.532 & 0.526 & 0.529 & 0.529 & 0.532 & 0.531 & \textbf{0.534} \\
 & D-2 $\rightarrow$ D-3 & 0.516 & 0.514 & 0.512 & 0.509 & 0.511 & 0.510 & \textbf{0.518} & 0.515 & 0.505 \\
 & D-3 $\rightarrow$ D-1 & 0.510 & 0.511 & 0.515 & 0.512 & 0.503 & \textbf{0.524} & 0.512 & 0.512 & 0.517 \\
 & D-3 $\rightarrow$ D-2 & 0.534 & 0.537 & 0.542 & 0.539 & 0.542 & \textbf{0.547} & 0.541 & 0.545 & 0.545 \\
 & D-1+D-2 $\rightarrow$ D-3 & 0.501 & 0.508 & 0.495 & 0.513 & 0.500 & \textbf{0.514} & 0.508 & 0.509 & 0.511 \\
 & D-1+D-3 $\rightarrow$ D-2 & 0.530 & 0.527 & 0.529 & 0.532 & 0.523 & 0.531 & 0.528 & \textbf{0.538} & 0.519 \\
 & D-2+D-3 $\rightarrow$ D-1 & 0.539 & 0.537 & 0.537 & 0.537 & 0.536 & 0.532 & 0.536 & \textbf{0.543} & 0.538 \\
\addlinespace[0.2em]
valence & D-1 $\rightarrow$ D-2 & 0.536 & 0.536 & \textbf{0.544} & 0.540 & 0.539 & 0.528 & 0.533 & 0.535 & 0.540 \\
 & D-1 $\rightarrow$ D-3 & 0.539 & 0.539 & 0.540 & \textbf{0.541} & 0.534 & 0.533 & 0.536 & 0.534 & 0.536 \\
 & D-2 $\rightarrow$ D-1 & 0.539 & 0.540 & 0.522 & \textbf{0.540} & 0.539 & 0.530 & 0.537 & 0.537 & 0.521 \\
 & D-2 $\rightarrow$ D-3 & 0.551 & 0.551 & 0.542 & 0.552 & \textbf{0.554} & 0.543 & 0.548 & 0.550 & 0.528 \\
 & D-3 $\rightarrow$ D-1 & 0.529 & 0.530 & 0.533 & 0.527 & 0.534 & 0.535 & 0.531 & \textbf{0.537} & 0.532 \\
 & D-3 $\rightarrow$ D-2 & 0.543 & 0.543 & 0.542 & 0.540 & 0.541 & \textbf{0.552} & 0.538 & 0.542 & 0.539 \\
 & D-1+D-2 $\rightarrow$ D-3 & 0.547 & 0.540 & 0.545 & 0.552 & 0.550 & 0.533 & \textbf{0.553} & 0.528 & 0.530 \\
 & D-1+D-3 $\rightarrow$ D-2 & 0.536 & 0.531 & 0.544 & 0.542 & 0.528 & \textbf{0.549} & 0.549 & 0.548 & 0.539 \\
 & D-2+D-3 $\rightarrow$ D-1 & 0.541 & 0.540 & 0.522 & 0.527 & 0.533 & 0.532 & \textbf{0.542} & 0.526 & 0.537 \\
\bottomrule
\end{tabular}
\end{table}

\begin{table}[!htbp]
\centering
\scriptsize
\setlength{\tabcolsep}{4pt}
\renewcommand{\arraystretch}{1.05}
\caption{Full cross-wave AUROC results for UDA models. Best per row in \textbf{bold}. CGDM was not evaluated in the cross-wave benchmark in this release.}
\label{tab:appendix_tier_c_da_results}
\resizebox{\linewidth}{!}{%
\begin{tabular}{@{}llcccccccccc@{}}
\toprule
Label & Train $\rightarrow$ Test & DANN & CDAN & DAN & CORAL & MCC & ADDA & MCD & JAN & SHOT & CBST \\
\midrule
arousal & D-1 $\rightarrow$ D-2 & 0.535 & \textbf{0.547} & 0.536 & 0.498 & 0.538 & 0.511 & 0.497 & 0.544 & 0.479 & 0.481 \\
 & D-1 $\rightarrow$ D-3 & 0.477 & 0.546 & 0.515 & 0.530 & 0.475 & 0.531 & 0.499 & \textbf{0.548} & 0.518 & 0.508 \\
 & D-2 $\rightarrow$ D-1 & 0.525 & 0.521 & 0.518 & 0.528 & 0.504 & 0.518 & 0.499 & \textbf{0.554} & 0.508 & 0.526 \\
 & D-2 $\rightarrow$ D-3 & 0.533 & 0.526 & 0.545 & 0.512 & 0.503 & 0.517 & 0.522 & \textbf{0.546} & 0.540 & 0.543 \\
 & D-3 $\rightarrow$ D-1 & \textbf{0.558} & 0.535 & 0.551 & 0.541 & 0.512 & 0.538 & 0.492 & 0.537 & 0.554 & 0.551 \\
 & D-3 $\rightarrow$ D-2 & 0.527 & 0.498 & 0.520 & \textbf{0.541} & 0.535 & 0.518 & 0.506 & 0.513 & 0.511 & 0.523 \\
 & D-1+D-2 $\rightarrow$ D-3 & 0.490 & 0.533 & 0.536 & 0.545 & 0.504 & 0.499 & 0.523 & 0.547 & \textbf{0.552} & 0.498 \\
 & D-1+D-3 $\rightarrow$ D-2 & \textbf{0.532} & 0.530 & 0.531 & 0.522 & 0.523 & 0.520 & 0.512 & 0.502 & 0.515 & 0.493 \\
 & D-2+D-3 $\rightarrow$ D-1 & 0.521 & 0.534 & 0.536 & 0.549 & 0.512 & 0.540 & 0.490 & \textbf{0.551} & 0.520 & 0.532 \\
\addlinespace[0.2em]
disturbance & D-1 $\rightarrow$ D-2 & 0.534 & 0.528 & \textbf{0.549} & 0.533 & 0.511 & 0.544 & 0.507 & 0.515 & 0.542 & 0.516 \\
 & D-1 $\rightarrow$ D-3 & 0.512 & 0.519 & 0.522 & 0.503 & 0.504 & \textbf{0.529} & 0.501 & 0.487 & 0.516 & 0.513 \\
 & D-2 $\rightarrow$ D-1 & 0.537 & 0.538 & 0.537 & 0.546 & 0.498 & \textbf{0.551} & 0.499 & 0.508 & 0.542 & 0.510 \\
 & D-2 $\rightarrow$ D-3 & 0.528 & \textbf{0.535} & 0.529 & 0.520 & 0.502 & 0.507 & 0.504 & 0.492 & 0.526 & 0.502 \\
 & D-3 $\rightarrow$ D-1 & \textbf{0.532} & 0.523 & 0.521 & 0.509 & 0.497 & 0.508 & 0.489 & 0.509 & 0.501 & 0.500 \\
 & D-3 $\rightarrow$ D-2 & 0.493 & 0.514 & \textbf{0.534} & 0.521 & 0.515 & 0.514 & 0.529 & 0.480 & 0.488 & 0.467 \\
 & D-1+D-2 $\rightarrow$ D-3 & 0.508 & \textbf{0.533} & 0.515 & 0.527 & 0.510 & 0.521 & 0.493 & 0.513 & 0.528 & 0.502 \\
 & D-1+D-3 $\rightarrow$ D-2 & 0.545 & 0.531 & \textbf{0.547} & 0.529 & 0.523 & 0.518 & 0.533 & 0.540 & 0.546 & 0.541 \\
 & D-2+D-3 $\rightarrow$ D-1 & 0.530 & \textbf{0.536} & 0.533 & 0.535 & 0.529 & 0.499 & 0.520 & 0.502 & 0.535 & 0.529 \\
\addlinespace[0.2em]
stress & D-1 $\rightarrow$ D-2 & 0.500 & 0.517 & 0.512 & 0.512 & 0.477 & 0.529 & 0.494 & 0.531 & 0.533 & \textbf{0.534} \\
 & D-1 $\rightarrow$ D-3 & 0.490 & 0.497 & 0.510 & 0.503 & 0.505 & 0.510 & \textbf{0.513} & 0.511 & 0.506 & 0.511 \\
 & D-2 $\rightarrow$ D-1 & 0.506 & 0.535 & \textbf{0.537} & 0.529 & 0.505 & 0.533 & 0.507 & 0.517 & 0.535 & 0.535 \\
 & D-2 $\rightarrow$ D-3 & 0.489 & 0.510 & 0.502 & 0.507 & 0.511 & 0.508 & 0.506 & \textbf{0.515} & 0.506 & 0.502 \\
 & D-3 $\rightarrow$ D-1 & 0.517 & 0.515 & 0.499 & 0.512 & 0.503 & 0.515 & 0.498 & \textbf{0.526} & 0.509 & 0.507 \\
 & D-3 $\rightarrow$ D-2 & 0.538 & 0.544 & 0.542 & 0.542 & 0.540 & 0.540 & 0.506 & 0.529 & 0.541 & \textbf{0.545} \\
 & D-1+D-2 $\rightarrow$ D-3 & 0.510 & 0.505 & 0.512 & 0.513 & 0.499 & \textbf{0.516} & 0.506 & 0.497 & 0.512 & 0.503 \\
 & D-1+D-3 $\rightarrow$ D-2 & 0.534 & 0.536 & 0.531 & 0.527 & 0.520 & 0.508 & 0.500 & \textbf{0.544} & 0.514 & 0.525 \\
 & D-2+D-3 $\rightarrow$ D-1 & \textbf{0.539} & 0.539 & 0.535 & 0.536 & 0.509 & 0.530 & 0.516 & 0.527 & 0.538 & 0.531 \\
\addlinespace[0.2em]
valence & D-1 $\rightarrow$ D-2 & 0.556 & 0.547 & \textbf{0.561} & 0.529 & 0.533 & 0.534 & 0.501 & 0.508 & 0.506 & 0.494 \\
 & D-1 $\rightarrow$ D-3 & 0.503 & 0.502 & 0.531 & \textbf{0.545} & 0.532 & 0.517 & 0.512 & 0.538 & 0.527 & 0.517 \\
 & D-2 $\rightarrow$ D-1 & 0.536 & 0.522 & 0.514 & \textbf{0.541} & 0.522 & 0.518 & 0.513 & 0.538 & 0.510 & 0.517 \\
 & D-2 $\rightarrow$ D-3 & 0.511 & 0.514 & 0.523 & 0.519 & 0.499 & 0.509 & 0.484 & 0.501 & \textbf{0.535} & 0.516 \\
 & D-3 $\rightarrow$ D-1 & 0.528 & 0.501 & 0.508 & 0.501 & 0.497 & 0.502 & 0.510 & \textbf{0.553} & 0.527 & 0.528 \\
 & D-3 $\rightarrow$ D-2 & 0.528 & 0.519 & 0.525 & 0.522 & 0.490 & \textbf{0.545} & 0.507 & 0.468 & 0.523 & 0.506 \\
 & D-1+D-2 $\rightarrow$ D-3 & 0.509 & 0.517 & 0.522 & \textbf{0.532} & 0.499 & 0.489 & 0.508 & 0.522 & 0.520 & 0.513 \\
 & D-1+D-3 $\rightarrow$ D-2 & 0.520 & 0.528 & 0.522 & 0.512 & \textbf{0.535} & 0.515 & 0.510 & 0.522 & 0.528 & 0.516 \\
 & D-2+D-3 $\rightarrow$ D-1 & \textbf{0.537} & 0.521 & 0.517 & 0.524 & 0.497 & 0.507 & 0.499 & 0.504 & 0.516 & 0.508 \\
\bottomrule
\end{tabular}
}
\end{table}

\begin{table}[!htbp]
\centering
\footnotesize
\setlength{\tabcolsep}{4pt}
\renewcommand{\arraystretch}{1.05}
\caption{Cross-wave family-best comparison by generalization setting. Each cell reports the best-performing model within a family as \textit{Model: AUROC} (averaged across directions in that setting). Bold marks the best family per (label, setting).}
\label{tab:appendix_tier_c_family_comparison}
\begin{tabular}{@{}lcccccc@{}}
\toprule
& \multicolumn{3}{c}{Single-source ($n=6$)} & \multicolumn{3}{c}{Multi-source ($n=3$)} \\
\cmidrule(lr){2-4} \cmidrule(lr){5-7}
Label & Baseline & DG & DA & Baseline & DG & DA \\
\midrule
arousal & FTT: 0.550 & \textbf{IRM: 0.565} & JAN: 0.540 & \textbf{LGB: 0.563} & VREx: 0.561 & CORAL: 0.539 \\
disturbance & DCN: 0.531 & \textbf{CSD: 0.538} & DAN: 0.532 & \textbf{XGB: 0.554} & IRM: 0.536 & SHOT: 0.536 \\
stress & LGB: 0.523 & \textbf{MASF: 0.525} & ADDA: 0.523 & LGB: 0.529 & \textbf{CSD: 0.530} & DANN: 0.528 \\
valence & XGB: 0.533 & \textbf{MLDG: 0.540} & DAN: 0.527 & \textbf{LGB: 0.559} & Fish: 0.548 & CORAL: 0.523 \\
\midrule
Mean $\pm$ Std & 0.534 $\pm$ 0.011 & \textbf{0.542 $\pm$ 0.017} & 0.531 $\pm$ 0.008 & \textbf{0.551 $\pm$ 0.015} & 0.544 $\pm$ 0.014 & 0.532 $\pm$ 0.007 \\
\bottomrule
\end{tabular}
\end{table}

\begin{table}[!htbp]
\centering
\scriptsize
\setlength{\tabcolsep}{4pt}
\renewcommand{\arraystretch}{0.95}
\caption{Cross-wave model ranking by mean test AUROC. Each models score is the mean of its per-row test AUROC across all 36 cross-wave (label, transfer-direction) cells; $\pm$ is the standard deviation across those rows.}
\label{tab:appendix_tier_c_model_ranking}
\begin{tabular}{rllc}
\toprule
Rank & Model & Family & Test AUROC \\
\midrule
1 & IRM & DG & 0.540 $\pm$ 0.019 \\
2 & CSD & DG & 0.540 $\pm$ 0.018 \\
3 & Fish & DG & 0.539 $\pm$ 0.019 \\
4 & MixStyle & DG & 0.538 $\pm$ 0.018 \\
5 & MLDG & DG & 0.538 $\pm$ 0.017 \\
6 & VREx & DG & 0.537 $\pm$ 0.018 \\
7 & GDRO & DG & 0.537 $\pm$ 0.017 \\
8 & MASF & DG & 0.536 $\pm$ 0.018 \\
9 & XGB & Base & 0.535 $\pm$ 0.020 \\
10 & LGB & Base & 0.533 $\pm$ 0.023 \\
11 & FTT & Base & 0.531 $\pm$ 0.017 \\
12 & SAINT & Base & 0.531 $\pm$ 0.017 \\
13 & SagNet & DG & 0.531 $\pm$ 0.018 \\
14 & DCN & Base & 0.528 $\pm$ 0.013 \\
15 & DAN & DA & 0.526 $\pm$ 0.014 \\
16 & ResNet & Base & 0.526 $\pm$ 0.014 \\
17 & MLP & Base & 0.524 $\pm$ 0.012 \\
18 & CORAL & DA & 0.524 $\pm$ 0.014 \\
19 & CDAN & DA & 0.523 $\pm$ 0.013 \\
20 & TabTr & Base & 0.523 $\pm$ 0.012 \\
21 & DANN & DA & 0.522 $\pm$ 0.017 \\
22 & SHOT & DA & 0.521 $\pm$ 0.016 \\
23 & JAN & DA & 0.520 $\pm$ 0.021 \\
24 & ADDA & DA & 0.518 $\pm$ 0.014 \\
25 & CBST & DA & 0.513 $\pm$ 0.016 \\
26 & MCC & DA & 0.511 $\pm$ 0.016 \\
27 & MCD & DA & 0.508 $\pm$ 0.011 \\
28 & TabNet & Base & 0.505 $\pm$ 0.008 \\
\bottomrule
\end{tabular}
\end{table}

% Camera-ready: cross-wave paired bootstrap comparisons
\begin{table}[!htbp]
\centering
\scriptsize
\setlength{\tabcolsep}{4pt}
\caption{Cross-wave paired contrasts against the MLP backbone on re-executed runs. Each cell is $\Delta$AUROC with its participant-level 95\% CI in brackets from the joint bootstrap of Appendix~C.1.2. Family representatives are fixed per label (the model with the highest mean AUROC over the nine directions of that label), and Sup includes MLP. Bold marks a CI that excludes 0. Per-direction intervals are not multiplicity adjusted. The last rows aggregate over the nine directions of each label and over all 36 directions, where $^{*}$ marks significance after Holm adjustment over the 24 per-label contrasts and $p_{\text{Holm}}$ is computed over the 12 principal pooled contrasts.}
\label{tab:appendix_tier_c_per_direction_contrasts}
\begin{tabular}{@{}lccc@{}}
\toprule
Train $\rightarrow$ Test & Best DG $-$ MLP & Best UDA $-$ MLP & Best Sup $-$ MLP \\
\midrule
\multicolumn{4}{@{}l}{\textit{arousal}} \\
D-1 $\rightarrow$ D-2 & \textbf{+.0276 [+.0083, +.0464]} & \textbf{+.0323 [+.0130, +.0506]} & \textbf{+.0303 [+.0091, +.0505]} \\
D-1 $\rightarrow$ D-3 & \textbf{+.0300 [+.0162, +.0439]} & \textbf{+.0717 [+.0481, +.0955]} & \textbf{+.0369 [+.0138, +.0593]} \\
D-2 $\rightarrow$ D-1 & +.0165 [$-$.0069, +.0411] & +.0034 [$-$.0157, +.0255] & +.0147 [$-$.0071, +.0367] \\
D-2 $\rightarrow$ D-3 & +.0024 [$-$.0121, +.0172] & $-$.0066 [$-$.0211, +.0069] & \textbf{+.0399 [+.0181, +.0620]} \\
D-3 $\rightarrow$ D-1 & +.0022 [$-$.0081, +.0133] & +.0112 [$-$.0106, +.0347] & +.0044 [$-$.0227, +.0310] \\
D-3 $\rightarrow$ D-2 & \textbf{+.0150 [+.0037, +.0264]} & +.0181 [$-$.0006, +.0370] & +.0211 [$-$.0023, +.0455] \\
D-1+D-2 $\rightarrow$ D-3 & +.0092 [$-$.0012, +.0195] & \textbf{+.0288 [+.0133, +.0430]} & \textbf{+.0542 [+.0336, +.0737]} \\
D-1+D-3 $\rightarrow$ D-2 & +.0055 [$-$.0054, +.0162] & \textbf{+.0176 [+.0083, +.0265]} & \textbf{+.0395 [+.0213, +.0579]} \\
D-2+D-3 $\rightarrow$ D-1 & +.0016 [$-$.0129, +.0155] & $-$.0041 [$-$.0135, +.0045] & \textbf{+.0342 [+.0121, +.0574]} \\
\addlinespace[0.3em]
\multicolumn{4}{@{}l}{\textit{disturbance}} \\
D-1 $\rightarrow$ D-2 & \textbf{+.0199 [+.0058, +.0346]} & \textbf{+.0188 [+.0036, +.0350]} & \textbf{$-$.0324 [$-$.0571, $-$.0076]} \\
D-1 $\rightarrow$ D-3 & +.0040 [$-$.0106, +.0201] & +.0054 [$-$.0048, +.0153] & $-$.0034 [$-$.0302, +.0229] \\
D-2 $\rightarrow$ D-1 & \textbf{+.0212 [+.0020, +.0408]} & +.0033 [$-$.0140, +.0214] & +.0101 [$-$.0145, +.0344] \\
D-2 $\rightarrow$ D-3 & +.0050 [$-$.0175, +.0272] & +.0101 [$-$.0162, +.0346] & $-$.0001 [$-$.0409, +.0366] \\
D-3 $\rightarrow$ D-1 & +.0121 [$-$.0226, +.0453] & +.0011 [$-$.0208, +.0242] & +.0028 [$-$.0413, +.0477] \\
D-3 $\rightarrow$ D-2 & +.0102 [$-$.0041, +.0246] & \textbf{+.0205 [+.0000, +.0416]} & +.0279 [$-$.0059, +.0640] \\
D-1+D-2 $\rightarrow$ D-3 & +.0115 [$-$.0148, +.0389] & +.0044 [$-$.0239, +.0332] & +.0158 [$-$.0162, +.0470] \\
D-1+D-3 $\rightarrow$ D-2 & $-$.0016 [$-$.0172, +.0125] & +.0070 [$-$.0075, +.0206] & +.0076 [$-$.0154, +.0286] \\
D-2+D-3 $\rightarrow$ D-1 & +.0175 [$-$.0112, +.0464] & $-$.0083 [$-$.0327, +.0162] & +.0222 [$-$.0017, +.0462] \\
\addlinespace[0.3em]
\multicolumn{4}{@{}l}{\textit{stress}} \\
D-1 $\rightarrow$ D-2 & +.0128 [$-$.0031, +.0289] & +.0184 [$-$.0006, +.0379] & +.0080 [$-$.0215, +.0354] \\
D-1 $\rightarrow$ D-3 & $-$.0010 [$-$.0115, +.0098] & $-$.0109 [$-$.0322, +.0092] & $-$.0083 [$-$.0333, +.0157] \\
D-2 $\rightarrow$ D-1 & +.0009 [$-$.0196, +.0203] & +.0028 [$-$.0087, +.0139] & $-$.0051 [$-$.0294, +.0192] \\
D-2 $\rightarrow$ D-3 & +.0081 [$-$.0062, +.0234] & \textbf{+.0129 [+.0008, +.0252]} & \textbf{+.0302 [+.0051, +.0571]} \\
D-3 $\rightarrow$ D-1 & +.0004 [$-$.0160, +.0169] & \textbf{$-$.0098 [$-$.0195, $-$.0002]} & +.0095 [$-$.0180, +.0362] \\
D-3 $\rightarrow$ D-2 & $-$.0083 [$-$.0269, +.0097] & +.0001 [$-$.0124, +.0122] & $-$.0182 [$-$.0392, +.0027] \\
D-1+D-2 $\rightarrow$ D-3 & +.0110 [$-$.0041, +.0257] & +.0034 [$-$.0092, +.0160] & +.0105 [$-$.0152, +.0374] \\
D-1+D-3 $\rightarrow$ D-2 & +.0118 [$-$.0054, +.0292] & +.0113 [$-$.0052, +.0289] & +.0103 [$-$.0089, +.0311] \\
D-2+D-3 $\rightarrow$ D-1 & +.0040 [$-$.0156, +.0231] & +.0021 [$-$.0087, +.0129] & +.0024 [$-$.0187, +.0243] \\
\addlinespace[0.3em]
\multicolumn{4}{@{}l}{\textit{valence}} \\
D-1 $\rightarrow$ D-2 & +.0028 [$-$.0132, +.0189] & +.0288 [$-$.0033, +.0602] & $-$.0103 [$-$.0532, +.0373] \\
D-1 $\rightarrow$ D-3 & $-$.0148 [$-$.0385, +.0078] & \textbf{$-$.0264 [$-$.0490, $-$.0027]} & +.0133 [$-$.0184, +.0462] \\
D-2 $\rightarrow$ D-1 & \textbf{+.0152 [+.0009, +.0306]} & +.0071 [$-$.0156, +.0299] & +.0173 [$-$.0306, +.0616] \\
D-2 $\rightarrow$ D-3 & +.0040 [$-$.0100, +.0180] & +.0093 [$-$.0109, +.0296] & +.0035 [$-$.0304, +.0351] \\
D-3 $\rightarrow$ D-1 & +.0006 [$-$.0126, +.0136] & $-$.0112 [$-$.0512, +.0363] & +.0328 [$-$.0092, +.0769] \\
D-3 $\rightarrow$ D-2 & +.0027 [$-$.0086, +.0132] & $-$.0043 [$-$.0326, +.0224] & $-$.0084 [$-$.0383, +.0204] \\
D-1+D-2 $\rightarrow$ D-3 & +.0008 [$-$.0200, +.0213] & +.0091 [$-$.0144, +.0332] & \textbf{+.0311 [+.0035, +.0600]} \\
D-1+D-3 $\rightarrow$ D-2 & +.0093 [$-$.0051, +.0232] & $-$.0092 [$-$.0289, +.0093] & +.0199 [$-$.0143, +.0549] \\
D-2+D-3 $\rightarrow$ D-1 & +.0097 [$-$.0121, +.0307] & +.0106 [$-$.0109, +.0303] & \textbf{+.0493 [+.0122, +.0878]} \\
\midrule
\multicolumn{4}{@{}l}{\textit{Aggregates over directions}} \\
arousal (all 9) & \textbf{+.0122$^{*}$ [+.0074, +.0172]} & \textbf{+.0192$^{*}$ [+.0129, +.0252]} & \textbf{+.0306$^{*}$ [+.0219, +.0395]} \\
disturbance (all 9) & \textbf{+.0111 [+.0027, +.0202]} & +.0069 [$-$.0024, +.0167] & +.0056 [$-$.0066, +.0179] \\
stress (all 9) & +.0044 [$-$.0008, +.0100] & +.0034 [$-$.0024, +.0090] & +.0044 [$-$.0050, +.0144] \\
valence (all 9) & +.0034 [$-$.0034, +.0099] & +.0015 [$-$.0097, +.0123] & \textbf{+.0165 [+.0017, +.0311]} \\
Pooled (all 36) & \textbf{+.0078 [+.0044, +.0113]} & \textbf{+.0077 [+.0036, +.0121]} & \textbf{+.0143 [+.0085, +.0204]} \\
$p_{\text{Holm}}$ (pooled) & 0.012 & 0.018 & 0.012 \\
\bottomrule
\end{tabular}
\end{table}

\begin{table}[!htbp]
\centering
\scriptsize
\setlength{\tabcolsep}{4pt}
\caption{Cross-wave paired contrasts of the family-best DG and UDA models against the strongest supervised model and against each other. Layout and notation as in Table~\ref{tab:appendix_tier_c_per_direction_contrasts}.}
\label{tab:appendix_tier_c_per_direction_contrasts_sup}
\begin{tabular}{@{}lccc@{}}
\toprule
Train $\rightarrow$ Test & Best DG $-$ best Sup & Best UDA $-$ best Sup & Best DG $-$ best UDA \\
\midrule
\multicolumn{4}{@{}l}{\textit{arousal}} \\
D-1 $\rightarrow$ D-2 & $-$.0027 [$-$.0226, +.0177] & +.0020 [$-$.0128, +.0177] & $-$.0047 [$-$.0223, +.0162] \\
D-1 $\rightarrow$ D-3 & $-$.0069 [$-$.0264, +.0142] & \textbf{+.0348 [+.0186, +.0510]} & \textbf{$-$.0417 [$-$.0607, $-$.0217]} \\
D-2 $\rightarrow$ D-1 & +.0018 [$-$.0267, +.0306] & $-$.0113 [$-$.0293, +.0064] & +.0131 [$-$.0101, +.0388] \\
D-2 $\rightarrow$ D-3 & \textbf{$-$.0375 [$-$.0631, $-$.0127]} & \textbf{$-$.0465 [$-$.0655, $-$.0268]} & +.0090 [$-$.0075, +.0259] \\
D-3 $\rightarrow$ D-1 & $-$.0022 [$-$.0298, +.0240] & +.0067 [$-$.0097, +.0236] & $-$.0089 [$-$.0326, +.0134] \\
D-3 $\rightarrow$ D-2 & $-$.0060 [$-$.0265, +.0146] & $-$.0030 [$-$.0224, +.0180] & $-$.0031 [$-$.0244, +.0189] \\
D-1+D-2 $\rightarrow$ D-3 & \textbf{$-$.0450 [$-$.0657, $-$.0246]} & \textbf{$-$.0254 [$-$.0427, $-$.0069]} & \textbf{$-$.0196 [$-$.0328, $-$.0064]} \\
D-1+D-3 $\rightarrow$ D-2 & \textbf{$-$.0340 [$-$.0503, $-$.0184]} & \textbf{$-$.0219 [$-$.0375, $-$.0072]} & \textbf{$-$.0121 [$-$.0214, $-$.0028]} \\
D-2+D-3 $\rightarrow$ D-1 & \textbf{$-$.0325 [$-$.0575, $-$.0079]} & \textbf{$-$.0383 [$-$.0600, $-$.0173]} & +.0057 [$-$.0071, +.0187] \\
\addlinespace[0.3em]
\multicolumn{4}{@{}l}{\textit{disturbance}} \\
D-1 $\rightarrow$ D-2 & \textbf{+.0523 [+.0228, +.0815]} & \textbf{+.0512 [+.0231, +.0813]} & +.0011 [$-$.0116, +.0141] \\
D-1 $\rightarrow$ D-3 & +.0074 [$-$.0175, +.0331] & +.0088 [$-$.0226, +.0393] & $-$.0014 [$-$.0185, +.0164] \\
D-2 $\rightarrow$ D-1 & +.0111 [$-$.0131, +.0349] & $-$.0068 [$-$.0361, +.0227] & +.0180 [$-$.0128, +.0466] \\
D-2 $\rightarrow$ D-3 & +.0051 [$-$.0255, +.0399] & +.0103 [$-$.0168, +.0403] & $-$.0052 [$-$.0245, +.0148] \\
D-3 $\rightarrow$ D-1 & +.0093 [$-$.0186, +.0414] & $-$.0016 [$-$.0367, +.0348] & +.0109 [$-$.0091, +.0312] \\
D-3 $\rightarrow$ D-2 & $-$.0178 [$-$.0483, +.0118] & $-$.0074 [$-$.0381, +.0220] & $-$.0104 [$-$.0261, +.0056] \\
D-1+D-2 $\rightarrow$ D-3 & $-$.0043 [$-$.0262, +.0202] & $-$.0115 [$-$.0319, +.0106] & +.0071 [$-$.0053, +.0196] \\
D-1+D-3 $\rightarrow$ D-2 & $-$.0093 [$-$.0327, +.0166] & $-$.0007 [$-$.0232, +.0240] & $-$.0086 [$-$.0235, +.0076] \\
D-2+D-3 $\rightarrow$ D-1 & $-$.0047 [$-$.0348, +.0260] & $-$.0305 [$-$.0604, +.0007] & +.0258 [$-$.0012, +.0511] \\
\addlinespace[0.3em]
\multicolumn{4}{@{}l}{\textit{stress}} \\
D-1 $\rightarrow$ D-2 & +.0047 [$-$.0180, +.0273] & +.0104 [$-$.0186, +.0396] & $-$.0057 [$-$.0228, +.0123] \\
D-1 $\rightarrow$ D-3 & +.0072 [$-$.0156, +.0311] & $-$.0026 [$-$.0280, +.0218] & +.0099 [$-$.0088, +.0292] \\
D-2 $\rightarrow$ D-1 & +.0059 [$-$.0179, +.0282] & +.0078 [$-$.0159, +.0303] & $-$.0019 [$-$.0168, +.0129] \\
D-2 $\rightarrow$ D-3 & $-$.0220 [$-$.0471, +.0031] & $-$.0172 [$-$.0397, +.0060] & $-$.0048 [$-$.0168, +.0077] \\
D-3 $\rightarrow$ D-1 & $-$.0091 [$-$.0354, +.0182] & $-$.0193 [$-$.0459, +.0072] & +.0102 [$-$.0025, +.0233] \\
D-3 $\rightarrow$ D-2 & +.0099 [$-$.0129, +.0327] & +.0183 [$-$.0071, +.0459] & $-$.0084 [$-$.0264, +.0070] \\
D-1+D-2 $\rightarrow$ D-3 & +.0005 [$-$.0271, +.0266] & $-$.0072 [$-$.0302, +.0158] & +.0076 [$-$.0086, +.0230] \\
D-1+D-3 $\rightarrow$ D-2 & +.0016 [$-$.0207, +.0219] & +.0011 [$-$.0224, +.0238] & +.0005 [$-$.0136, +.0140] \\
D-2+D-3 $\rightarrow$ D-1 & +.0015 [$-$.0226, +.0252] & $-$.0003 [$-$.0239, +.0223] & +.0019 [$-$.0149, +.0187] \\
\addlinespace[0.3em]
\multicolumn{4}{@{}l}{\textit{valence}} \\
D-1 $\rightarrow$ D-2 & +.0131 [$-$.0349, +.0579] & +.0390 [$-$.0137, +.0892] & $-$.0259 [$-$.0595, +.0070] \\
D-1 $\rightarrow$ D-3 & $-$.0280 [$-$.0675, +.0082] & \textbf{$-$.0397 [$-$.0764, $-$.0050]} & +.0116 [$-$.0098, +.0332] \\
D-2 $\rightarrow$ D-1 & $-$.0021 [$-$.0480, +.0474] & $-$.0102 [$-$.0570, +.0370] & +.0081 [$-$.0134, +.0303] \\
D-2 $\rightarrow$ D-3 & +.0005 [$-$.0315, +.0347] & +.0057 [$-$.0256, +.0376] & $-$.0053 [$-$.0215, +.0109] \\
D-3 $\rightarrow$ D-1 & $-$.0322 [$-$.0770, +.0113] & $-$.0439 [$-$.1013, +.0175] & +.0117 [$-$.0378, +.0536] \\
D-3 $\rightarrow$ D-2 & +.0111 [$-$.0166, +.0398] & +.0041 [$-$.0240, +.0303] & +.0070 [$-$.0164, +.0313] \\
D-1+D-2 $\rightarrow$ D-3 & \textbf{$-$.0303 [$-$.0584, $-$.0037]} & $-$.0220 [$-$.0483, +.0028] & $-$.0084 [$-$.0210, +.0049] \\
D-1+D-3 $\rightarrow$ D-2 & $-$.0106 [$-$.0466, +.0254] & $-$.0291 [$-$.0610, +.0014] & +.0186 [$-$.0044, +.0424] \\
D-2+D-3 $\rightarrow$ D-1 & \textbf{$-$.0396 [$-$.0815, $-$.0014]} & \textbf{$-$.0387 [$-$.0783, $-$.0023]} & $-$.0009 [$-$.0202, +.0205] \\
\midrule
\multicolumn{4}{@{}l}{\textit{Aggregates over directions}} \\
arousal (all 9) & \textbf{$-$.0183$^{*}$ [$-$.0258, $-$.0108]} & \textbf{$-$.0114$^{*}$ [$-$.0177, $-$.0051]} & \textbf{$-$.0069 [$-$.0118, $-$.0018]} \\
disturbance (all 9) & +.0055 [$-$.0044, +.0164] & +.0013 [$-$.0085, +.0121] & +.0042 [$-$.0021, +.0108] \\
stress (all 9) & +.0000 [$-$.0096, +.0087] & $-$.0010 [$-$.0105, +.0075] & +.0011 [$-$.0042, +.0071] \\
valence (all 9) & $-$.0131 [$-$.0291, +.0015] & $-$.0150 [$-$.0311, +.0003] & +.0018 [$-$.0081, +.0116] \\
Pooled (all 36) & \textbf{$-$.0065 [$-$.0122, $-$.0009]} & \textbf{$-$.0065 [$-$.0125, $-$.0009]} & +.0000 [$-$.0034, +.0036] \\
$p_{\text{Holm}}$ (pooled) & 0.120 & 0.120 & 0.955 \\
\bottomrule
\end{tabular}
\end{table}
\FloatBarrier

\flushbottom

\newpage
\section {Data Hosting, Licensing, Ethics, and Governance}
\label{appendix:hosting_ethics}

\subsection*{D.1 Data Hosting, Licensing, and Maintenance Plan}

\textbf{Host.} The dataset is hosted on Harvard Dataverse at \url{https://doi.org/10.7910/DVN/QGRVU8}. Due to the sensitive nature of the data -- including physiological signals, location traces, and communication metadata -- the dataset is released under open credentialed access. Researchers must agree to the data use agreement before downloading any files.

\textbf{Project website.}
The official EmoPhone project website is available at
\url{https://emophone.iclab.dev/}, providing a centralized
overview of the dataset and links to associated research resources.

\textbf{Code repository.} All benchmark code, preprocessing scripts, EDA notebooks, and dataset documentation are openly available at \url{https://github.com/Kaist-ICLab/EmoPhone}.

\textbf{License and access terms.} The benchmark code is released under Apache-2.0. Dataset documentation, metadata, and supporting materials are released under CC BY 4.0. Because the dataset contains sensitive human-subject mobile and wearable sensing data, the data files are released under credentialed access and governed by a dataset-specific Data Use Agreement. In the Dataverse record, CC BY 4.0 applies to the documentation and metadata, whereas access to restricted data files is controlled by the credentialed-access terms and Data Use Agreement, which prohibit re-identification attempts, redistribution to unauthorized parties, and non-research misuse.

\textbf{Repository structure.} The repository is organised as follows:

% \begin{tcolorbox}[
%   colback=gray!6,
%   colframe=gray!40,
%   arc=4pt,
%   boxrule=0.5pt,
%   left=8pt, right=8pt, top=6pt, bottom=6pt
% ]
% \small
% \dirtree{%
%   .1 CrossUserDataset/.
%   .2 README.md \DTcomment{}.
%   .2 DATASHEET.md \DTcomment{}.
%   .2 LICENSE \DTcomment{}.
%   .2 CITATION.cff \DTcomment{}.
%   .2 metadata/.
%   .3 croissant.json \DTcomment{}.
%   .2 data/.
%   .3 README.md \DTcomment{}.
%   .3 schema.md \DTcomment{}.
%   .2 preprocessing/.
%   .3 README.md \DTcomment{}.
%   .3 pipeline\_decisions.md \DTcomment{}.
%   .2 benchmark/.
%   .3 README.md \DTcomment{}.
%   .3 utils/ \DTcomment{}.
%   .3 setting\_a/ \DTcomment{}.
%   .3 setting\_b/ \DTcomment{}.
%   .3 setting\_c/ \DTcomment{}.
%   .2 eda/.
%   .3 README.md \DTcomment{}.
% }
% \end{tcolorbox}
\begin{tcolorbox}[
  colback=gray!6,
  colframe=gray!40,
  arc=4pt,
  boxrule=0.5pt,
  left=8pt, right=8pt, top=6pt, bottom=6pt
]
\small
\dirtree{%
  .1 EmoPhone/.
  .2 README.md \DTcomment{overview, benchmark ladder, data access}.
  .2 DATASHEET.md \DTcomment{datasheet for datasets}.
  .2 LICENSE \DTcomment{Apache-2.0 (code)}.
  .2 LICENSE-DATA.md \DTcomment{data access terms and DUA}.
  .2 CITATION.cff \DTcomment{machine-readable citation}.
  .2 AUTHORS.md, RESPONSIBILITY.md, MAINTENANCE.md \DTcomment{authorship and governance}.
  .2 CHANGELOG.md, CONTRIBUTING.md \DTcomment{release history and contribution guide}.
  .2 requirements.txt, environment.yml \DTcomment{pinned Python dependencies}.
  .2 docs/ \DTcomment{dataset overview, feature alignment, ethics, consent form, checklist}.
  .2 data/.
  .3 README.md \DTcomment{Dataverse download and loading examples}.
  .3 schema.md \DTcomment{column-level reference}.
  .2 metadata/ \DTcomment{Croissant metadata and checksums}.
  .3 croissant\_rai\_EmoPhone\_A\_Multi-Wave\_Dataset\_for\_In-the-Wild\_Mob.json.
  .3 checksums.md5.
  .2 preprocessing/ \DTcomment{raw sensor CSVs to feature matrices}.
  .3 README.md, pipeline\_decisions.md, feature\_alignment.md.
  .3 01\_preprocessing/ \DTcomment{app categorisation and ESM cleanup notebooks}.
  .3 02\_feature\_extraction/ \DTcomment{per-wave feature and label extraction notebooks}.
  .3 common/ \DTcomment{shared sliding-window utilities}.
  .2 benchmark/ \DTcomment{protocol specifications and result tables}.
  .3 README.md, utils/, setting\_a/, setting\_b/, setting\_c/.
  .3 results/ \DTcomment{per-setting result CSVs}.
  .2 basemodel-benchmarking/ \DTcomment{Setting A/B/C runner: baselines and tabular NNs}.
  .3 benchmark.py, cross\_dataset.py, data\_loader.py, hparams\_registry.py.
  .3 models/ \DTcomment{XGBoost, LightGBM, MLP, ResNet, TabNet, SAINT, and others}.
  .3 scripts/ \DTcomment{run\_setting\_a.sh, run\_setting\_b.sh, run\_setting\_c.sh}.
  .2 domain\_adaptation/ \DTcomment{DG and DA algorithms on a shared MLP backbone}.
  .3 models/dg/, models/da/.
  .2 EDA/ \DTcomment{dataset characterisation notebooks and loader utilities}.
  .2 images/ \DTcomment{figures used in the documentation}.
}
\end{tcolorbox}

\noindent\textbf{Reproduction entry points.} Benchmark runs are launched with the shell drivers in \texttt{basemodel-benchmarking/scripts/} (\texttt{run\_setting\_a.sh}, \texttt{run\_setting\_b.sh}, \texttt{run\_setting\_c.sh}), which call \texttt{benchmark.py} for Settings A and B and \texttt{cross\_dataset.py} for Setting C. The DG and DA families are implemented in \texttt{domain\_adaptation/models/}. Split definitions, model-selection rules, and committed result tables are documented in \texttt{benchmark/README.md} and \texttt{benchmark/results/}. The feature matrices are regenerated from the raw Dataverse archives with the notebooks in \texttt{preprocessing/}, following \texttt{preprocessing/README.md}.

\textbf{Long-term preservation.} The dataset will be maintained by the authorship team via Harvard Dataverse, which provides DOI-based persistent access. The GitHub repository will be actively maintained alongside the dataset, with updates applied as benchmark code evolves. Error corrections and documentation updates will be versioned and communicated through the repository's release notes.

\subsection*{D.2 Author Responsibility Statement}

All authors collectively bear full responsibility for the content of this paper, the EmoPhone dataset, the benchmark code, and all associated materials, including the accuracy of reported results, the completeness of dataset documentation, and the validity of the benchmark design and evaluation protocol. 

\vspace{0.5em}

We confirm that the dataset does not contain content that violates privacy rights, intellectual property rights, or other legal rights. All raw data were processed with privacy-preserving transformations prior to release: contact numbers in call and message logs were one-way hashed using MD5, Wi-Fi BSSID and Bluetooth MAC addresses were replaced with UUIDs, and GPS longitude values were randomly displaced per participant to preserve relative behavioral structure without exposing precise locations. Privacy-sensitive identifiers were transformed and access is credentialed.

\vspace{0.5em}

We confirm that all three waves of data collection (D-1, D-2, and D-3, conducted in 2020, 2021, and 2022 respectively) were conducted under protocols approved by the relevant Institutional Review Board (IRB). All participants were informed of the study's purpose, procedures, and data usage prior to enrollment, and provided written informed consent before participation. The consent form, including a clause on voluntary public dataset release, is included in the supplementary materials.

\vspace{0.5em}

We are committed to the long-term availability and maintenance of both the dataset and the benchmark code. The dataset and benchmark code will be maintained by the authorship team as described in Section D.1, with versioned updates communicated through the repository's release notes.

\subsection*{D.3 Ethics and Governance}

All data collection procedures were conducted under institutional review and informed consent. Participants joined voluntarily, received compensation, and retained the right to withdraw. To improve transparency, we include an English translation of the participant consent form in the supplementary materials. The release is intended to support responsible academic research in affective computing, with clear attention to participant protection and appropriate limits on downstream use.

\subsection*{D.4 Privacy Protection and Release}

The release follows privacy-preserving transformations at the raw-data level. Contact numbers in call and message logs were one-way hashed, Wi-Fi and Bluetooth identifiers were replaced with UUIDs, and GPS longitude was randomly displaced per participant so that relative behavioral structure is preserved without exposing precise locations. Quality-control filtering was also applied before modeling and release, including the removal of structurally invalid rows and exclusion of data streams with inadequate collection rates.

%%%%%%%%%%%%%%%%%%%%%%%%%%%%%%%%%%%%%%%%%%%%%%%%%%%%%%%%%%%%

\newpage
\section*{NeurIPS Paper Checklist}

\begin{enumerate}

\item {\bf Claims}
    \item[] Question: Do the main claims made in the abstract and introduction accurately reflect the paper's contributions and scope?
    \item[] Answer: \answerYes{} % Replace by \answerYes{}, \answerNo{}, or \answerNA{}.
    \item[] Justification: The abstract and introduction state the paper’s main contributions as a three-wave in-the-wild affect dataset, a characterization-first analysis of cross-wave and cross-user heterogeneity, and a three-setting benchmark. These claims are directly supported by the dataset overview, benchmark design, and results sections.
    % \item[] Guidelines:
    % \begin{itemize}
    %     \item The answer \answerNA{} means that the abstract and introduction do not include the claims made in the paper.
    %     \item The abstract and/or introduction should clearly state the claims made, including the contributions made in the paper and important assumptions and limitations. A \answerNo{} or \answerNA{} answer to this question will not be perceived well by the reviewers. 
    %     \item The claims made should match theoretical and experimental results, and reflect how much the results can be expected to generalize to other settings. 
    %     \item It is fine to include aspirational goals as motivation as long as it is clear that these goals are not attained by the paper. 
    % \end{itemize}

\item {\bf Limitations}
    \item[] Question: Does the paper discuss the limitations of the work performed by the authors?
    \item[] Answer: \answerYes{} % Replace by \answerYes{}, \answerNo{}, or \answerNA{}.
    \item[] Justification:  The paper includes an explicit limitations section discussing the student-centered cohort, Android-only sensing, imperfect cross-wave comparability, partial modality overlap, and the fact that richer affect labels are available only in D-3. These limitations are framed as boundaries on the claims the dataset can support.

\item {\bf Theory assumptions and proofs}
    \item[] Question: For each theoretical result, does the paper provide the full set of assumptions and a complete (and correct) proof?
    \item[] Answer: \answerNA{} % Replace by \answerYes{}, \answerNo{}, or \answerNA{}.
    \item[] Justification: This paper is empirical in nature and does not present new theoretical results requiring formal assumptions or proofs.
    % \item[] Guidelines:
    % \begin{itemize}
    %     \item The answer \answerNA{} means that the paper does not include theoretical results. 
    %     \item All the theorems, formulas, and proofs in the paper should be numbered and cross-referenced.
    %     \item All assumptions should be clearly stated or referenced in the statement of any theorems.
    %     \item The proofs can either appear in the main paper or the supplemental material, but if they appear in the supplemental material, the authors are encouraged to provide a short proof sketch to provide intuition. 
    %     \item Inversely, any informal proof provided in the core of the paper should be complemented by formal proofs provided in appendix or supplemental material.
    %     \item Theorems and Lemmas that the proof relies upon should be properly referenced. 
    % \end{itemize}

    \item {\bf Experimental result reproducibility}
    \item[] Question: Does the paper fully disclose all the information needed to reproduce the main experimental results of the paper to the extent that it affects the main claims and/or conclusions of the paper (regardless of whether the code and data are provided or not)?
    \item[] Answer: \answerYes{} % Replace by \answerYes{}, \answerNo{}, or \answerNA{}.
    \item[] Justification: The paper specifies the benchmark settings, tasks, splits, preprocessing, feature alignment, normalization, model families, training protocol, and hyperparameter search spaces in the main text and Appendix~\ref{appendix:benchmark}. The hosted dataset and benchmark code are also described in the release section.

\item {\bf Open access to data and code}
    \item[] Question: Does the paper provide open access to the data and code, with sufficient instructions to faithfully reproduce the main experimental results, as described in supplemental material?
    \item[] Answer: \answerYes{} % Replace by \answerYes{}, \answerNo{}, or \answerNA{}.
    \item[] Justification: The paper provides a hosted dataset release and a code repository, and describes the repository structure and long-term maintenance plan. The supplementary materials and dataset documentation provide the supporting release details needed to access and use the assets.
    % \item[] Guidelines:
    % \begin{itemize}
    %     \item The answer \answerNA{} means that paper does not include experiments requiring code.
    %     \item Please see the NeurIPS code and data submission guidelines (\url{https://neurips.cc/public/guides/CodeSubmissionPolicy}) for more details.
    %     \item While we encourage the release of code and data, we understand that this might not be possible, so \answerNo{} is an acceptable answer. Papers cannot be rejected simply for not including code, unless this is central to the contribution (e.g., for a new open-source benchmark).
    %     \item The instructions should contain the exact command and environment needed to run to reproduce the results. See the NeurIPS code and data submission guidelines (\url{https://neurips.cc/public/guides/CodeSubmissionPolicy}) for more details.
    %     \item The authors should provide instructions on data access and preparation, including how to access the raw data, preprocessed data, intermediate data, and generated data, etc.
    %     \item The authors should provide scripts to reproduce all experimental results for the new proposed method and baselines. If only a subset of experiments are reproducible, they should state which ones are omitted from the script and why.
    %     \item At submission time, to preserve anonymity, the authors should release anonymized versions (if applicable).
    %     \item Providing as much information as possible in supplemental material (appended to the paper) is recommended, but including URLs to data and code is permitted.
    % \end{itemize}

\item {\bf Experimental setting/details}
    \item[] Question: Does the paper specify all the training and test details (e.g., data splits, hyperparameters, how they were chosen, type of optimizer) necessary to understand the results?
    \item[] Answer: \answerYes{} % Replace by \answerYes{}, \answerNo{}, or \answerNA{}.
    \item[] Justification: The paper specifies the split design for all three benchmark settings, model families, validation protocol, normalization, feature alignment, optimizer, epoch budget, early stopping, and detailed hyperparameter search spaces in Appendix~\ref{appendix:benchmark}.
    % \item[] Guidelines:
    % \begin{itemize}
    %     \item The answer \answerNA{} means that the paper does not include experiments.
    %     \item The experimental setting should be presented in the core of the paper to a level of detail that is necessary to appreciate the results and make sense of them.
    %     \item The full details can be provided either with the code, in appendix, or as supplemental material.
    % \end{itemize}

\item {\bf Experiment statistical significance}
    \item[] Question: Does the paper report error bars suitably and correctly defined or other appropriate information about the statistical significance of the experiments?
    \item[] Answer: \answerYes{} % Replace by \answerYes{}, \answerNo{}, or \answerNA{}.
    \item[] Justification: Main tables report variability as stated in each caption. Appendix~C.1.2 describes a participant-level paired bootstrap ($B=2{,}000$) used for 95\% intervals of AUROC and paired $\Delta$AUROC and for Holm-adjusted $p$-values. Appendix~C.3 reports these intervals together with Brier score and expected calibration error. All runs use one training seed.

    % \item[] Guidelines:
    % \begin{itemize}
    %     \item The answer \answerNA{} means that the paper does not include experiments.
    %     \item The authors should answer \answerYes{} if the results are accompanied by error bars, confidence intervals, or statistical significance tests, at least for the experiments that support the main claims of the paper.
    %     \item The factors of variability that the error bars are capturing should be clearly stated (for example, train/test split, initialization, random drawing of some parameter, or overall run with given experimental conditions).
    %     \item The method for calculating the error bars should be explained (closed form formula, call to a library function, bootstrap, etc.)
    %     \item The assumptions made should be given (e.g., Normally distributed errors).
    %     \item It should be clear whether the error bar is the standard deviation or the standard error of the mean.
    %     \item It is OK to report 1-sigma error bars, but one should state it. The authors should preferably report a 2-sigma error bar than state that they have a 96\% CI, if the hypothesis of Normality of errors is not verified.
    %     \item For asymmetric distributions, the authors should be careful not to show in tables or figures symmetric error bars that would yield results that are out of range (e.g., negative error rates).
    %     \item If error bars are reported in tables or plots, the authors should explain in the text how they were calculated and reference the corresponding figures or tables in the text.
    % \end{itemize}

\item {\bf Experiments compute resources}
    \item[] Question: For each experiment, does the paper provide sufficient information on the computer resources (type of compute workers, memory, time of execution) needed to reproduce the experiments?
    \item[] Answer: \answerYes{} % Replace by \answerYes{}, \answerNo{}, or \answerNA{}.
    \item[] Justification: The paper reports the hardware used for neural and tree-based models, including GPU type and memory.

    % \item[] Guidelines:
    % \begin{itemize}
    %     \item The answer \answerNA{} means that the paper does not include experiments.
    %     \item The paper should indicate the type of compute workers CPU or GPU, internal cluster, or cloud provider, including relevant memory and storage.
    %     \item The paper should provide the amount of compute required for each of the individual experimental runs as well as estimate the total compute. 
    %     \item The paper should disclose whether the full research project required more compute than the experiments reported in the paper (e.g., preliminary or failed experiments that didn't make it into the paper). 
    % \end{itemize}
    
\item {\bf Code of ethics}
    \item[] Question: Does the research conducted in the paper conform, in every respect, with the NeurIPS Code of Ethics \url{https://neurips.cc/public/EthicsGuidelines}?
    \item[] Answer: \answerYes{} % Replace by \answerYes{}, \answerNo{}, or \answerNA{}.
    \item[] Justification: The paper states that all three collection waves were conducted under IRB-approved protocols with informed consent, voluntary participation, compensation, and the right to withdraw. It also documents privacy-preserving transformations, credentialed access, and governance measures for release.

    % \item[] Guidelines:
    % \begin{itemize}
    %     \item The answer \answerNA{} means that the authors have not reviewed the NeurIPS Code of Ethics.
    %     \item If the authors answer \answerNo, they should explain the special circumstances that require a deviation from the Code of Ethics.
    %     \item The authors should make sure to preserve anonymity (e.g., if there is a special consideration due to laws or regulations in their jurisdiction).
    % \end{itemize}

\item {\bf Broader impacts}
    \item[] Question: Does the paper discuss both potential positive societal impacts and negative societal impacts of the work performed?
    \item[] Answer: \answerYes{} % Replace by \answerYes{}, \answerNo{}, or \answerNA{}.
    \item[] Justification: The paper discusses intended research value for affective computing and responsible academic use, while also acknowledging privacy, comparability, and misuse-related concerns through the ethics, governance, privacy-protection, and limitations discussions. These sections clarify both the potential utility of the resource and the need for bounded claims and responsible downstream use.

\item {\bf Safeguards}
    \item[] Question: Does the paper describe safeguards that have been put in place for responsible release of data or models that have a high risk for misuse (e.g., pre-trained language models, image generators, or scraped datasets)?
    \item[] Answer: \answerYes{} % Replace by \answerYes{}, \answerNo{}, or \answerNA{}.
    \item[] Justification: The release uses credentialed access rather than unrestricted public download, and the paper documents privacy-preserving transformations applied to GPS, call/message identifiers, Wi-Fi/Bluetooth identifiers, and quality-control filtering before release. These measures are intended to reduce misuse risk while preserving research utility.

    % \item[] Guidelines: 
    % \begin{itemize}
    %     \item The answer \answerNA{} means that the paper poses no such risks.
    %     \item Released models that have a high risk for misuse or dual-use should be released with necessary safeguards to allow for controlled use of the model, for example by requiring that users adhere to usage guidelines or restrictions to access the model or implementing safety filters. 
    %     \item Datasets that have been scraped from the Internet could pose safety risks. The authors should describe how they avoided releasing unsafe images.
    %     \item We recognize that providing effective safeguards is challenging, and many papers do not require this, but we encourage authors to take this into account and make a best faith effort.
    % \end{itemize}

\item {\bf Licenses for existing assets}
    \item[] Question: Are the creators or original owners of assets (e.g., code, data, models), used in the paper, properly credited and are the license and terms of use explicitly mentioned and properly respected?
    \item[] Answer: \answerNA{} % Replace by \answerYes{}, \answerNo{}, or \answerNA{}.
    \item[] Justification: The paper does not introduce or redistribute third-party assets as part of its contribution. Prior datasets, methods, and software packages are cited for comparison, implementation reference, or background, but are not released as assets with this paper.

    % \item[] Guidelines:
    % \begin{itemize}
    %     \item The answer \answerNA{} means that the paper does not use existing assets.
    %     \item The authors should cite the original paper that produced the code package or dataset.
    %     \item The authors should state which version of the asset is used and, if possible, include a URL.
    %     \item The name of the license (e.g., CC-BY 4.0) should be included for each asset.
    %     \item For scraped data from a particular source (e.g., website), the copyright and terms of service of that source should be provided.
    %     \item If assets are released, the license, copyright information, and terms of use in the package should be provided. For popular datasets, \url{paperswithcode.com/datasets} has curated licenses for some datasets. Their licensing guide can help determine the license of a dataset.
    %     \item For existing datasets that are re-packaged, both the original license and the license of the derived asset (if it has changed) should be provided.
    %     \item If this information is not available online, the authors are encouraged to reach out to the asset's creators.
    % \end{itemize}

\item {\bf New assets}
    \item[] Question: Are new assets introduced in the paper well documented and is the documentation provided alongside the assets?
    \item[] Answer: \answerYes{} % Replace by \answerYes{}, \answerNo{}, or \answerNA{}.
    \item[] Justification: The paper releases a new dataset resource and benchmark code, and documents them through the main paper, Appendix A, the dataset documentation file, schema descriptions, supplementary study documents, and the repository/data-hosting plan. The paper also states that informed consent included a clause on voluntary public dataset release.
    % \item[] Guidelines:
    % \begin{itemize}
    %     \item The answer \answerNA{} means that the paper does not release new assets.
    %     \item Researchers should communicate the details of the dataset\slash code\slash model as part of their submissions via structured templates. This includes details about training, license, limitations, etc. 
    %     \item The paper should discuss whether and how consent was obtained from people whose asset is used.
    %     \item At submission time, remember to anonymize your assets (if applicable). You can either create an anonymized URL or include an anonymized zip file.
    % \end{itemize}

\item {\bf Crowdsourcing and research with human subjects}
    \item[] Question: For crowdsourcing experiments and research with human subjects, does the paper include the full text of instructions given to participants and screenshots, if applicable, as well as details about compensation (if any)? 
    \item[] Answer: \answerYes{} % Replace by \answerYes{}, \answerNo{}, or \answerNA{}.
    \item[] Justification: The supplementary materials include translated participant-facing study documents, including the consent form and orientation materials, and the paper states that participants received compensation. The study procedures were conducted under IRB-approved protocols with informed consent. 

    % \item[] Guidelines:
    % \begin{itemize}
    %     \item The answer \answerNA{} means that the paper does not involve crowdsourcing nor research with human subjects.
    %     \item Including this information in the supplemental material is fine, but if the main contribution of the paper involves human subjects, then as much detail as possible should be included in the main paper. 
    %     \item According to the NeurIPS Code of Ethics, workers involved in data collection, curation, or other labor should be paid at least the minimum wage in the country of the data collector. 
    % \end{itemize}

\item {\bf Institutional review board (IRB) approvals or equivalent for research with human subjects}
    \item[] Question: Does the paper describe potential risks incurred by study participants, whether such risks were disclosed to the subjects, and whether Institutional Review Board (IRB) approvals (or an equivalent approval/review based on the requirements of your country or institution) were obtained?
    \item[] Answer: \answerYes{} % Replace by \answerYes{}, \answerNo{}, or \answerNA{}.
    \item[] Justification: The paper states that all three waves were conducted under IRB-approved protocols, that participants provided informed consent, and that privacy-sensitive data were transformed before release. The ethics and governance sections also describe the release risks and the measures used to mitigate them.

    % \item[] Guidelines:
    % \begin{itemize}
    %     \item The answer \answerNA{} means that the paper does not involve crowdsourcing nor research with human subjects.
    %     \item Depending on the country in which research is conducted, IRB approval (or equivalent) may be required for any human subjects research. If you obtained IRB approval, you should clearly state this in the paper. 
    %     \item We recognize that the procedures for this may vary significantly between institutions and locations, and we expect authors to adhere to the NeurIPS Code of Ethics and the guidelines for their institution. 
    %     \item For initial submissions, do not include any information that would break anonymity (if applicable), such as the institution conducting the review.
    % \end{itemize}

\item {\bf Declaration of LLM usage}
    \item[] Question: Does the paper describe the usage of LLMs if it is an important, original, or non-standard component of the core methods in this research? Note that if the LLM is used only for writing, editing, or formatting purposes and does \emph{not} impact the core methodology, scientific rigor, or originality of the research, declaration is not required.
    %this research? 
    \item[] Answer: \answerNA{} % Replace by \answerYes{}, \answerNo{}, or \answerNA{}.
    \item[] Justification: LLMs are not an important, original, or non-standard component of the dataset construction, benchmark design, or empirical methodology reported in this paper.

    % \item[] Guidelines:
    % \begin{itemize}
    %     \item The answer \answerNA{} means that the core method development in this research does not involve LLMs as any important, original, or non-standard components.
    %     \item Please refer to our LLM policy in the NeurIPS handbook for what should or should not be described.
    % \end{itemize}

\end{enumerate}

\end{document}